\documentclass[final,3p,times]{elsarticle}

\usepackage{amssymb}
\usepackage{amsthm}
\usepackage{amsmath}
\usepackage{longtable}
\usepackage{graphicx}
\usepackage{epsfig}
\usepackage{amsfonts}
\usepackage{amsmath}
\usepackage{bm}
\usepackage{setspace}
\usepackage{graphicx}
\usepackage{overpic}
\usepackage{color}
\usepackage{multirow}
\usepackage{url}
\usepackage{array}
\usepackage{braket}
\usepackage{threeparttable}
\usepackage{booktabs}
\usepackage{makecell}
\usepackage{longtable}
\usepackage{nomencl}
\usepackage{longtable}
\usepackage[nonumberlist,toc,section,acronym]{glossaries}
\usepackage{natbib}

\newacronym{dof}{DOF}{degrees of freedom}
\newacronym{stm}{STM}{scanning tunneling microscope}
\newacronym{afm}{AFM}{atomic force microscope}
\newacronym{mcbj}{MCBJ}{mechnically controlled break junction}
\newacronym{qd}{QD}{quantum dot}
\newacronym{fdt}{FDT}{fluctuation-dissipation theorem}
\newacronym{dos}{DOS}{density of states}
\newacronym{si}{SI}{International System of Units}
\newacronym{negf}{NEGF}{non-equilibrium Green's function}
\newacronym{qme}{QME}{quantum master equation}
\newacronym{heom}{HEOM}{hierarchical equations of motion}
\newacronym{seom}{SEOM}{stochastic equation of motion}
\newacronym{dft}{DFT}{density functional theory}
\newacronym{tddft}{TDDFT}{time-dependent density functional theory}
\newacronym{tdcdft}{TDCDFT}{time-dependent current-density functional theory}
\newacronym{dftb}{DFTB}{density-functional tight-binding}
\newacronym{zcc}{ZCC}{zero-current condition}
\newacronym{fdr}{FDR}{fluctuation-dissipation relation}
\newacronym{sthm}{SThM}{scanning thermal microscopy}
\newacronym{mpc}{MPC}{minimal perturbation condition}
\newacronym{sem}{SEM}{scanning electron microscope}
\newacronym{stem}{STEM}{scanning tunneling electron microscopy}
\newacronym{uhv}{UHV}{ultra-high-vacuum}
\newacronym{sjem}{SJEM}{scanning Joule expansion microscopy}
\newacronym{eels}{EELS}{electron energy loss spectroscopy}
\newacronym{bls}{BLS}{Brillouin light scattering}
\newacronym{yig}{YIG}{yttrium-iron-garnet}
\newacronym{qsd}{QSD}{quantum state diffusion}
\newacronym{mosfet}{MOSFET}{metal-oxide-semiconductor field-effect transistor}
\newacronym{bec}{BEC}{Bose-Einstein condensation}

\newcommand{\PreserveBackslash}[1]{\let\temp=\\#1\let\\=\temp}
\newcolumntype{C}[1]{>{\PreserveBackslash\centering}p{#1}}
\newcolumntype{R}[1]{>{\PreserveBackslash\raggedleft}p{#1}}
\newcolumntype{L}[1]{>{\PreserveBackslash\raggedright}p{#1}}
\newcommand{\Sec}[1]{Sec.\,\ref{#1}}

\newcommand{\be}{\begin{equation}}
\newcommand{\ee}{\end{equation}}
\newcommand{\bea}{\begin{eqnarray}}
\newcommand{\eea}{\end{eqnarray}}
\newcommand{\Fig}[1]{Fig.~\ref{#1}}
\newcommand{\Figure}[1]{Figure~\ref{#1}}

\newcommand{\Eq}[1]{Eq.\,(\ref{#1})}
\newcommand{\Eqs}[1]{Eqs.\,(\ref{#1})}

\newcommand{\la}{\langle}
\newcommand{\ra}{\rangle}

\newcommand{\dg}{\dagger}

\newcommand{\w}{\omega}
\newcommand{\ep}{\epsilon}

\newcommand{\tC}{\tilde{C}}

\newcommand{\Hs}{H_{_{\rm S}}}
\newcommand{\Hb}{H_{_{\rm B}}}
\newcommand{\Hsb}{H_{_{\rm SB}}}
\newcommand{\Ht}{H_{_{\rm T}}}

\newcommand{\trs}{{\rm tr}_{_{\rm S}}}
\newcommand{\trb}{{\rm tr}_{_{\rm B}}}
\newcommand{\trt}{{\rm tr}_{_{\rm T}}}
\newcommand{\rhos}{\rho_{_{\rm S}}}

\newcommand{\rhob}{\rho_{_{\rm B}}}
\newcommand{\rhot}{\rho_{_{\rm T}}}
\newcommand{\subb}{{_{\rm B}}}
\newcommand{\subs}{{_{\rm S}}}
\newcommand{\subt}{{_{\rm T}}}

\newcommand{\hF}{\hat{F}}
\newcommand{\hA}{\hat{A}}
\newcommand{\hB}{\hat{B}}

\journal{Physics Reports}

\makeglossaries

\begin{document}

\title{Local Temperatures Out of Equilibrium}

\author[label1]{Daochi Zhang}


\author[label1]{Xiao Zheng}
\ead{xz58@ustc.edu.cn}

\author[label2]{Massimiliano Di Ventra}

\address[label1]{Hefei National Laboratory for Physical Sciences at the Microscale
\& Synergetic Innovation Center of Quantum Information and Quantum Physics
\& CAS Center for Excellence in Nanoscience,
University of Science and Technology of China, Hefei, Anhui 230026, China}

\address[label2]{Department of Physics, University of California, San Diego, 9500 Gilman Drive, La Jolla, CA 92093-0319}

\cortext[cor1]{Corresponding author at: Hefei National Laboratory for Physical Sciences at the Microscale,
University of Science and Technology of China, Hefei, Anhui 230026, China}


\begin{abstract}
The temperature of a physical system is operationally defined in physics as ``that quantity which is measured by a thermometer'' weakly coupled to, and at equilibrium with the system. This definition is unique only at global equilibrium in view of the zeroth law of thermodynamics: when the system and the thermometer have reached equilibrium, the ``thermometer degrees of freedom'' can be traced out and the temperature read by the thermometer can be uniquely assigned to the system.  Unfortunately, such a procedure cannot be straightforwardly extended to a system out of equilibrium, where local excitations may be spatially inhomogeneous and the zeroth law of thermodynamics does not hold. With the advent of several experimental techniques that attempt to extract a single parameter characterizing the degree of local excitations of a (mesoscopic or nanoscale) system out of equilibrium, this issue is making a strong comeback to the forefront of research. In this paper, we will review the difficulties to define a unique temperature out of equilibrium, the majority of definitions that have been proposed so far, and discuss both their advantages and limitations.
We will then examine a variety of experimental techniques developed for measuring the non-equilibrium local temperatures under various conditions.
Finally we will discuss the physical implications of the notion of local temperature, and present the practical applications of such a concept in a variety of nanosystems out of equilibrium.
\end{abstract}

\begin{keyword}

local temperature; nanosystem; non-equilibrium; thermodynamics

\end{keyword}

\date{\today}

\maketitle

\tableofcontents

\section{Introduction} \label{sec:Intro}

\subsection{Plan of the review} \label{subsec:pla}

\subsubsection{Scope of the review} \label{subsubsec:scope}

Temperature is one of the most fundamental properties of physical systems.
In fact, the concept of temperature reaches far beyond the realm of physics,
and profoundly influences our daily life in many aspects.
The importance of temperature for living systems (such as the human body) is self-evident.
For example, studies on biological processes at the molecular level have demonstrated
sensitive dependence of protein function on temperature \cite{Zei04871,He1626737}.
A small fluctuation in temperature may then significantly affect the protein structure,
causing it to lose its normal bioactivity.

Biosystems and living organisms are certainly interesting. However, what really motivates this review is the
tremendous advancements, we have witnessed in the past decade or so, in the experimental techniques for fabricating \cite{Rac15271,Sun15980,Kop14780},
measuring \cite{Mad157249}, and manipulating \cite{Kel17336,Lap041135,Lee17} objects at the nanometer scale. These techniques allow nanoscale systems to be probed out of equilibrium, often far from equilibrium, and attempt to characterize the state of non-equilibrium with one or a few parameters.
One such parameter is the ``local temperature'' of the system. However, unlike its equilibrium counterpart, it is clear that
the notion of temperature in nanoscale systems, hence its local definition, and in contexts in which this concept is not so solidly grounded, requires
further investigation and a critical assessment.

The scope of this review is precisely to provide an overview of this concept, its
various experimental incarnations, its strengths and
limitations. Although many of the ideas discussed in this review apply to general
systems out of equilibrium, for the sake of definiteness we will mostly consider nanoscale systems as test beds.
Therefore, to complete this introduction, let us provide a bird's-eye view of typical physical situations in which the notion of a temperature out of equilibrium may emerge in these systems.

Let us first recall that one of the main goals of nanotechnology is to design and manufacture devices at the nanoscale with desirable
properties and functions \cite{Ven04}. Typically, the specific function of a nano-device is realized
via the response of a certain local physical property to an external field,
which drives the nanosystem out of equilibrium.
Therefore, the action of the external field on the nanosystem results in a certain form of work
that serves a practical purpose.
However, besides the ``desired'' response, other internal \gls{dof}
may also be excited by the external field or through the coupling between the internal \gls{dof},
which inevitably leads to accumulation of heat within the system.
If the locally produced heat cannot be dissipated efficiently into the surrounding environment,
the ``undesired'' local motion may hamper the function of the nanosystem
or even imperil its structural stability.

\begin{figure}[htbp]
  \centering
  \includegraphics[width=0.75\columnwidth]{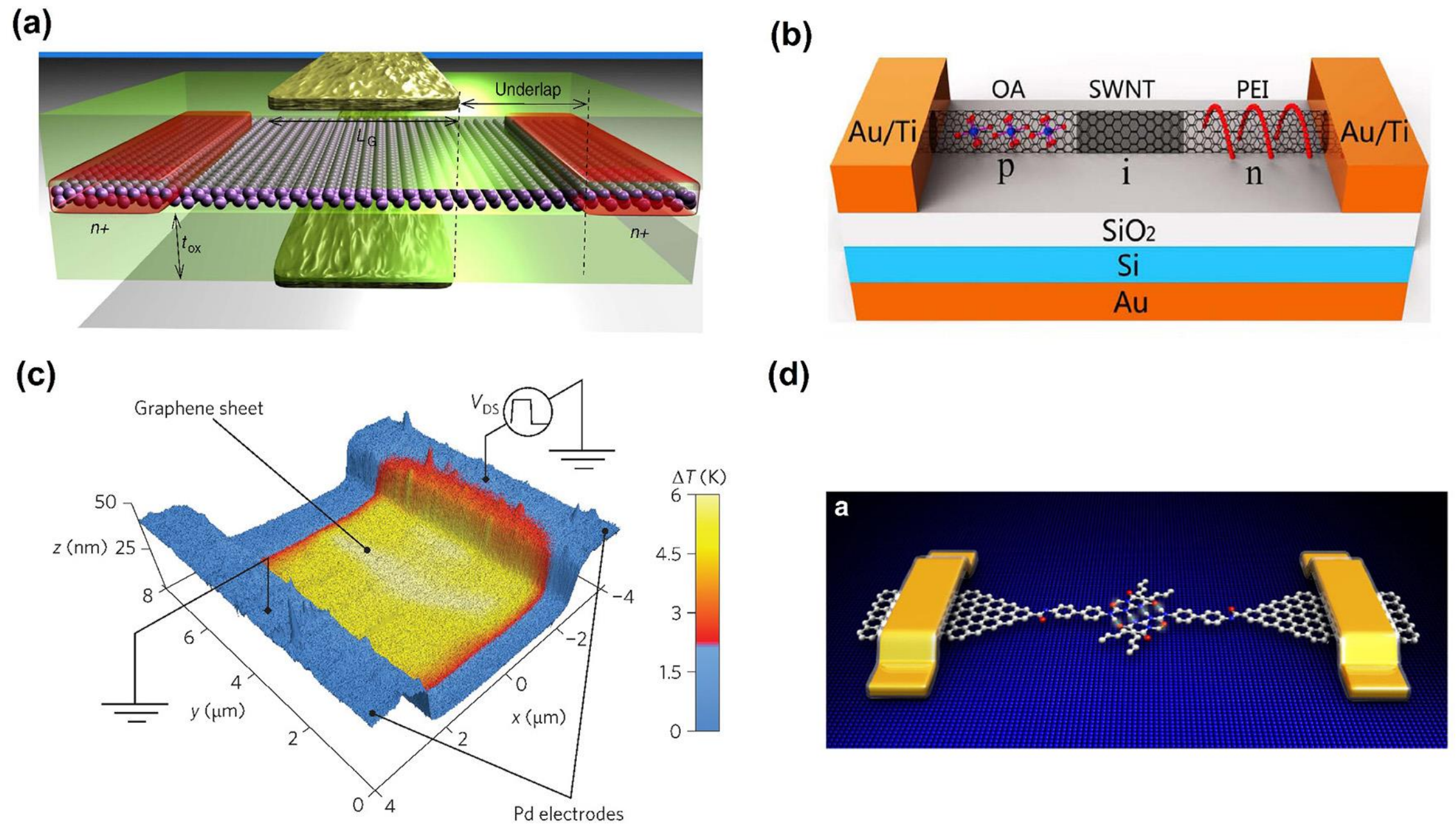}
  \caption
  {
  Schematic representation of various types of nanoelectronic devices with different nano-sized materials
   as the electron conduction channel.
   (a) A \gls{mosfet}. Reprinted with permission from \cite{Piz1612585}. Copyright 2016 Springer Nature.
   (b) A carbon-nanotube-based transistor. Reprinted with permission from \cite{Che17280}. Copyright 2017 Elsevier.
   (c) A graphene-sheet-based transistor. Reprinted with permission from \cite{Gro11287}. Copyright 2011 Springer Nature.
   (d) A molecular transistor. Reprinted with permission from \cite{Zho18807}. Copyright 2018 Springer Nature.
   }
  \label{fig:nanojunction}
\end{figure}

In order to provide the reader with a practical example where these effects are particularly pressing we consider the important case of our modern
computers built out of a collection of transistors \cite{Pat9751}.
The increasing demand for computational performance has spurred the development
and proposal of novel materials and techniques to increase the switch frequency of transistors in computer processors.
Nowadays, the miniaturization of transistors has reached the size of a few nanometers.
Under such a small scale, the local physical properties of a material may differ drastically
from those of its bulk counterpart \cite{Mor17195162}, and quantum effects may be prominent as the size of
the transistor is comparable to the mean free path of conduction electrons.
Moreover, the construction of novel types of transistors has gone beyond silicon-based materials \cite{Yam171589,Mar15640}.
For instance, electronic devices based on single molecules have also been proposed \cite{Rat13378,Avi74277}.
Related experiments can be traced back to the conductivity measurements on monolayers of cadmium salts of fatty acids
conducted by Kuhn and coworkers in the 1970s \cite{Man714398}.
A nanoelectronic or molecular electronic device usually has the geometry of a nanojunction,
which consists of a nano-sized material that provides the electron conduction channel,
and two or more macroscopic leads that play the role of electron reservoir and heat bath.
Figure\,\ref{fig:nanojunction} depicts various types of nanoelectronic devices,
including a \gls{mosfet} \cite{Piz1612585},
a carbon-nanotube-based transistor \cite{Che17280}, a graphene-sheet-based transistor \cite{Ins1185,Xia11179,Gro11287},
and a molecular transistor \cite{Ree97252,Tao06173,Ara13399,Zho18807}.

Although these devices vary in size and structure, one expects that due to the large current densities that flow in the device,
local heating may develop~\cite{Di08}.
Physically, local heating arises due to the electron-electron and electron-phonon scattering processes
both within the nano-sized conduction channel and at the contact interfaces between the channel and the leads.
In particular, when a nanoelectronic device is operating under an applied voltage,
the local phonon modes are excited by the conduction electrons via electron-phonon scattering.
Consequently, the device experiences local Joule heating.
For a computer chip that integrates billions of nanoscale transistors, the local Joule heating
becomes a major bottleneck that limits the performance and efficiency of the device.
Moreover, the intensive current-induced heating also threatens the stability of the device,
and may even lead to breakdown of the nanostructure if the local heating is too severe \cite{Hua061240}.

A current-carrying nanoelectronic device is a typical non-equilibrium nanosystem,
yet the concept of ``local temperature'' has been adopted, in an intuitive sense,
to describe or characterize the local heating effect.
For instance, it has been argued that a smaller external force is needed to break a nanojunction
with a higher local temperature.
Based on this idea, experimental measurements on the local heating in nanojunctions
have been conducted \cite{Eva01105,Smi04S472,Tsu055188}.
For instance, Huang \emph{et~al.} have investigated the local heating in a single molecule
covalently bonded to two electrodes by a \gls{mcbj} \cite{Hua061240}.
By measuring the average breakdown force of the \gls{mcbj}, the local temperature was estimated.
These authors found that at a bias voltage of about $1$\,V, the temperature could be raised to $\sim 30$\,K above the ambient room temperature.
Moreover, the development of \gls{stm} and \gls{afm} techniques has
greatly enriched the experimental means of probing the local temperature.
In an AFM, an atomically sharp tip mounted at the free end of a cantilever
can sense the thermal expansion of a sample,
which provides the technical basis for a scanning Joule expansion microscope \cite{Var9837}.
Thermal imaging has also been realized by measuring the variation of Joule expansion signal
with the cantilever position.
Using an AFM-based instrument and techniques, Grosse \emph{et~al.} have been able to measure the
spatial distribution of local temperature in a graphene-sheet-based transistor \cite{Gro11287}.
More experimental works will be introduced in Section\,\ref{sec:Experiment} of this review.

Local heating is also important for other types of nanosystems.
The 2016's Nobel Prize in chemistry was awarded to Sauvage, Stoddart and Feringa,
for the design and synthesis of molecular machines \cite{Nob16}.
Like a traditional electric motor, a molecular motor operates in an out-of-equilibrium state
and converts a certain form of energy into mechanical work.
However, because of the small size of molecules and limited energy relaxation channels,
it may be rather challenging to efficiently dissipate the heat generated during the operation of the motor
into the environment, so as to sustain the functionality of the molecular machine.

Besides local heating, external sources (voltage, temperature gradient,
electromagnetic fields, mechanical force, etc.) may affect the physical
properties and functions of a nanosystem in many other aspects.
Being out of equilibrium, the local properties of a nanosystem may
deviate significantly from their equilibrium values,
which are likely manifested by a change in local temperature.
In particular, the quantum features of a nanosystem,
including quantum coherence, quantum entanglement and quantum correlation,
are profoundly influenced by the local thermal excitations in the system.
Therefore, local temperature, which by intuition measures the degree of
local thermal excitations, would also be an ideal quantity for characterizing
the local properties of quantum origin.

In the past decade, there have been enormous experimental efforts on realizing the
precise control and manipulation of local quantum states in nanoscopic materials.
These efforts have opened new horizons for nanotechnology, which may lead to
important applications in spintronics \cite{Wol011488,Aws131174,Mic104463,Man15871},
quantum storage \cite{Lvo09706,Wol17060502},
quantum information processing \cite{Van1734,Mau1613575,Vel171766},
and quantum computation \cite{Van1734}.

\begin{figure}[htbp]
  \centering
  \includegraphics[width=0.8\columnwidth]{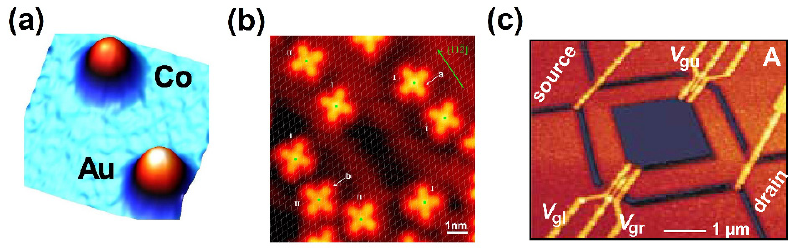}
  \caption { Various forms of QDs:
  (a) Au and Co atoms adsorbed on a Cu(100) surface. Reprinted with permission from \cite{Chi12266803}. Copyright 2012 American Physical Society.
  (b) FePc molecules adsorbed on an Au(111) surface. Reprinted with permission from \cite{Gao07106402}. Copyright 2007 American Physical Society.
  (c) Two-dimensional electron gas formed at the interface of an AlGaAs/GaAs heterostructure. Reprinted with permission from \cite{Wie002105}. Copyright 2000 AAAS.
  }
  \label{fig:QD}
\end{figure}

Take the \gls{qd} as an example. A \gls{qd} is an artificial object that has discretized energy levels, and which is typically coupled to macroscopic leads which serve as electron reservoirs.
Experimentally, a \gls{qd} can be realized by a magnetic atom \cite{Chi12266803}, an organometallic molecule with transition metal centers \cite{Gao07106402},
and a two-dimensional electron gas at the interface of semiconductors \cite{Kou0133} (see Figure~\ref{fig:QD}).
The local charge or spin state in a \gls{qd} can be utilized as a quantum bit (qubit) \cite{Los98120},
a fundamental element of quantum computers.
Because of the finite size of a \gls{qd}, the on-dot electrons are subjected to strong Coulomb repulsion.
The electron-electron interaction and the dot-lead coupling give rise to diversified exotic phenomena,
such as Coulomb blockade \cite{Par02722}, spin excitations \cite{Han071217}, quantum memristive effect \cite{Per11145}, and Kondo effect \cite{Hew93}.
Some of these phenomena, like the Kondo effect, occur only at sufficiently low temperatures.

\begin{figure}[t]
  \centering
  \includegraphics[width=0.6\columnwidth]{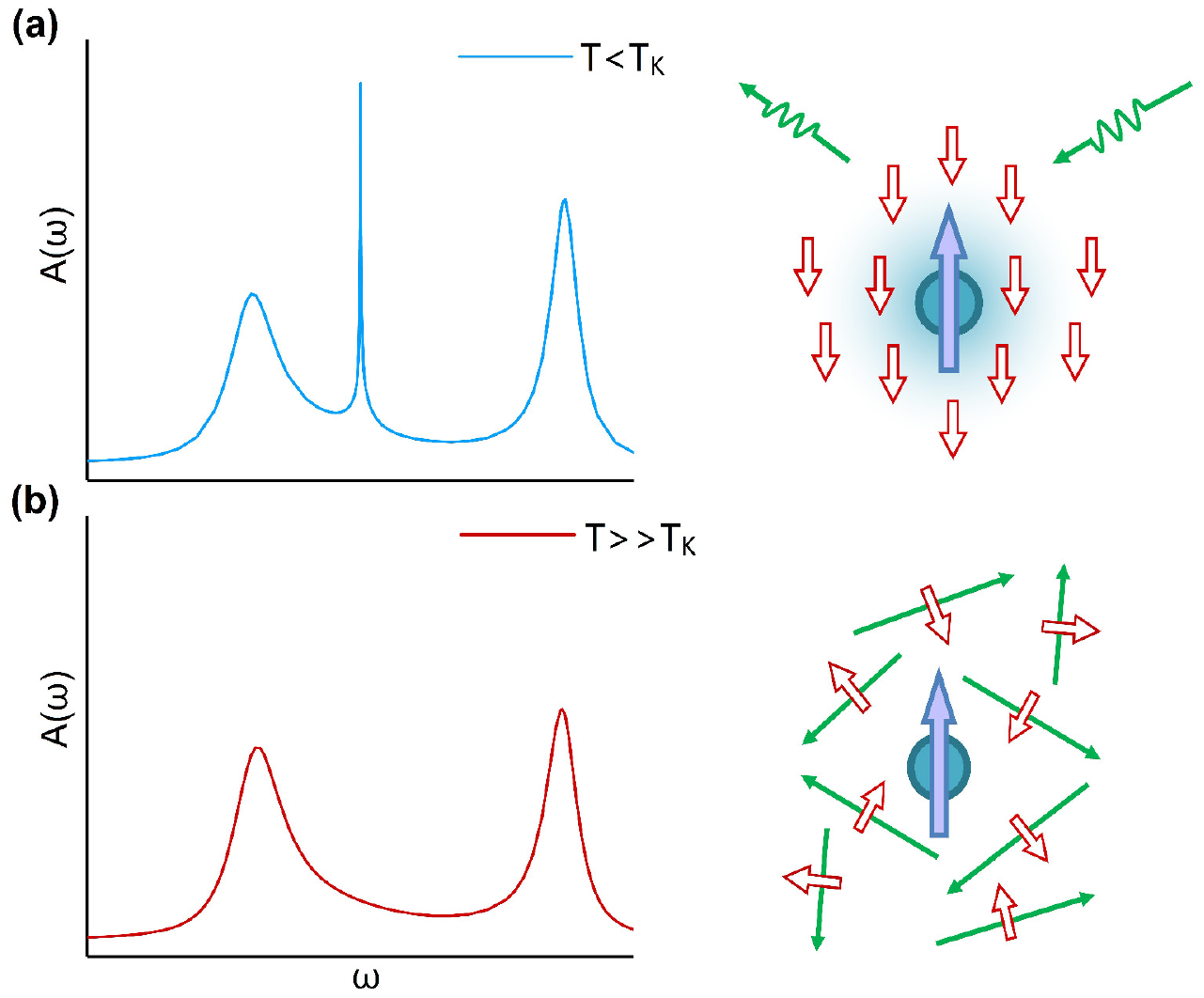}
  \caption {
  (a) Impurity spectral function $A(\w)$ of a single-level Anderson impurity model (sketched on the left panel)
  at a temperature lower than the characteristic Kondo temperature ($T < T_{\rm K}$).
  The sharp peak centered at the Fermi energy of the leads is the Kondo resonance peak.
  Schematic illustration on the right shows that the Kondo state is formed
  with the local spin moment at the impurity, screened by the spins of conduction electrons in the leads.
  (b) $A(\w)$ of the same model (left panel) at a temperature that is much higher than the Kondo temperature ($T \gg T_{\rm K}$),
  and the Kondo resonance peak is absent.
  As illustrated in the right panel, the enhanced thermal motion of conduction electrons destroys the Kondo state.}
  \label{fig-kondo2}
\end{figure}

Kondo phenomena are a characteristic signature of strong electron correlations.
Theoretically, the Kondo state in a \gls{qd} can be described by a quantum impurity model
introduced by Anderson \cite{And6141}, in which the \gls{qd} is represented
by a number of discretized energy levels and the metallic leads are approximated by reservoirs of free electrons.
When the background temperature is lower than the Kondo temperature ($T < T_{\rm K}$),
the spectral function of the \gls{qd} exhibits a sharp resonant peak at the Fermi energy of the leads,
corresponding to the Kondo state that originates from the screening of the local spin
by the spins of the surrounding free electrons; see Figure\,\ref{fig-kondo2}(a).
In contrast, at a high temperature ($T > T_{\rm K}$) the central resonant peak is absent,
because the Kondo screening is destroyed by the enhanced thermal fluctuations; see Figure\,\ref{fig-kondo2}(b).
It has been found both experimentally \cite{Gro08246601} and theoretically \cite{Gol985225} that
the relation between the zero-bias conductance of a Kondo \gls{qd}, $G$,
and the equilibrium temperature $T$ obeys the following scaling relation:
\be
 G(T) = G_0 \left[ 1 + \left(\frac{T}{T_{\rm K}}\right)^2 \left(2^{1/s} - 1 \right) \right]^{-s},
 \label{G-Kondo-1}
\ee
where $G_0$ is the conductance at $T=0\,$K, and $s$ is a parameter depending only on the
total spin of the \gls{qd}.

For both practical and fundamental reasons, it is important to learn how the quantum features of nanosystems vary
in response to external perturbations.
For instance, for a \gls{qd}, it has been observed in experiments that the Kondo correlation is suppressed
by an applied magnetic field \cite{Cro98540} or a bias voltage \cite{Kog04166602,Ama05045308},
leading to broadening or splitting of the Kondo peak in the measured differential conductance spectra.
It has also been proposed that the non-equilibrium properties of a Kondo \gls{qd} can be
described by universal scaling relations \cite{Wan13035129}.
For instance, Rosch \emph{et~al.} \cite{Ros01156802} have discovered that for a \gls{qd} under a high bias voltage $V$,
the Kondo physics is governed by the scales $V$ and $\gamma$, where $\gamma \sim V/\ln^2(V/T_{\rm K})$
is the non-equilibrium decoherence rate induced by the voltage-driven current.
In view of the simple form of \Eq{G-Kondo-1}, it is intriguing to ask whether it is possible to
define a local temperature (denoted as $T^\ast$) for a non-equilibrium \gls{qd}, so that
it can be used in lieu of the equilibrium temperature $T$ in \Eq{G-Kondo-1},
to achieve a unified and preferably more convenient description of non-equilibrium properties.

From a statistical perspective, a properly defined local temperature $T^\ast$ should be able to
characterize the magnitude of local excitations and the local distribution of particles
inside a non-equilibrium nanosystem, and thus quantify its deviation from equilibrium.
This is particularly important for the studies on precise control or fine-tuning of local quantum states in nanostructures.
For instance, in the past few years, the competition between Kondo resonance and local spin excitations
in magnetic molecular QDs has become a focus of experimental efforts \cite{Wan182418}.
Parks \emph{et~al.} have demonstrated that, by stretching a molecular \gls{qd} embedded in an \gls{mcbj},
the Kondo conductance peak is gradually split by enhancing the spin-orbit coupling  \cite{Par101370}.
Hiraoka \emph{et~al.} have used an \gls{stm} tip as a probe to tune the local quantum states
in an adsorbed FePc molecule \cite{Hir1716012}. By varying the position of the tip, they have
realized a crossover between the Kondo-dominant regime and the spin-excitation-dominant regime.
These studies have paved the way for the realization and design of more sophisticated quantum-state control techniques.

{\it Quantum thermodynamics} is expected to be a useful method to
explore the underlying relations between different response functions.
However, it is still a developing field with a considerable gap between theoretical predictions and experiments \cite{Mer1720}.
The concept of local temperature is thus of critical importance for bridging this gap,
provided that such a quantity can be measured reliably in experiments.

The above examples, which represent only a small fraction of a large body of studies, clearly show that it would be highly desirable to extend the notion of temperature from thermal equilibrium to out of equilibrium.
However, such an extension is not at all trivial, because the zeroth law of thermodynamics that defines the
temperature for equilibrium systems cannot be applied directly to non-equilibrium situations.
Moreover, the ``unique'' aspects of nanosystems, such as their small sizes, inhomogeneous structures, quantum features,
and the embedded environment, all add to the challenge of defining a local temperature.

The goal of this review is thus to survey the existing theoretical and experimental efforts
to do just so, with the focus on the following basic questions:

\begin{enumerate}
  \item
     How do we define a local temperature for systems out of equilibrium?

  \item
     How do we measure it?

  \item
     What is its physical meaning?
\end{enumerate}

Logically speaking, a question 0 before the above three should be addressed:
does a local temperature even exist for a non-equilibrium system?

\subsubsection{Organization of the review} \label{subsubsec:org}

In the remainder of Section\,\ref{sec:Intro} we shall give a historical account on the measurement of temperature
for systems in thermal equilibrium,
which will be followed by addressing the challenges in the definition and measurement of local temperature in non-equilibrium systems.
We will finish this section by further elaborating on the physical motivations to extend
the concept of local temperature out of equilibrium.

In Section\,\ref{sec:Def}, we will review the existing theoretical definitions of local temperatures for nanoscopic systems out of equilibrium.
We will begin by describing the basic ideas and physical considerations behind these definitions,
which generally fall into several categories.
We will then elaborate on a number of theoretical definitions by discussing their advantages and limitations,
particularly focusing on the practical feasibility of the experimental realization of these definitions.
We will finish this section by discussing the uniqueness/non-uniqueness of the local temperature
in non-equilibrium situations, as well as the underlying connections, if they exist, among the different definitions.

In Section\,\ref{sec:Experiment}, we will review the experimental techniques that have been proposed or implemented to determine the local temperatures of non-equilibrium systems. We will start with the general principles of the experimental strategies. In particular, we will discuss the difficulties in realizing a nano-sized thermometer via measuring the heat current flow. Then we will review a number of representative experimental works that have been carried out on measuring the local temperature of different types of systems and for various non-equilibrium situations. The strengths and limitations of these works will be analyzed.

In Section\,\ref{sec:Imp&App}, we will focus on the practicality and usefulness of the concept of local temperature.
We will first discuss the physical interpretations of the measured local temperatures, along with their implications on the extension of the thermodynamic laws to non-equilibrium scenarios, and on the quantitative correspondence between the equilibrium and non-equilibrium systems.
We will then review the applications of local temperature in various emerging fields.

In Section\,\ref{sec:Sum}, we will conclude the review by summarizing the contents presented in the previous sections.
Finally, we will provide future perspectives by discussing the ongoing activities in related fields,
addressing the remaining open questions, and suggesting possible directions for future research.

\subsection{Temperature at equilibrium} \label{subsec:equ}

\subsubsection{The measurement of temperature} \label{subsubsec:MeaofTeq}

In early times the standard of temperature was simply our bodily sensation of ``hot'' and ``cold'' \cite{Cha04}.
In the 16th and 17th centuries, a more quantitative way of measuring this sensation was established. It is believed that Galileo Galilei invented the first ``thermoscope'',
a tubular device in which a liquid rises and falls as the temperature changes due to
the variation in liquid density \cite{Mul07,Cha04}.
Using the thermoscope, scientists were then able to understand basic thermal phenomena,
such as the distinction between physiological senses and physical properties \cite{Mul07}.
The thermoscope then gradually evolved into the modern thermometer,
with the addition of a scale in the early 17th century, and its standardization through the 17th and 18th centuries \cite{Ben84}.

The early temperature measurements depended on various empirical physical phenomena,
and different temperature scales had been introduced to determine the quantitative value of temperature \cite{Cha04}.
The second law of thermodynamics selects the definition of an absolute thermodynamic temperature,
unique up to an arbitrary scale factor \cite{Atk06}.
Until 2018, the kelvin, the base unit of thermodynamic temperature in the \gls{si},
was defined as the fraction $1/273.16$ of the thermodynamic temperature of the triple point of water.
On November~16, 2018, a new definition was adopted,  by which the Boltzmann's constant is fixed
to the exact value of $k_{\rm B} = 1.380\,649\times10^{-23}\,{\rm J}\cdot{\rm K}^{-1}$.
The advantage of the new definition is obvious: it is independent of any particular substance,
and thus provides a universal standard for any measurement method \cite{Sto19022001}.

The thermodynamical meaning of temperature is given by the zeroth law of thermodynamics
proposed by Fowler and Guggenheim \cite{Fow39} in 1939, as follows:
\begin{quote}
 \textsl{If two assemblies are each in thermal equilibrium with a third assembly, they are in thermal equilibrium with each other.}
\end{quote}
The zeroth law states the transitivity of thermal equilibrium and hence provides a solid physical basis for a thermometer.
From an operational point of view, {\it temperature is defined as that quantity measured by
a thermometer coupled to the system with which it reaches thermal equilibrium}.
Precisely because the thermometer plus the system reach a state of global equilibrium when coupled,
the quantity that is measured by the thermometer is then attributed to the system in the limit of weak coupling and
negligible heat capacity of the thermometer.

Because of the inextricable connection between the concept of temperature and that of heat in thermodynamics,
the measurement of the two quantities is closely related to each other.
In practice, the heat production $Q$ is usually determined by counting the amount of latent heat released
during a phase transition process, or by tracing the temperature change $\Delta T$
of a substance with known heat capacity $C$, \emph{i.e.},
\be
  C = \frac{Q}{\Delta T},   \label{Q-T-1}
\ee
where $C$ is assumed to be constant if the temperature change $\Delta T$ is sufficiently small \cite{Atk06}.
However, this assumption must be used with caution for nanosystems with a small number of \gls{dof},
because the heat capacity may depend rather sensitively to the microscopic details
(such as the electronic structure) of the system and may thus vary significantly upon the small change in temperature \cite{Lin08075133}.

In the conventional experiments carried out on bulk materials
it is reasonable to assume that the measurement is conducted in an adiabatic manner, \emph{i.e.},
the heat dissipated into the surrounding environment is negligibly small as compared to the heat collected by the
thermometer or calorimeter.
However, the situation can be quite different for nanosystems.
This is because the number of energy transfer channels diminishes with the reduced size or dimension of the system,
while the energy dissipation between the system and the environment may be enhanced by quantum coherence. 
In this sense, energy dissipation has been regarded as a main characteristic that distinguishes quantum from classical phenomena \cite{Hal16407}.
Such a distinct difference should be carefully taken into account in the definition and actual measurement of a local temperature.

More detailed discussions on thermometers at the nanoscale and experimental measurement of local temperature
will be given in \Sec{subsec:mueasurement-local-temp}.

\subsubsection{Finite-size effects} \label{subsubsec:eff}

Small is different \cite{Hod07639}.
In fact, because of the finite \gls{dof} of nanosystems, a careful definition of local thermodynamical quantities is important,
to preserve as much as possible their original thermodynamical meaning in macroscopic bulk systems,
and to retain a correct asymptotic behavior as the system approaches the thermodynamical limit.
Experiments also face new challenges. The thermal probe or, in general, the detector of local observables should be made
comparable in size to the target object to ensure the measurement is truly local,
and the coupling between the probe and the system should be kept sufficiently weak,
to avoid any significant perturbation to the system's intrinsic structure or thermal state.
Moreover, as the system size diminishes, the fluctuations of the local observables increase as $1/\sqrt{N}$ \cite{Kha03},
with $N$ being the number of \gls{dof}.
Thus, an appropriate time average or ensemble average
is needed to minimize the uncertainty of the measured quantity.

\begin{figure}[htbp]
  \centering
  \includegraphics[width=0.95\columnwidth]{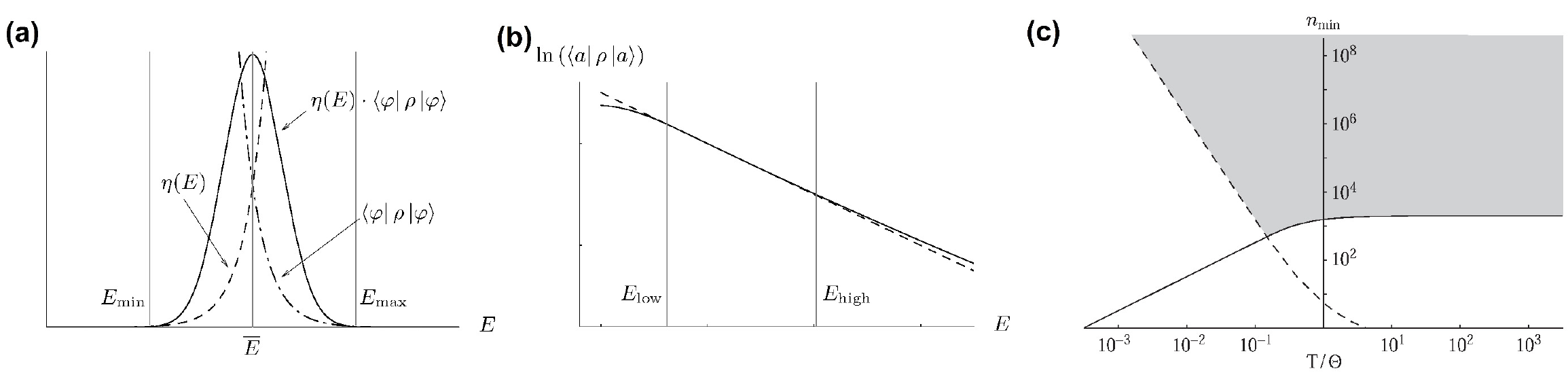}
  \caption
  {
  (a) The product of the density of states $\eta(E)$ times the global distribution function
   $\la \varphi |\rho|\varphi \ra$ forms a strongly pronounced peak at the expectation value of the global system energy $\bar{E}$.
  (b) The logarithm of the local distribution function $\la a | \rho | a \ra$  (solid line)
  and the logarithm of a canonical distribution (dashed line) with the same local temperature for a harmonic chain.
  (a) and (b) are reprinted with permission from \cite{Har04066148}. Copyright 2004 American Physical Society;
  (c) Minimal scale $n_{\rm min}$ as a function of temperature $T$ for a harmonic chain.
  Local temperature exists in the shaded regions in (c).
  $\Theta$ is the Debye temperature, a characteristic constant of the respective material.
  Reprinted with permission from \cite{Har0566}. Copyright 2005 Elsevier.
  }
  \label{fig:Tloc-1}
\end{figure}

So, can a system be too small to even define a local temperature?
To address this question,
from the perspective of statistical mechanics,
Hartmann \emph{et~al.} have investigated the minimal length scales for the existence of a local temperature
in a periodic lattice involving short-range interactions \cite{Har04613,Har04080402,Har04066148,Har05579,Har0689}.
They divided a large quantum system in global thermal equilibrium into a group of equivalent subsystems,
and analyzed at what size a local temperature exists in each individual subsystem.
In the global system, the \gls{dos} exponentially grows as a function of energy \cite{Har0689},
while the canonical distribution function $\la \varphi | \rho | \varphi \ra$ (with $|\varphi \ra$ being an eigenstate of the global Hamiltonian and the global equilibrium state $\rho$)
is an exponentially decaying function of energy. The product of these two functions forms
a strongly pronounced peak at the expectation value of the global energy $\bar{E}$; see Figure\,\ref{fig:Tloc-1}(a).

According to Hartmann \emph{et~al.}, the only requirement for the existence of local temperature
is that the local distribution function $\la a | \rho | a \ra$
(with $|a\ra$ being a tensor product of eigenstates of the individual subsystems) is canonically distributed
in an appropriate energy range around $\bar{E}$; see Figure\,\ref{fig:Tloc-1}(b).
In other words, the local distribution function should reproduce the line shape of the global distribution function,
so that any local observable that can be evaluated from the former will agree with the true value extracted from the latter.
%
Using this criterion, these authors have estimated the minimal length scale over which a local temperature exists
in a harmonic chain.
Multiplying $n_{\rm min}$ by a lattice constant $a_0$, namely the distance between the neighboring atoms,
the minimal length scale on which  temperature can exist in real materials is given by $l_{\rm min} = n_{\rm mmin} a_0$.
The results of two representative examples, crystalline silicon and carbon nanotubes,
have been shown in Figures (8) and (9) of \cite{Har0689}.
%
From their estimation the minimal length can be quite large at a somewhat low global temperature.
Clearly, Hartmann \emph{et~al.}'s definition of local temperature relies on the existence of a
local Boltzmann distribution. It is applicable to systems in a global thermal equilibrium state,
with possible extension to situations in which temperature gradients are small.

From the perspective of thermodynamics,
Kliesch \emph{et al.} have attempted to determine the minimal length scales for the existence of a local temperature
in a spin- and fermionic-lattice with short-range interactions \cite{Kli14031019}.
For a large system in a global equilibrium state divided into a group of subsystems with various length scales,
the zeroth law shows that \cite{San15085007}:
(1) there exists a thermal equilibrium state which is characterized by a single parameter called temperature;
(2) the temperature is local, namely, each subsystem of the global system is in a thermal state with its own local temperature;
(3) the temperature is an intensive quantity, meaning that if the global system is in equilibrium, all the subsystems have the same temperature.
Nevertheless, when the interactions present in a global system are non-negligible,
the state of a subsystem will be strongly perturbed by its environment, namely, the rest of the global system,
and will deviate from its equilibrium thermal state with a specific value of the global temperature.
Thus, temperature is not intensive for a global system with strong interactions.

To address under what scale temperature is an intensive property,
Kliesch \emph{et al.} focused on a subsystem with a specific length scale.
According to Kliesch \emph{et al.}'s work,
temperature is an intensive quantity in the subsystem with a given length scale,
if and only if the interactions are negligible compared to $k_BT^*$ with $T^*$ being the local temperature of the subsystem.
In other word,
the minimal scale of local temperatures is related to the decay of the interactions.
When the interactions between the subsystem and its environment become vanishingly weak,
the subsystem is homogenous inside, and is nearly independent of the rest of the global system.
Kliesch \emph{et~al.}'s condition for the existence of local temperature is applicable to systems with sufficiently weak interactions.

De Pasquale \emph{et al.} have defined a size-dependent quantity, the local quantum thermal susceptibility, 
to quantify the accuracy of local temperature measurement on quantum systems \cite{Pas1612782,Pal17052115}. 
The local quantum thermal susceptibility increases with the size of the measured subsystem. 
When the subsystem coincides with the whole system, 
it reaches the maximum value and thus corresponds to the highest achievable accuracy for the local temperature measurement.

If the size of a system is comparable to or even smaller than the wavelength or mean free path of its constituent particles,
quantum coherence effects will come into play, and the motion of a particle is more appropriately described by the propagation of its wave-function.
This gives rise to important consequences.

First, the wave-function of a particle in the system may extend into the surrounding environment.
Because of the system-environment entanglement, the local distribution function of the particles often
deviates significantly from the global distribution function, \emph{i.e.},
the equilibrium state of the system may not be described by a local Boltzmann distribution.
In such cases, a rigorous treatment of the system-environment entanglement is essential for
defining a local temperature and understanding the thermodynamic behavior of the system.

Second, as already mentioned in \Sec{subsubsec:MeaofTeq},
energy dissipation is one important characteristic that distinguishes quantum from classical phenomena.
In classical systems, energy dissipation is always one-way and irreversible;
whereas in the quantum regime, the system-environment entanglement may greatly prolong the time scale
of energy dissipation, and occasionally even results in temporary revival of dissipated energy back into the system.
Further remarks on the influence of the environment will be given in \Sec{subsubsec:flu}.

Third, the microscopic mechanism of energy exchange largely determines how the energy is distributed over
the constituent particles or internal \gls{dof} of the system.
It is well-known that, for a classical system in the linear response regime,
the electric current density $i$ through the system driven by an electric field $E$ obeys Ohm's law \cite{Gri99},
\be
  i = \sigma E,  \label{ohm-law}
\ee
with $\sigma$ being the electrical conductivity.
Similarly, the heat current density $j$
driven by a temperature gradient $\nabla T$ follows the Fourier's law \cite{Lie11},
\be
  j = -\lambda \nabla T, \label{fourier-law}
\ee
with $\lambda$ being the thermal conductivity.

The physical mechanism behind Ohm's law is the incoherent or
phase-breaking scattering processes between the conduction electrons
and the local phonons~\cite{Di08}.
The same incoherent scattering events also lead to Joule heating in a conductor \cite{Di08}.
Moreover, the conventional mechanisms that lead to the Fourier's law, such as collisions between particles
and vibrational motions of atoms, are also incoherent.

However, in the case of a nano- or molecular conductor whose size is smaller than the
electron coherence length (mean free path for the quantum phase),
the electron transport process can be fully coherent due to the absence of
phase-breaking scattering events inside the conductor~\cite{Lan57223,But861761}.
In the coherent limit, both the Ohm's law and Fourier's law are no longer valid.
Interestingly, it has been found that Fourier's law can be reconstructed by introducing
disorders in the conductor, because disorder leads to dephasing \cite{Dub09115415}.
The crossover from the fully coherent regime (invalid Fourier's law) to
incoherent (valid Fourier's law) regime is characterized by a thermal length scale,
which is directly related to the profile of the local temperature \cite{Dub09042101}.


\subsubsection{Influence of the environment } \label{subsubsec:flu}

When a nanosystem is embedded in or adsorbed onto a macroscopic bulk material, the latter serves as an environment of the former.
For instance, in the case of a \gls{qd} coupled to a metal lead, the lead plays the role of heat bath and electron reservoir of the \gls{qd}.
The environment usually possesses a much larger number of \gls{dof} than the system. Therefore, when the two are put into contact
and form a composite system, the thermalization process will ultimately lead to a global thermal equilibrium state,
with the global temperature $T$ and global chemical potential $\mu$ determined largely by those of the environment.
The spontaneous thermal fluctuations within the environment always exist and they have crucial influence on the dynamics of the system \cite{Kub66255}.
Such influence can be understood by considering the classical Brownian motion of pollen grains in solution \cite{Ein05549}.
The thermal fluctuations of the surrounding solvent molecules result in random impacts on a pollen grain.
These random impacts generally cause two kinds of effects at the same time:
firstly, they act as a random driving force to maintain the irregular motion of the pollen grain;
and secondly, they give rise to a frictional force to dissipate away the energy of the pollen grain and thus slow down its motion.
In thermal equilibrium, these two opposite effects of environmental fluctuations reach a certain form of balance,
which is usually described by a quantitative relation, the \gls{fdt} \cite{Kub66255}, we will discuss in the following.

Similar effects are also generally found in open quantum systems, in which the environmental fluctuations
govern the system's dissipative dynamics~\cite{Gra88115,Wei08}. Besides the exchange of particles and energy,
the environmental fluctuations also cause decoherence through the phase-breaking system-environment interactions,
which makes the quantum system behave more classical-like.

For a general system-plus-environment composite, the total Hamiltonian $\Ht$ consists of three parts:
\be
  \Ht = \Hs + \Hb + \Hsb,  \label{Htotal-1}
\ee
where $\Hs$, $\Hb$ and $\Hsb$ are the Hamiltonian of the system, the Hamiltonian of the environment (heat bath or particle reservoir),
and the system-environment coupling Hamiltonian, respectively.
The specific formulation of $\Hs$, $\Hb$ and $\Hsb$ depends on the details of the composite system under study.

The dynamics of the system in the presence of the environment is generally characterized by the reduced density matrix
$\rhos(t) \equiv \trb[\rhot(t)]$.
The time evolution of $\rhos$ generally follows the Nakajima-Zwanzig \gls{qme}
as follows \cite{Nak58948,Zwa601338,Zwa641109} (hereafter, we adopt the atomic units, \emph{i.e.}, $\hbar \equiv 1$ and $k_{\rm B} \equiv 1$):
\begin{align}
\dot\rhos(t) = -i [\Hs ,\rhos(t)] - \int^{t}_{t_0} d\tau \tC (t - \tau) \rhos(\tau).
\label{rhos-Nakajima-Zwanzig}
\end{align}
Here, the first term on the right-hand side describes the
intrinsic system dynamics,
while the second term accounts for the influence of the environment on the system dynamics.
Clearly, the second term has the form of a convolution integral.
Consequently, $\rhos$ at the present time $t$ depends explicitly on $\rhos(\tau < t)$, namely the whole history of its evolution.

The two-time environment correlation function $\tC(t-\tau) \equiv \la \hF^\dag(t) \hF(\tau) \ra_\subb$
serves as the non-Markovian memory kernel.
Here,  $\la \hat{O} \ra_\subb \equiv \trb ( \hat{O}\rhob^{\rm eq})$
represents expectation value taken for the equilibrium bath,
with $\trb$ denoting the trace over all the environment's \gls{dof},
and $\hat F(t) \equiv e^{i \Hb t} \hat F e^{-i \Hb t}$ with $\hat F$ being a bath operator.
For bath environments that satisfy Gaussian statistics,
such as harmonic oscillator baths, non-interacting phonon baths
or non-interacting electron reservoirs,
the two-time correlation function $\tC(t-\tau)$ completely characterizes
the influence of the environment \cite{Yan05187}.

If the longest correlation time scale of the bath dynamics $\tau_\subb$ is shorter than the characteristic time scale of the system Hamiltonian $\tau_\subs$
\emph{i.e.}, $\tau_\subb \ll \tau_\subs$, the influence of the environment fluctuations on the reduced system dynamics
is almost instantaneous. This is referred to as the Markovian limit, in which the memory effect of the environment is negligible \cite{Spo80569}.
On the contrary, if $\tau_\subb > \tau_\subs$ holds, the system's state at the present time will depend strongly on the history
of the dynamics, and such a situation is called non-Markovian \cite{Bre16021002}.
In the non-Markovian regime, the memory effect of the environment is important and quantum coherence effects become prominent.
The non-Markovian memory affects both the transient behavior and the asymptotic long-time limit of the reduced system dynamics \cite{Bre16021002}.

Traditionally, the derivation of a \gls{qme} often involves
various approximations to the dissipative process between the system and the environment.
For instance, the Born approximation gives a perturbative treatment of $\Hsb$,
while the Markovian approximation neglects any long-time component of $\tC (t-\tau)$ \cite{Yan05187,Wei08}.
These approximations have led to simplified \gls{qme}s, such as the Redfield theory \cite{Red5719,Red651}.
Another frequently used \gls{qme} is the Lindblad equation \cite{Lin76119,Gor76821},
which has the advantage that the resulting $\rhos$ is positive semidefinite \cite{Ali87}.

In the past two decades, formally exact quantum dissipative theories have been developed.
These include the deterministic \gls{heom} \cite{Tan89101,Tan906676,Yan04216,Xu05041103,Ye16608}
and the stochastic theories such as the quantum state diffusion theory \cite{Dio882885,Dio97569,Gis925677,Per99}
and the stochastic equation of motion theory \cite{Sto02170407,Sha045053,Moi13134106,Han19050601}.
Interested readers may refer to review articles, such as Refs. \cite{Bre16021002,Ye16608,Veg17025001}.

The memory kernel is of central importance to the influence of the environment,
as it directly reflects the \gls{fdt}.
For a boson bath at thermal equilibrium, the memory kernel satisfies the following relationship \cite{Yan05187}:
\be
  \tC(t - \tau) = \frac{1}{\pi}\int_{-\infty}^\infty d\w \, e^{- i \w (t-\tau) } f(\w) J(\w).
  \label{fdt-fermi}
\ee
Here, $f(\w) \equiv 1 + f^{\rm BE}(\w)$, with $f^{\rm BE}(\w) = 1/ [e^{\, \beta(\w - \mu)} - 1 ]$ being the Bose--Einstein distribution function,
$\beta = 1/T$ is the inverse temperature,
$\mu$ is the chemical potential of the environment, and $J(\w)$ is the bath linewidth or hybridization function.

Equation \eqref{fdt-fermi} is a manifestation of the \gls{fdt}, because $\tC (t-\tau)$ on the left-hand side
quantifies the statistical features of the random fluctuations in the bath \cite{Sha045053,Sto982657,Han19050601},
while the linewidth function on the right-hand side characterizes the rate of energy dissipation
between the system and the bath for any energy $\w$.
If the environment is a fermionic bath, such as a reservoir of free electrons,
an \gls{fdt} that is formally analogous to \Eq{fdt-fermi} can be derived \cite{Yan05187}.

%

\begin{figure}[t]
  \centering
  \includegraphics[width=0.6\columnwidth]{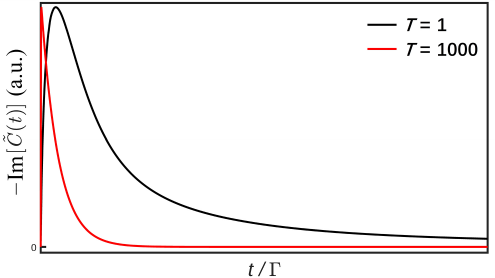}
  \caption{
   The imaginary part of $\tC(t)$ of a free-electron bath at high (red) and low (black) temperatures.
   The two lines are scaled to have the same maximum. The bath hybridization function assumes a Lorentzian form of
   $J(\w) = \Gamma W^2/(\w^2 + W^2)$ with $W = 20$.
   $\Gamma$ is the strength of system-bath coupling, and $W$ is the bandwidth of the electron bath.
   All energies are in units of $\Gamma$.
   } \label{fig:corr}
\end{figure}

The memory time of the environment $\tau_\subb$ as well as the non-Markovianity of the reduced system dynamics
depend substantially on the background temperature $T$.
For a fermionic environment such as a metal lead, it is the imaginary part of $\tC(t)$
that varies with $T$.
Figure\,\ref{fig:corr} depicts ${\rm Im}[\tC(t)]$ of a free-electron reservoir.
As it is evident, the reservoir correlation function exhibits a much slower decay at low temperatures.
From the Fourier transform relation of \Eq{fdt-fermi}, the long-time behavior of $\tC(t)$
is closely related to the line shape of the Fermi--Dirac distribution function in the low-energy range.
Recently, Cui \emph{et~al.} have shown that $\tau_\subb \propto 1/T$ at low temperatures \cite{Cui19}.
In the zero-temperature limit, $\tau_\subb \rightarrow \infty$, and overall
$\tC(t)$ decays as $1/t$ in the long-time regime.
This implies that the non-Markovian memory is crucially important for many exotic quantum phenomena
in nanosystems, such as the Kondo resonance and spin excitations in systems,
which are prominent only at sufficiently low temperatures.
%


\subsubsection{Local equilibrium conditions}  \label{subsubsec:localEquilibrium}

The concept of local equilibrium can be traced back to non-equilibrium bulk materials.
For a nanosystem coupled to a macroscopic environment, the definition of local temperature is
usually based on the assumption that the nanosystem is in some form of local equilibrium state,
even when the whole composite is out of equilibrium \cite{Eng811151}.
For a macroscopic system in a global thermal equilibrium state,
there is no net flow of energy or particles anywhere inside the system,
and so the temperature and chemical potential are uniform
in any subregion of the total system.

For near-equilibrium systems, a useful assumption is that in each sufficiently small region of space
the gradients of intensive thermodynamic functions are small
and their local values vary relatively slowly compared with those of the global system \cite{hor09}.
This idea leads to a local equilibrium condition, which conceptually states that
a large system can be spatially and temporally divided into subsystems of small sizes, and within a subsystem the net flow of energy or particles is negligibly small.
Consequently, a common local temperature and a common local chemical potential can be defined for each subsystem.

It should be emphasized that if the local subsystem is allowed to exchange particles with its surrounding,
it requires both the local temperature and local electrochemical potential
to characterize the local equilibrium state as well as the associated local thermodynamics functions.
In 1981 Engquist and Anderson \cite{Eng811151} have proposed a local equilibrium condition from the perspective of the zeroth law of thermodynamics,
by putting forward a definition of ideal potentiometer and thermometer.
They considered a local subregion of a current-carrying system
that is weakly coupled to a macroscopic probe. The local temperature
and local electrochemical potential of the subsystem are well-defined
only when the net electrical and heat currents flowing through the probe are zero:
%
\begin{align}
  I_p &= 0,  \label{zcc-i} \\
  J_p &= 0,  \label{zcc-j}
\end{align}
where $I_p$ and $J_p$ are the net electrical current and heat current through the macroscopic probe, respectively. Therefore, the probe serves as an ideal potentiometer/thermometer.
Equations \eqref{zcc-i} and \eqref{zcc-j} emphasize that in local equilibrium there is no net energy and particle flow
between the local subsystem and its environment.
The zero-current conditions have been extensively used to determine the local temperatures
and local electrochemical potentials of various types of non-equilibrium systems.
We will provide a more extensive discussion in \Sec{subsubsec:zcc}.

From the statistical point of view, the local equilibrium condition for a classical system means that
the statistical distribution of canonical states inside a subsystem closely approaches the Boltzmann distribution
and hence classical thermodynamic functions and equations remain valid \cite{Kei87,Dem14}.

In \Sec{subsubsec:eff}, we have given an example on the minimal size of a subsystem required for
the existence of a local Boltzmann distribution; see Figure\,\ref{fig:Tloc-1}.
However, we should point out that there exist a vast amount of nanosystems,
such as molecules, nanotubes and few-electron \gls{qd}s,
whose sizes are distinctly smaller than that required to establish a local Boltzmann distribution.
So, what are the theoretical criteria and experimental means that can be used
to confirm the existence of local equilibrium in such ``ultrasmall'' systems?

The canonical density matrix of a system-plus-environment composite in its global equilibrium state is
\be
 \rhot^{\rm eq}(\beta) = \frac{e^{-\beta\Ht}}{Z},
 \label{rhot-eq-1}
\ee
where $Z = \trt ( e^{-\beta\Ht})$ is the partition function of the composite, and $\trt$ denotes the trace over all the system-plus-bath \gls{dof}.
In general, the system reduced density matrix does not follow the local Boltzmann distribution \cite{Con0615},
\be
 \rhos^{\rm eq} = \frac{ \trb\left(e^{-\beta\Ht}\right)}{Z} \neq \frac{e^{-\beta \Hs} }{ \trs\left({e^{-\beta \Hs}}\right) },
 \label{rhos-eq-2}
\ee
even at the weak system-environment coupling limit.
The same inequity in \Eq{rhos-eq-2} also holds for a grand-canonical total density matrix $\rhos^{\rm eq}(\beta, \, \mu)$.
This is because the system and environment Hamiltonians, $\Hs$ and $\Hb$,
usually do not commute, and $\rhot$ cannot be described as a product
of independent system and environment states due to the system-environment entanglement.
For instance, Moix \emph{et~al.} \cite{Moi12115412} have applied a stochastic
quantum dissipation theory approach to compute $\rhos^{\rm eq}$
of several exciton-transfer-related complexes, including the Fenna--Matthews--Olson protein
and the light harvesting complex of purple bacteria, LH2. They have found that
the reduced density matrix, $\rhos^{\rm eq}$, of an LH2 complex at $100\,$K is almost completely delocalized,
whereas the coherence length of the exact $\rhos^{\rm eq}$ is drastically reduced due to the coupling
to the thermal bath. The difference becomes more important when either the temperature is lowered or the system-bath
coupling increases.

Although local equilibrium cannot be verified by $\rhos$, its existence can still be revealed
by examining the detailed balance relation via the system's dynamical variables \cite{Ver94301,Ale111538,Ego00101,Fow55}.
In practice, this idea has been adopted in some high-energy physics experiments
to measure the temperature of a plasma by using x-ray Thomson scattering \cite{Dop09182,Val188432},
and it may be potentially useful for nanosystems of our primary interest as well.

Let us consider a fundamental question for non-equilibrium systems:
how does an external field affect the dynamical variables of the systems?
In the linear-response regime, the response of a system observable $\hA$ at time $t$
to an external perturbation $\hB$ at prior time $\tau$ is generally characterized by the response function
\begin{align}
\tilde\chi_{AB}(t-\tau) \equiv i \la [\hA(t),\hB(\tau)] \ra_\subt = i \left\{ \la \hA(t)\hB(\tau) \ra_\subt-\la\hB(\tau)\hA(t) \ra_\subt \right\} =i\left\{ \tC_{AB}(t-\tau)-\tC_{BA}(\tau-t) \right\}.
\end{align}
Here, $\hat O(t) \equiv e^{i\Ht t} \hat O e^{-i\Ht t}$,
and $\tC_{AB}(t-\tau) \equiv \la \hA(t) \hB(\tau) \ra_\subt = \trt[\hA(t) \hB(\tau)\rhot^{\rm eq}]$ is the two-time
correlation function of the unperturbed system at thermal equilibrium.

Response and correlation functions are of great importance to spectroscopy \cite{Muk95}
and many-body physics \cite{Mah00}. For instance,
the emission and absorption of light and the transport of charge carriers are often described
in terms of Green's functions, which are essentially correlation functions between particle creation and annihilation operators.

Let $C_{AB}(\w) \equiv \frac{1}{2}\int_{-\infty}^\infty dt\,e^{i\w t} \tC_{AB}(t)$. This quantity satisfies the detailed-balance relation as follows \cite{Yan05187}:
\begin{align}
  \frac{C_{BA}(-\omega)}{C_{AB}(\omega)} = e^{-\beta \omega}.    \label{eqn:balance-1}
\end{align}
Equation \eqref{eqn:balance-1} can be interpreted as a characterization of the statistical relation
between two dynamical processes of an equilibrium system.
Here, $C_{AB}(\omega)$ is regarded as the rate of particle excitation by absorbing an energy $\w$,
and $C_{BA}(-\omega)$ is the rate of the de-excitation process by releasing the energy $\w$.
The ratio of these two rates is precisely controlled by the temperature of the system.
For systems out of equilibrium, \Eq{eqn:balance-1} provides a criterion for
the existence of local equilibrium within a small subsystem.

For equilibrium systems, $C_{AB}(\omega)$ and the anti-Hermitian response function \cite{Yan05187}
$\chi_{AB}^{(-)}(\w) \equiv \frac{1}{2i}\int_{-\infty}^\infty dt\, e^{i\w t} \tilde\chi_{AB}(t)$
satisfy the \gls{fdt}:
\be
\tC_{AB}(t) = \frac{1}{\pi}\int_{-\infty}^\infty d\w \, e^{- i \w t } \, f(\w) \, \chi_{AB}^{(-)}(\w).
\label{FDT-system}
\ee
Equation \eqref{FDT-system} is formally analogous to \Eq{fdt-fermi}.
It is possible to acquire  $C_{AB}(\w)$ experimentally by measuring the response property $\chi_{AB}^{(-)}(\w)$,
and then check the validity of \Eq{eqn:balance-1} to confirm the existence of local equilibrium.
Furthermore, in equilibrium systems, the occupations of bosonic and fermionic
particles on microstates obey the Bose-Einstein and Fermi-Dirac distributions, respectively.
If the distribution is preserved approximately within a small subregion of a non-equilibrium system,
a local equilibrium state is considered to exist for the small subsystem.
Note that the particle distribution function involves both the local temperature and local electrochemical potential.
For instance, Yan \emph{et al.} have shown that the local chemical potential is essential for describing
the non-equilibrium magnon density distribution in a Heisenberg spin chain \cite{Yan17024417}.

\subsection{Physical motivations to extend the concept of temperature out of equilibrium} \label{subsec:mot}

Generally speaking, extension of thermodynamic and statistical approaches for
equilibrium systems to non-equilibrium situations faces fundamental difficulties.
For systems out of equilibrium, the local equilibrium conditions are in principle approximate,
and some important relations such as the detailed-balance relation and the \gls{fdt} may no longer hold true,
though their non-equilibrium versions are being investigated \cite{Zha12170402,Sei0910007}.

For macroscopic and classical systems out of equilibrium,
the concept of local temperature has been adopted.
Casas and Jou have provided a general overview of the definitions of temperature in non-equilibrium systems,
as well as a wide range of applications of these definitions to different systems \cite{Cas031937}.
They have concluded that under the local equilibrium condition,
different definitions give the same temperature,
while in a non-equilibrium state beyond local equilibrium,
the values of temperature defined by different approaches
may deviate from one another.
Puglisi \emph{et al}. have reviewed different definitions of temperature in classical systems
and discussed their relevance to energy fluctuations \cite{Pug171}.
Particularly, a generalized \gls{fdr} was introduced,
which provides a statistical basis for extending the definition
of temperature to far-from-equilibrium classical systems.

For non-equilibrium nanosystems,
the quantum coherence effects and influence of the environment can be conspicuous.
For instance, temperature oscillations may appear in nanosystems
under bias voltages or temperature gradients \cite{Dub0997,Cas10041301,Cas11165419,Mea14035407},
which clearly contradict the conventional Fourier's law for classical systems \cite{Dub11131}.
For the case of a nanowire coupled to two metal leads at different temperatures,
it was indeed predicted that the local temperature of the wire may exhibit spatial oscillations at certain lead-wire coupling strengths \cite{Dub0997}.
Similar oscillations have also been predicted to occur in a thermally-driven armchair graphene sheet \cite{Mea14035407}.
Further calculations have revealed the relation between the wavelength of temperature oscillations and that of the Friedel oscillations \cite{Ber15125407},
a ubiquitous oscillation of charge density in electronic systems.
These works indicate that the temperature oscillation is a non-classical property
and reflects the quantum coherent nature of the system \cite{Dub11131}.

From thermodynamic relations, we have the following definition of temperature at fixed volume $V$ and number of particles $N$:
\be
  T = \left( \frac{\partial E}{\partial S} \right)_{V,N}, \label{def-TbyS}
\ee
where $E$ and $S$ are the system's internal energy and entropy, respectively.
The direct application of \Eq{def-TbyS} to open quantum systems is difficult \cite{Bus0543,Pug171},
because different definitions of entropy do not lead to a consistent value
even at the same temperature \cite{Hor081161,All01056117}.
For example, the von Neumann entropy of the reduced system, $S_{\rm\! VN} \equiv -\trs(\rhos \ln \rhos)$,
has been used to define the local temperature via \Eq{def-TbyS} \cite{Ali88918,Joh09041119,Ali1635568}.
However, unlike the thermodynamical entropy,
$S_{\rm\! VN}$ remains finite even at zero background temperature \cite{All01056117}.

For systems out of equilibrium,
the definition of basic thermodynamic quantities, such as entropy and free energy,
is still under debate \cite{Hor132059,Skr144185,Uzd15031044}.
In addition, the equipartition theorem of energy among the system's different \gls{dof} has to be applied carefully \cite{Esf051,Bia1915LT01}.
For instance, in a nanoelectronic device driven by a constant bias,
the electron-electron interactions and electron-phonon interactions are expected
to induce electronic and ionic temperatures that are different from each other \cite{Pea69488,Bel08221507,Di08,Ins1185}.
Such difference is particularly significant in elastic electron transport,
because the motion of electrons and ions is completely decoupled \cite{Ber09245125}.


Enormous efforts have been made to combine thermodynamics and quantum theory \cite{Ali79L103,Kos841625,Lie991,Kie04140403,Lev12061126},
leading to the development of {\it quantum thermodynamics} \cite{Kos132100},
which closely relates the laws of thermodynamics to their quantum origin.
In 1959, Scovil \emph{et~al.} have established the equivalence between a three-level maser and a Carnot heat engine \cite{Sco59262,Geu67343}.
Later, the Lindblad--Gorini--Kossakowski--Kossakowski semi-group master equation \cite{Lin76119,Gor76821} has been employed
to address quantum thermodynamics from a dynamic perspective \cite{Ali79L103,Cur7522}
and to explore non-equilibrium thermodynamic processes such as heat flow and entropy production \cite{Uzd15031044,Kos132100,Mar1712447}.
Despite this progress, there are still many unsolved or controversial problems in quantum thermodynamics,
and experimental verifications of theoretical predictions are far from adequate \cite{Clo16170401,Mon1859,Har1512953}.
At any rate, the problem of defining a local temperature stands again at the heart of
quantum thermodynamics.

Summarizing the above, we see that extending the notion of temperature from equilibrium
to non-equilibrium situations should at least have the following implications:

\begin{enumerate}

\item
  Existence of a local temperature indicates the local equilibrium approximation is valid.
  In view of the zeroth law of thermodynamics, this means that the net flow of particle or energy
  between the system and its environment should reduce to a minimal extent.

\item
  Local temperature offers a characteristic energy scale for the magnitude of local excitations and fluctuations,
  and it also determines the distribution of particles among the relevant microscopic states.
  Comparing the local temperature to the background temperature provides a quantitative description of
  how far the system is away from a global equilibrium state.

\item
  Local temperature is also useful for characterizing the variation of the system's intrinsic dynamical and
  thermodynamic properties under an external bias. This is particularly important
  for investigating systems involving strong electron correlations.

\end{enumerate}

In addition, an ideal definition of local temperature should also be universal
(so that it may be applied to as many non-equilibrium situations as possible),
unique (so that it predicts one and only one parameter),
feasible (so that it can be measured by experimental means),
and asymptotically correct (so that it can retrieve the correct thermodynamic
temperature in a global equilibrium state).

We end this section by some remarks on a pertinent concept that shows up frequently
in the literature: the ``effective temperature'' $T_{\rm eff}$ \cite{Pek04056804,Cas10041301,Cas11165419,Cas12266,Hua061240,Hua07698,War1133}.
The physical meaning of effective temperature varies according to the different contexts in which it is employed.
Sometimes it is used simply as an alias of local temperature \cite{Pek04056804,Cas10041301,Cas11165419,Hsu11041404,Cas12266};
whereas in certain circumstances, effective temperature merely provides an equivalent
expression for a specific value of energy, such as Fermi temperature and Kondo temperature \cite{Kri751101,Pau07026402}.
In the latter scenario, $T_{\rm eff}$ is just an energy scale,
and it hardly conveys any thermodynamic or statistical meaning.
In the experimental works, the term ``effective temperature $T_{\rm eff}$ '' is often adopted
(see \Sec{sec:Experiment}), and it usually provides a quantitative characterization of
the magnitude of local heating or local excitation in a nanosystem.
Therefore, in the rest of this review, we shall distinguish the two concepts
only when $T_{\rm eff}$ refers to an energy scale rather than a physical temperature.

\section{Theoretical definitions of local temperature out of equilibrium} \label{sec:Def}

\subsection{General Remarks} \label{subsec:gen}

Numerous theoretical approaches have been proposed to define local temperature of non-equilibrium nanosystems.
In light of their different underlying physical assumptions, most of the existing theoretical approaches
can be ascribed to three general categories, by making connection to the physical implications of local temperature as described in \Sec{subsec:mot}.
In the following, we shall begin by giving a brief overview on the three categories of theoretical definitions,
and then provide an in-depth review of these definitions at a more detailed level in \Sec{subsec:def}.

First, from the thermodynamic point of view, the global thermal equilibration provides the conceptual and practical basis for the definition of a thermometer.
In nanosystems out of equilibrium, the local equilibration of energy and particles has been taken by many authors
as ``local equilibrium condition'' for determining the local temperature \cite{Hes98,Che031691,Har0689,Ye15205106,Ye16245105}.
However, in this review the above condition will be termed as ``zero-current condition'', since the latter more precisely describes
the consequence of local equilibration processes --- vanishing of particle and heat currents in a local subregion of a system.
Moreover, thermodynamic relations connect different intrinsic system properties.
For instance, \Eq{def-TbyS} links the system energy $E$ to the entropy $S$ via the temperature $T$.
Extending thermodynamic relations such as \Eq{def-TbyS} to non-equilibrium situations under the local equilibrium approximation
also gives rise to quantitative definitions of local temperature out of equilibrium.

Second, from the statistical perspective, temperature determines the distribution of particles among microscopic states.
The concept of particle distribution function has been extended to non-equilibrium situations, and determination of
local temperature from a non-equilibrium distribution function has been proposed \cite{Pek04056804,Sch9814978,Her913720,Nes13022121}.
Moreover, for systems far from equilibrium, the \gls{fdt} which characterizes the balance between the environment fluctuations and dissipation
is no longer rigorous.
Instead, various forms of approximate \gls{fdr} have been proposed \cite{Cug973898,Cug984979,Foi12P09011,Gal06045314,Nes14045409,Int14050101,Foi17052116},
from which a local temperature can be extracted \cite{Cug973898,Cug984979,Foi12P09011,Cug11483001,Pug171,Foi17052116}.
For instance, Cugliandolo \cite{Cug11483001} and Puglisi \emph{et al.} \cite{Pug171} have determined the local temperatures of non-equilibrium
classical systems by scrutinizing the deviation of non-equilibrium \gls{fdr} from equilibrium \gls{fdt}.
In the rest of this section, we shall instead focus on nanosystems in which quantum effects are prominent.

Third, a variety of physical properties sensitively depends on the local temperature \cite{Har0689}.
This has facilitated the design of practical schemes for measuring a local temperature of non-equilibrium systems.
Moreover, in practice it is ideal that a thermometer (such as a non-invasive thermal probe) should cause a minimal perturbation to the system
under study when the measurement is conducted.
Such a \gls{mpc} can be examined by monitoring the evolution of a certain thermosensitive property of the system \cite{Dub0997}.
As a consequence, operational protocols based on the \gls{mpc} have been proposed, with which
the local temperature can be determined quantitatively by presuming the local equilibrium approximation \cite{Ye15205106,Ye16245105}.

In \Sec{subsec:def} we will review the above three categories of theoretical definitions of local temperature.
The physical essence and operational protocol of these definitions will be exemplified by specific examples,
along with some remarks on their advantages and limitations.
The experimental feasibility of these theoretical definitions will also be addressed.

 \begin{figure}[htbp]
  \centering
  \includegraphics[width=0.5\columnwidth]{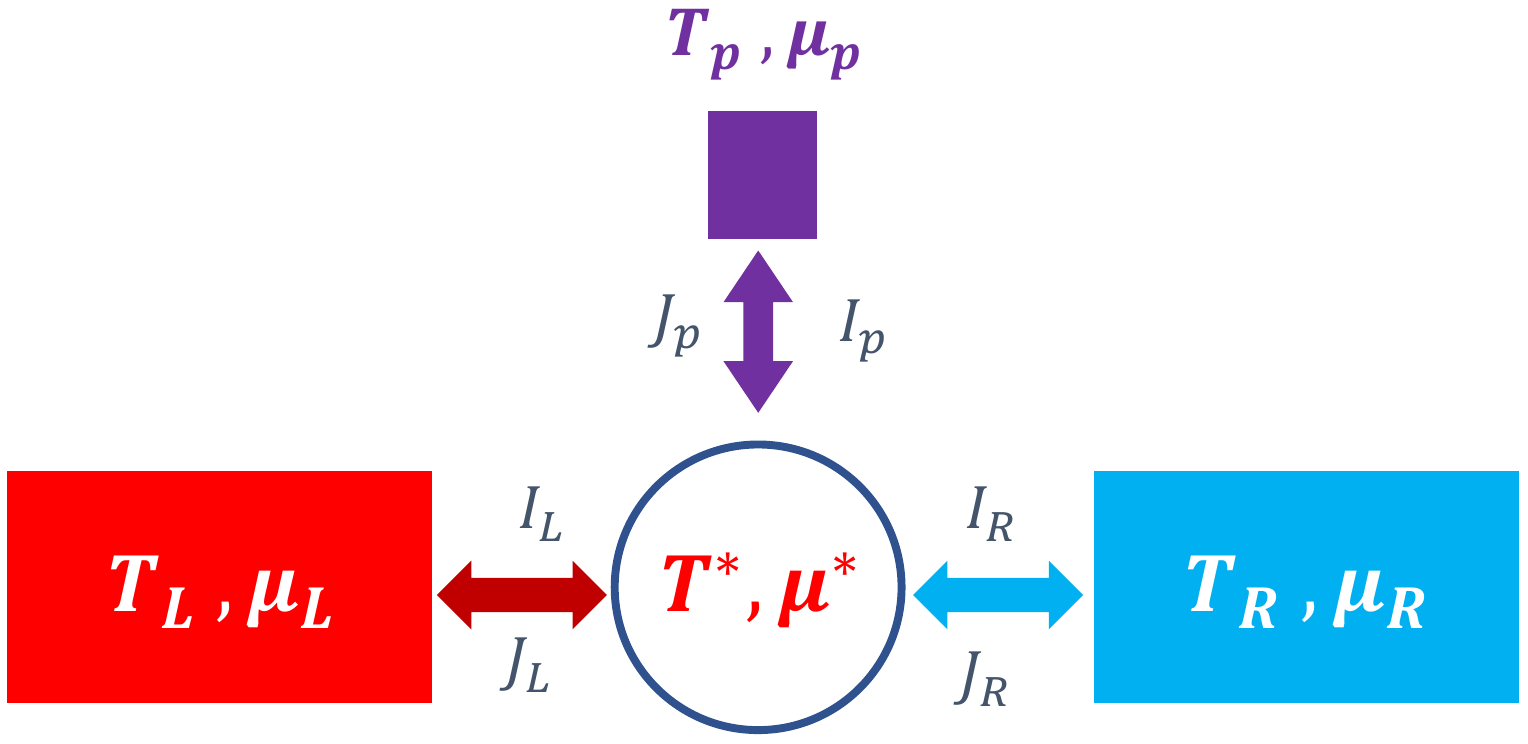}
  \caption
  {Schematic of a nanojunction.
   The central system (blue circle) is coupled to three leads -- left lead, right lead, and probe, which are labeled
   by $\alpha= L, R, p$, respectively.
   Each lead can have a distinct background temperature ($T_\alpha$) and electrochemical potential ($\mu_\alpha$).
   $I_\alpha$  is the electric current flowing into the $\alpha$-lead, and $J_\alpha$ is the resulting heat current.
   The central system is considered to have a local temperature $T^\ast$ and local electrochemical potential $\mu^\ast$.
   } \label{fig-model}
 \end{figure}

In the the rest of this section, we introduce some basic quantities which will be frequently used hereafter.
To start, let us consider a nanojunction sketched in \Fig{fig-model} as an example.
The junction consists of a central system (such as a nanoelectronic device) coupled
to a number of leads (labeled by $\alpha$). Each lead has its own background temperature $T_\alpha$
and electrochemical potential $\mu_\alpha$.
If $T_\alpha$ or $\mu_\alpha$ of a certain lead is different from the counterpart of any other lead,
the whole junction is out of equilibrium.
In many of the existing theoretical approaches, the determination of a local temperature $T^\ast$
requires the information of particle current and heat flux between the local system and the thermal probe.
In the nanojunction depicted in \Fig{fig-model}, the electric current flowing into the $\alpha$-lead and
the resulting heat flux are denoted by $I_\alpha$ and $J_\alpha$, respectively.

Consider $\{T_\alpha, \mu_\alpha\}$ to be all time-independent, so that the bias voltage and thermal gradient
across any pair of leads are constant, and the whole junction has reached a stationary state.
Under these conditions, the electric current flowing into $\alpha$-lead ($\alpha=L,R,p$) can be expressed by the following Landauer formula \cite{Mei922512}
\be
  I_\alpha = i \int d\epsilon\, {\rm tr} \left\{
  \bm\Gamma_\alpha(\ep) \left\{\bm G^<(\ep) + f_\alpha(\ep)\left[ \bm G^r(\ep) - \bm G^a(\ep)  \right]
  \right\}  \right\},  \label{electric-current}
\ee
and the associated heat current is expressed by \cite{Ber093072}
\be
  J_\alpha = i \int d\epsilon\, (\epsilon-\mu_\alpha)\, {\rm tr} \left\{
  \bm\Gamma_\alpha(\ep) \left\{\bm G^<(\ep) + f_\alpha(\ep)\left[ \bm G^r(\ep) - \bm G^a(\ep)  \right]
  \right\}  \right\}.  \label{thermal-current}
\ee
Here, $\bm G^r$, $\bm G^a$ and $\bm G^<$ are the retarded, advanced and lesser Green's functions, respectively, in the Keldysh, or \gls{negf}, formalism,
and the boldface indicates that these quantities are matrices in the Hilbert space of the central system.
In particular, $\bm G^r$ and $\bm G^a$ describe the propagation of electronic excitations within the central system,
while $\bm G^<$ describes the non-equilibrium electron distribution.
$f_\alpha(\ep) = 1/[1 + e^{(\ep - \mu_\alpha)/T_\alpha}]$ is the Fermi distribution function of the $\alpha$-lead,
and $\bm\Gamma_\alpha$ is the hybridization (or linewidth) matrix characterizing the coupling strength
between the central system and the $\alpha$-lead.
Note that \Eq{electric-current} and \Eq{thermal-current} are formally exact even when
the central system involves explicit electron-electron interactions \cite{Mei922512}.

For non-interacting or effectively non-interacting (mean-field) central systems, \Eq{electric-current} and \Eq{thermal-current} can be simplified as \cite{Mei922512,Ber093072}
\begin{align}
   I_\alpha &= \sum_{\gamma \neq \alpha} \int d\ep\, \mathcal{T}_{\alpha\gamma}(\ep) \,
   \left[ f_\alpha(\epsilon)-f_\gamma(\epsilon) \right], \label{non-interacting-electric-Landauer} \\
   J_\alpha &= \sum_{\gamma \neq \alpha} \int d\ep\, (\ep - \mu_\alpha) \,\mathcal{T}_{\alpha\gamma}(\ep) \,
   \left[ f_\alpha(\epsilon)-f_\gamma(\epsilon) \right], \label{non-interacting-thermal-Landauer}
\end{align}
where $\mathcal{T}_{\alpha\gamma} = {\rm tr}\{\bm \Gamma_\alpha \bm G^r \bm \Gamma_\gamma \bm G^a\}$
is the electron transmission function from $\alpha$-lead to $\gamma$-lead.
Equation~\eqref{non-interacting-electric-Landauer}
is the well-known Landauer formula for a multi-terminal nano-conductor \cite{Lan57223,Ryn16,Di08}.
The above equations are of fundamental importance to quantum electron and thermal transport in nanoscopic/mesoscopic systems.
It is important to point out, however, that the above equations do not include the influence of phonons.
To account for the contribution of electron-phonon interactions,
\Eqs{electric-current}--\eqref{non-interacting-thermal-Landauer} need to be modified,
e.g., by including the self-energies due to phonon baths. 

We now pay particular attention to the linear response regime because it is closely related to experiments.
If $I_\alpha$ and $J_\alpha$ are driven by a small bias voltage, $\Delta\mu=\mu_L-\mu_R$,
or a small temperature gradient, $\Delta{T}={T}_L-{T}_R$, across the left and right leads,
their expressions can be simplified as \cite{Don0211747}
\begin{align}
  I_\alpha &= -\mathcal{L}_{11} \frac{\Delta\mu}{T} - \mathcal{L}_{12} \frac{\Delta{T}}{T^2}, \label{I-linear-response} \\
  J_\alpha &= -\mathcal{L}_{21} \frac{\Delta\mu}{T} - \mathcal{L}_{22} \frac{\Delta{T}}{T^2}. \label{J-linear-response}
\end{align}
Here, $T$ is the background temperature,
and ${\mathcal{L}_{ij}}$ ($i,j=1,2$) are the linear response coefficients \cite{Don0211747,But904869},
which are closely related to the transport properties measured in experiments.
The electric current induced purely by a small bias voltage reflects the electric conductance
$G=-\frac{1}{T}\mathcal{L}_{11}$ \cite{Don0211747,Dub11131,But904869},
while the thermopower
$S= \lim_{\Delta{T}\to{0}}{\frac{\delta{V}}{\delta{T}}}|_{I_\alpha=0}=-\frac{1}{T}\frac{\mathcal{L}_{12}}{\mathcal{L}_{11}}$
measures the induced thermoelectric voltage in response to a temperature difference across the two leads \cite{Don0211747,Dub11131,Tur02115332}.
In the situation in which $I_\alpha = 0$, the thermal flux can be simplified as $J_\alpha=-\kappa\,\Delta{T}$, with $\kappa=\frac{1}{T^2}({\mathcal{L}}_{22}-\mathcal{L}_{22}^2/\mathcal{L}_{11})$ being the thermal conductance.
Moreover, the Onsager reciprocal relation $\mathcal{L}_{12}=\mathcal{L}_{21}$ has also been proved \cite{Don0211747}.
Particularly, the relation between the electronic contribution of $\kappa$ and $G$ of a metal (denoted $\kappa_e$ and $G_e$, respectively) is revealed by the Wiedemann-Franz law,
which states that the ratio of $\kappa_e$ to $G_e$ is proportional to the temperature of the system,
$\kappa_e / G_e = L T$, with the proportionality constant $L$, known as the Lorenz number \cite{Kit04}.

In addition to the \gls{negf} method, there are some other approaches that can be used to
describe quantum electron and thermal transport through nanojunctions.
These include the \gls{qme} approach and the \gls{heom} approach \cite{Li05205304,Har07235309,Jin08234703,Hu11244106,Wan14673,Ye16608}.
To simulate realistic nanojunctions from first principles, such as
single molecular junctions \cite{Tho18030901,Que143250,Liu1515386,Div00979}
and graphene nano-devices \cite{Ni1625914,Tes1623439,Ji1317883,Per18155432},
the above theoretical approaches must be combined with electronic structure methods.
For instance, single-particle scattering theory has been combined with \gls{dft} \cite{Koh65A1133,Par89}
to simulate quantum electron transport through atomic or molecular electronic devices \cite{Tay01245407,Heu02256803,Que073477,Div00979,Kim12226801}.
In this approach, the Landauer formula of \Eq{non-interacting-electric-Landauer} for non-interacting systems is employed,
since the Kohn--Sham reference system is effectively non-interacting.
However, driven by a bias voltage or thermal gradient, the nanosystem is not in its ground state,
and thus the use of ground-state \gls{dft} methods is only an approximation \cite{SaiDiVentra:05,Xue02151,Kur05035308}.

The time-dependent extension of \gls{dft}, \gls{tddft}, is capable of addressing
excited-state properties of electronic systems \cite{Run84997}, and, in principle,
provides an exact theory for the total current flowing across the device
\cite{Div048025}. \gls{tddft} methods for open systems have also been developed.
For instance, Burke \emph{et al.} have constructed a Markovian \gls{qme} including dissipation to phonons
for the time-dependent Kohn--Sham system \cite{Bur05146803}, while Di Ventra and D'Agosta have developed
a stochastic time-dependent current-density functional theory method
for open systems via the real-time propagation of a stochastic Schr\"{o}dinger equation \cite{Ven07226403,Dag08165105}.
Zheng \emph{et al.} have proved the existence of a rigorous \gls{tddft}
for open systems with variable number of electrons \cite{Zhe1114358},
and developed a practical \gls{tddft} scheme for open systems (\gls{tddft}--OS) \cite{Yam11245448,Zhe1426,Wan15144112}.
The \gls{tddft}--OS method has been applied to perform atomistic simulations of
real-time electron transport through nanojunctions \cite{Wan13205126}.

With the latest \gls{dft}/\gls{tddft} methods or their approximate variants,
such as the \gls{dftb} method \cite{Els987260},
it is now possible to model and simulate a nanojunction consisting of tens of thousands of atoms
at the quantum-mechanical level \cite{Yam151763}.
For instance, Yam \emph{et al.} have employed a \gls{dftb}+\gls{negf} method
to study the current-voltage characteristic of a junctionless silicon-nanowire
field-effect transistor with more than 15000 atoms in the core region \cite{Yam13062109}.

\subsection{Definitions of non-equilibrium local temperatures} \label{subsec:def}

\subsubsection{Definitions based on zero-current condition} \label{subsubsec:zcc}

 As we have already briefly mentioned, in an early study, Engquist and Anderson proposed a local equilibrium condition \cite{Eng811151}
 which has been employed extensively in the literature \cite{Arr11235436,Arr14125450,Ait13085122,Ven13256801}.
 In their work, an ideal potentiometer/thermometer (a voltage/temperature probe)
 in the form of a one-dimensional chain of non-interacting electrons is weakly coupled to the non-equilibrium system of interest.
 Then the temperature $T_p$ and electrochemical potential ${\mu}_p$ of the probe are adjusted until
 there is no net electric current $I_p$ and heat current $J_p$ flowing between the probe and the system.
 Such a condition is referred to as the \gls{zcc}.
 Under the \gls{zcc}, the local temperature $T^\ast$ and local electrochemical potential ${\mu}^\ast$ of the system are
 determined to be identical to those of the probe, respectively.
 Therefore, the \gls{zcc} can be expressed by the following form
 \begin{equation}
 \left\{
        \begin{array}{lr}
            I_p|_{T_p=T^\ast, {\mu}_p={\mu}^\ast} \!\! &= 0, \\
            J_p|_{T_p=T^\ast, {\mu}_p={\mu}^\ast} \!\! &= 0.
        \end{array}
 \right.
 \label{def-zcc}
 \end{equation}
 In some cases, for example, when the local probe exclusively couples to the vibrational \gls{dof},
 the zero electric current condition is not relevant, since the phonon number is not conserved \cite{Arr11235436,Ait13085122,Arr14125450}.
 %
 %
 %

\begin{figure}[htbp]
  \centering
  \includegraphics[width=0.75\columnwidth]{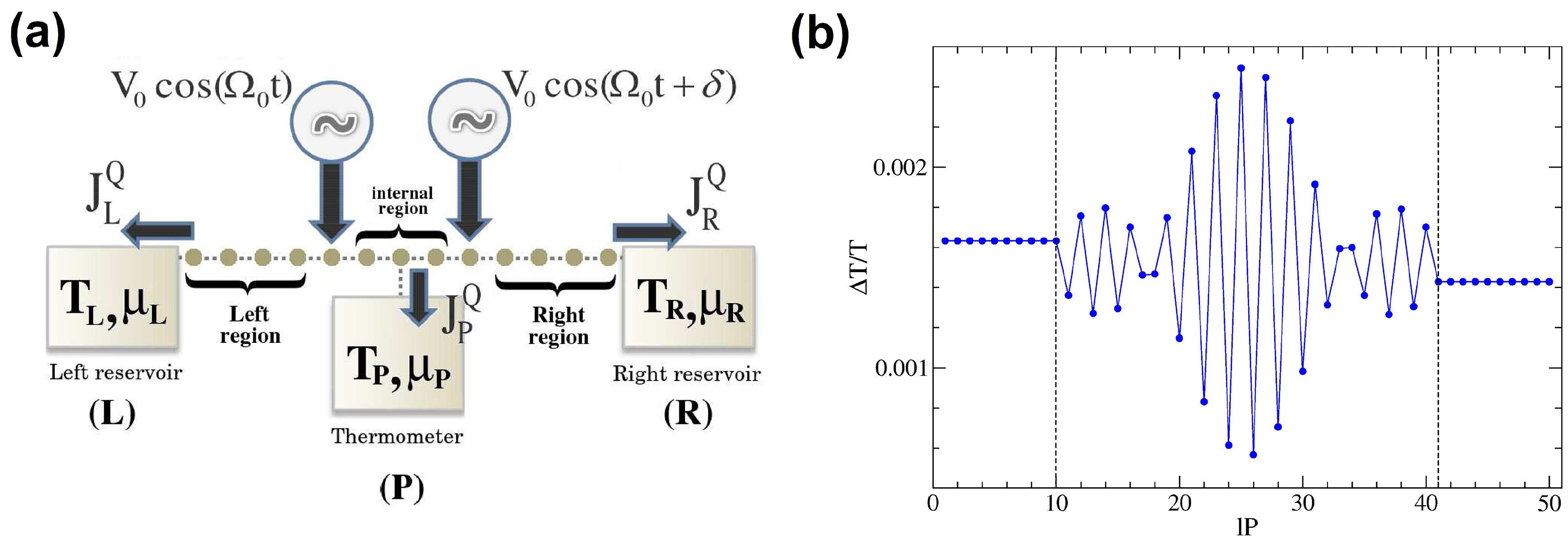}
  \caption
  {
   (a) Schematic of a quantum driven system.
   The central device is a nanowire connected to two reservoirs.
   The third reservoir $P$ represents a local thermometer, which consists of a macroscopic system weakly coupled to a given point of the central device.
   (b) Local temperature along a lattice with $N =50$ sites under two ac fields with a phase shift $\delta=\pi/2$ at the positions indicated by vertical dotted lines.
   Reprinted with permission from \cite{Cas11165419}. Copyright 2011 American Physical Society.
   }
  \label{figure2-zcc}
 \end{figure}

 Caso \emph{et al.} have applied the \gls{zcc} to study local temperatures of quantum systems driven out of equilibrium by external periodic fields  \cite{Cas10041301,Cas11165419,Cas12266}.
 Figure \ref{figure2-zcc}(a) displays the sketch of a quantum driven system.
 The central system is a nanowire modeled by a one-dimensional lattice, with its two ends connected to the left and right reservoirs.
 External fields which are periodic in time are applied to two separated sites of the nanowire.
 The periodic fields driving the electron and heat transport through the nanowire consist of both dc and ac components.
 The dc component creates a constant bias voltage which presents an energy barrier for electron transport \cite{Foi10125434,Foi09085430},
 while the two local ac fields have the identical amplitude and frequency but differ by a constant phase.
 In principle, the ac fields should lead to a local temperature which changes with time.
 To remove the time dependence, the authors investigated the local temperature in the low-frequency regime,
 where the periodic time of ac fields is much longer than the dwell time of electrons in the nanojunction
 so that the external field can be treated as an adiabatic perturbation.
 Therefore, the microscopic states of the system do not change significantly during the pumping process,
 and the \gls{zcc} results in a well-defined time-independent $T^*_l$ for a given site $l$.
 More specifically, $T^*_l$ is measured by monitoring the electric and thermal currents
 flowing into the third reservoir (the probe), which is weakly coupled to the $l$th site of the nanowire.

 Figure \ref{figure2-zcc}(b) shows the spatial distribution of local temperature along the nanowire
 determined by \Eq{def-zcc}.
 The value of $T^*_l$ is plotted for each site for the situation of $T_L=T_R=T$, $\mu_L=\mu_R=\mu$ and a low driving frequency.
 From the profile of $T^*_l$ one can easily recognize three regions in the nanowire.
 Clearly, the local temperature is almost constant in the left/right region.
 Note that the left end of the nanowire is slightly hotter than the right end
 because the presence of the phase shift $\delta$ breaks the original symmetry of the nanowire \cite{Cas11165419}.
 In the internal region between two energy barriers, the local temperature oscillates periodically in space.
 The Friedel-like oscillation has also been observed in other nanostructures \cite{Ber15125407}.
 In the nanowire the two energy barriers act as local impurities in the charge transport,
 and the oscillation is a consequence of quantum interference generated by scattering processes
 between conduction electrons and local impurities \cite{Cas10041301,Cas11165419,Cas12266}.
 The oscillations of local temperature have been observed in the left and right regions under a somewhat higher driving frequency \cite{Cas10041301}.

 The definition of local temperature based on the \gls{zcc} has the following advantages:
 (1) Recent theoretical research shows that the \gls{zcc}-defined $T^*$ for a quantum system in steady state, arbitrarily far from equilibrium, with arbitrary interaction within the system, is unique when the solution of \Eq{def-zcc} exists \cite{Sta16155433}.
 (2) The \gls{zcc} reflects the zeroth law of thermodynamics because the nanosystem and the thermometer reach a local equilibrium state with no electron and energy exchange.  As a result, the \gls{zcc} usually acts as a reference for other theoretical definitions.

 Such a macroscopic definition of local temperature, however, does not reflect the microscopic change of the system state in a non-equilibrium situation and
 its experimental realization is not straightforward. For instance, the \gls{zcc} may appear to provide an operational definition with the aid of a non-invasive probe that can monitor the heat current through the probe-nanosystem coupling. This is, however, not the case. This is because it is difficult for present calorimeters to measure the heat current through a nanosized sample without the {\it a priori} knowledge of the local temperature of the sample, which defies the purpose.
 This issue will be discussed in detail in \Sec{sec:Experiment}.

\subsubsection{Definitions based on balance of local heat generation and dissipation} \label{subsubsec:tec}

\begin{figure}[htbp]
  \centering
  \includegraphics[width=\columnwidth]{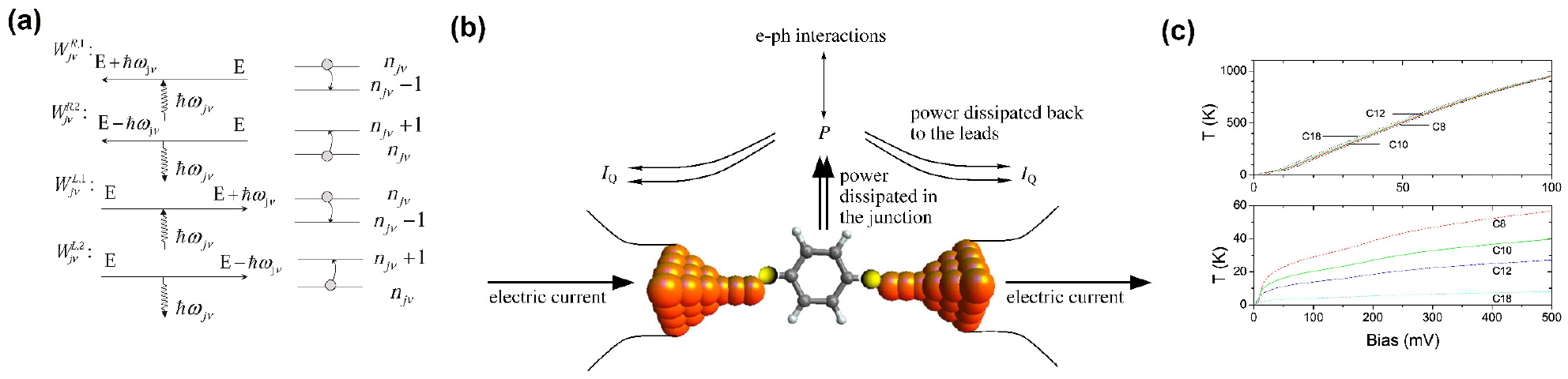}
  \caption
  {
   (a) Feynman diagrams of the four main electron-phonon scattering processes contributing to local heating of the junction. From top to bottom:
   (1) Cooling process due to absorption of a phonon from a left-moving electron;
   (2) heating process due to the emission of a phonon from a left-moving electron;
   (3) and (4) are the equivalent mechanisms corresponding to the right-moving electrons.
   Reprinted with permission from \cite{Che031691}. Copyright 2003 American Chemical Society.
   (b)  A schematic representation of the mechanism of local ionic heating in nanoscale junctions.
   The thermal power generation $P_{\rm e-p}$ induced by the electric current is balanced by the heat
   dissipated into the electrodes $I_Q = J^{\rm out} - J^{\rm in}$.
   The balance between the thermal power in the junction and the heat current flowing out of the junction determines the local ionic temperature of the junction.
   Reprinted with permission from \cite{Dub11131}. Copyright 2011 American Physical Society.
   (c) Local temperature calculated from \gls{dft} methods as a function of bias for various lengths of alkanethiols (color lines).
   Reprinted with permission from \cite{Che05621}. Copyright 2005 American Chemical Society.
   }
  \label{fig-dissipation}
 \end{figure}

 Chen \emph{et al.} have suggested an alternative local equilibrium condition for nanoscale junctions \cite{Che031691}.
 The idea is that local thermal equilibrium is established when the local heating process
 within the nanojunction balances the local cooling due to heat dissipation into the environment.
 Assuming there is no inelastic electron-electron scattering,
 local ionic heating occurs in a nanoscale conductor when electrons release energy to the ions via inelastic scattering with phonons.
 Figure \ref{fig-dissipation}(a) shows the four main electron-phonon scattering processes which contribute to local heating of the junction.
 These scattering processes occur in the central region of the junction as well as within a few atomic layers of the bulk electrodes; see \Fig{fig-dissipation}(b).
 Absorption and emission of energy through these scattering processes lead to the local heating and cooling effects, respectively;

Two scenarios have been considered in the literature \cite{Che031691}.
In the first scenario, the central system is presumed to be thermally isolated from the environment and
hence there is no heat current flow into the bulk electrodes.
Consequently, local thermal equilibrium is reached when the local heating exactly cancels the local
cooling by summing up the four electron-phonon scattering processes over all phonon modes in the central system.
Such a balance of local heating and cooling is expressed by \cite{Che031691,Che05621}
\be
 P_{\rm e-p}(T^*) = \sum_{j,\nu} \left\{ W^{R,2}_{j\nu}(T^*)+W^{L,2}_{j\nu}(T^*)-W^{R,1}_{j\nu}(T^*)-W^{L,1}_{j\nu}(T^*) \right\} =0.
 \label{def-thermal-power}
\ee
Here, $P_{\rm e-p}(T^*)$ is the thermal power generated within the central system due to electron-phonon interactions,
 $W^{\alpha,k}_{j\nu}$ is the contribution of $\nu$th coordinate component ($\nu=x,y,z$) of the
 $j$th phonon mode to the junction thermal power,
$\alpha=L,R$ represents the electrode from which the electron enters the junction, and
$k=1,2$ indicates the phonon loses/gains energy.
$W^{\alpha,k}_{j\nu}$ is a function of the local temperature $T^*$ of the central system.
This is because the value of $W^{\alpha,k}_{j\nu}$ depends on the occupation number of each phonon mode,
$\langle n_{j\nu} \rangle=1/(e^{E_{j\nu}/T^*}-1)$ \cite{Yan02041405,Ven00979}.
The local temperature is thus determined by tuning the value of $T^\ast$ until
\Eq{def-thermal-power} holds.

 In the second scenario, the energy stored locally in the junction is allowed to dissipate away into the bulk electrodes via elastic phonon scattering.
 Figure \ref{fig-dissipation}(b) shows the consequent thermal current $J^{\rm out}$ (denoted by $I_Q$ in the figure)
 flowing into both the left and right electrodes with a background temperature $T_0$.
 The heat is dissipated away into the bulk electrode at a rate of $J^{\rm out}={\kappa_{\rm th}}(T^*)^4$,
 with $\kappa_{\rm th}$ being the bulk thermal conductance that can be estimated from a microscopic model \cite{Ash76}. The fourth power comes from the assumption of the bulk form of lattice heat conduction.
 If the electrodes are at a finite temperature ($T_0 > 0$), there is also a heat current flowing into the central system, \emph{i.e.},
 $J^{\rm in} = \kappa_{\rm th} T_0^4$.
 In this scenario, local thermal equilibrium is established when the thermal power generation $P_{\rm e-p}$
 balances the net heat flux \cite{Dub11131}:
 \be
  P_{\rm e-p}(T^*) = J^{\rm out} - J^{\rm in} = \kappa_{\rm th}  \left[(T^*)^4 - T_0^4 \right].  \label{def-thermal-power-2}
 \ee
 The local temperature $T^\ast$ is determined by tuning $T^\ast$ until \Eq{def-thermal-power-2} is satisfied.

 Using \Eq{def-thermal-power-2}, Chen \emph{et al.} have studied the local temperature of
 alkanethiol chains of different lengths sandwiched between two electrodes  \cite{Che05621}.
 The calculated $T^\ast$ as a function of bias voltage are shown in \Fig{fig-dissipation}(c).
 The results indicate that local temperature is considerably lowered by heat transfer into the electrodes even for relatively large biases.
 It was also found that
 the molecular length has an important effect on $T^*$ when considering heat dissipation into the electrodes.
 Longer alkanethiols are more stable against current flow, which is consistent with experimental results on similar systems \cite{Wol015549}.

 In real systems, the thermal power generation $P_{\rm e-p}$ is a fraction of the total power of the entire circuit (nanojunction plus power source) \cite{Dub11131}.
 The latter is given by $P = V^2_{\rm bias} / R$, where $V_{\rm bias}$  is the bias voltage
 and $R$ the junction resistance (assuming no resistance of the external circuit).
 A minimal bias $V_c$ is necessary to excite the lowest-energy phonon mode of the nanostructure.
 $P_{\rm e-p}$ can thus be estimated as
\be
 P_{\rm e-p} = \Theta(V_{\rm bias} - V_c)\, \alpha P = \Theta (V_{\rm bias} - V_c) \left( \frac {\alpha} {R} \right) V^2_{\rm bias},
 \label{theraml-power-of-ep}
\ee
 where $\Theta(x)$ is the Heaviside step function.
 The factor $\alpha$ describes the fraction of $P$ dissipated into the ionic \gls{dof} of the junction due to electron-phonon interaction,
 and its value needs to be determined from a microscopic theory \cite{Tod98965}.
 Combining \Eq{def-thermal-power-2} and \Eq{theraml-power-of-ep} leads to \cite{Che031691}
\be
 {(T^*)^4=T_{0}^{4}+{\gamma^{4}_{\rm e-p}}V_{\rm{bias}}^{2}}, \label{local-temp-of-ep}
\ee
where $\gamma_{\rm e-p} = \Theta(V_{\rm bias} - V_c) (\alpha/ R\, \kappa_{\rm th})^{1/4}$
characterizes the contribution of the electron-phonon interaction to local ionic heating.

 In the presence of electron-electron interactions, some part of the thermal power generated
 in the junction ends up heating the electrons \cite{Dag062935}, and that part of the power
 is no longer available to induce local ionic heating.
 This changes the balance of local heat generation and dissipation to the following form:
 \be
  P_{\rm e-p}-P_{\rm e-e}=J^{\rm out}-J^{\rm in}.  \label{def-thermal-power-3}
 \ee
 In analogy to \Eq{local-temp-of-ep}, the variation of local ionic temperature $T^\ast$ with
 the bias voltage becomes \cite{Sch04145901,Dag062935}
 \be
 (T^*)^4=T_{0}^{4}+{\gamma^{4}_{\rm e-p}}V_{\rm{bias}}^{2}-{\gamma^{4}_{\rm e-e}}V_{\rm{bias}}^{4},
 \label{local-temp-of-ep-ee}
 \ee
 where $\gamma_{\rm e-e}$ describes the contribution of electron-electron interaction to local ionic heating.
 As the heated electrons carry energy away from the junction and eventually dissipate it to the bulk electrodes,
 the electron heating effect lowers the local temperature of the ions in the junction.
 Therefore, \Eq{local-temp-of-ep-ee} also predicts a local ionic cooling with increasing bias,
 which has been observed in alkanedithiol stretching experiments using the \gls{mcbj} setup \cite{Hua07698}.

 The method described above has the advantage of being able to treat realistic systems.
 However, it is not easily conducive to practical measurements.
 This is because tuning the temperature of a nanojunction implies that one has determined
 the local temperature prior to the temperature measurement.
 In addition, it is difficult to distinguish the thermal power $W^{\alpha,k}_{j\nu}$
 for each scattering mechanism and
 to figure out how $W^{\alpha,k}_{j\nu}$ depends on $T^\ast$ in practice.

\subsubsection{Definitions based on thermodynamic relations} \label{subsubsec:Ent}

  In classical thermodynamics, temperature measures the average kinetic energy of the particles in a system,
 $E_{\rm K} = \frac{3}{2} k_B T$.
 Such a relation has been employed widely in molecular dynamics simulations \cite{Sod03046702,Ong10155408,Hic16184311}.
 In the framework of \gls{dft},
 Ayers \emph{et al.} have defined a local kinetic energy, $E_{\rm K} \lbrack \Phi_{\rm KS}[\rho];{\bm r} \rbrack$,
 used to determine the local temperature $T^*({\bm r})$ of the electron gas \cite{Aye02309},
 \be
   E_{\rm k} \lbrack \Phi_{\rm KS}[\rho];{\bm r} \rbrack = \frac{3}{2} k_B T^*({\bm r}). \label{def-kinetic-deinition}
 \ee
 Here, the determinantal Kohn-Sham wavefunction $\Phi_{\rm KS}[\rho]$ describes
 a reference system with the electron density function $\rho(\bm r)$.
 The validity of such a kinetic definition relies on the equipartition theorem,
 which is strictly proven only in the thermodynamic limit for systems
 whose energy is quadratic in the particle momenta \cite{Dub11131}.

 To generalize the thermodynamic definition of \Eq{def-TbyS} to a non-equilibrium nanosystem,
 various thermodynamic quantities,
 particularly the energy $E_{\rm s}$ and entropy $S$ of the nanosystem,
 need to be properly defined within the framework of quantum mechanics.
 For a general dynamical variable $\hat{O}$,
 its expectation value has the form
 $O(t) = \langle{\hat{O}}\rangle = {\rm tr_s}\lbrack{\rho_{\rm s}(t)\hat{O}}\rbrack$.
 The energy $E_{\rm s}(t)$ of the system at time $t$ becomes
 \be
 E_{\rm s}(t) = {\rm tr_s} \lbrack \rho_{\rm s}(t) H_{\rm s} \rbrack.
 \label{internal-energy}
 \ee
 Although the definition of non-equilibrium entropy is still under debate \cite{Hor132059,Skr144185,Uzd15031044},
 the von Neumann entropy has been widely used in the literature \cite{Ali88918,Joh09041119,Ali1635568}. 
 The von Neumann entropy of the system is defined as
 \be
 S_{\rm VN}(t) = -k_{\rm B} {\rm tr_s} \lbrack \rho_{\rm s}(t) \ln{\rho_{\rm s}(t)} \rbrack,
 \label{entropy}
 \ee
 where $k_{\rm B}$ is the Boltzmann's constant.
 Clearly, both $E_{\rm s}$ and $S_{\rm VN}$ are functionals of $\rho_{\rm s}$.
 Since \Eqs{internal-energy} and (\ref{entropy}) extend the concept of energy and entropy to a time-dependent situation,
 a direct generalization of \Eq{def-TbyS} is expressed by \cite{Ali1635568,Ali88918,Joh09041119,Bar93196,Fel03016101,Wei0830008}
 \be
 T^*(t) = \left(\frac{\delta E_{\rm s}(t)}{\delta S_{\rm VN}(t)}\right)_{V,N},
 \label{def-direct-generlize-temp}
 \ee
 at a fixed ``volume'' $V$ and ``number of particles'' $N$.
 However, it is not always clear how to define the volume $V$ in generic quantum system,
 and the particle number $N$ of many-body systems is not always a conserved quantity \cite{Ali1635568}.

 %
 %
Di~Ventra and Dubi have proposed an alternative definition based on thermodynamic relations \cite{Ven0940004}.
 They have defined a dynamical quantity, ``information compressibility'' $K_I(t)$,
 to characterize the ability of an open system to change its entropy in response to a variation of its energy as
 \begin{align}
 K_I(t) = \frac{1}{k_B} \frac{\partial S_{\rm VN}}{\partial t'}\left( \frac{\partial E_{\rm s}}{\partial t'}\right)^{-1}\bigg|_{t=t'},
 \label{information-compressibility}
 \end{align}
 at the time $t$.
 Here, the energy $E_{\rm s}$ is given by \Eq{internal-energy}
 and the information entropy is the von Neumann one, $S_{\rm VN}$.
 A theoretical analysis has shown that \cite{Ven0940004} for a two-level system coupled to external baths with different temperatures,
 the information compressibility $K_I$ agrees exactly with the reciprocal of the local temperature $\beta^* \equiv 1/T^*$
 measured at any given time by a weakly-coupled ``temperature floating probe'' \cite{Dub0997},
 whose temperature is adjusted so that the system dynamics is minimally perturbed.
 The details of the latter approach to define a local temperature will be reviewed in \Sec{subsubsec:mpc}.

 Ali \emph{et al.} have studied nonequilibrium thermodynamics of a single-particle quantum system in contact with a heat bath \cite{Ali18}.
 Basic thermodynamic quantities, such as energy, entropy, and local temperature have been evaluated by \Eqs{internal-energy}, (\ref{entropy}), and (\ref{def-direct-generlize-temp}), respectively.
 Referring to the definition of the Helmholtz free energy,
 the non-equilibrium free energy $F(t)$ has been defend as
 \be
 F(t)=E_{\rm s}(t)-T^*(t)S_{\rm VN}(t).
 \label{free-energy}
 \ee
 Based on the non-equilibrium free energy $F(t)$,
 the non-equilibrium quantum thermodynamic work is defined,
 which is the time-dependent dissipative work done by the system,
 as the decrease of the non-equilibrium free energy of the system: $\delta W(t) = -dF(t)$.
 As a consequence of the first law of thermodynamics, the time-dependent heat $\delta Q(t)$ transferred into the system can be expressed by
 \be
 {\delta Q(t)=dE_{\rm s}(t)+\delta W(t)=dE_{\rm s}(t)-dF(t)},
 \label{heat-trans}
 \ee
 where $dE_{\rm s}(t)$ is the change in the system energy during the non-equilibrium process.

 \begin{figure}[htbp]
  \centering
  \includegraphics[width=0.6\columnwidth]{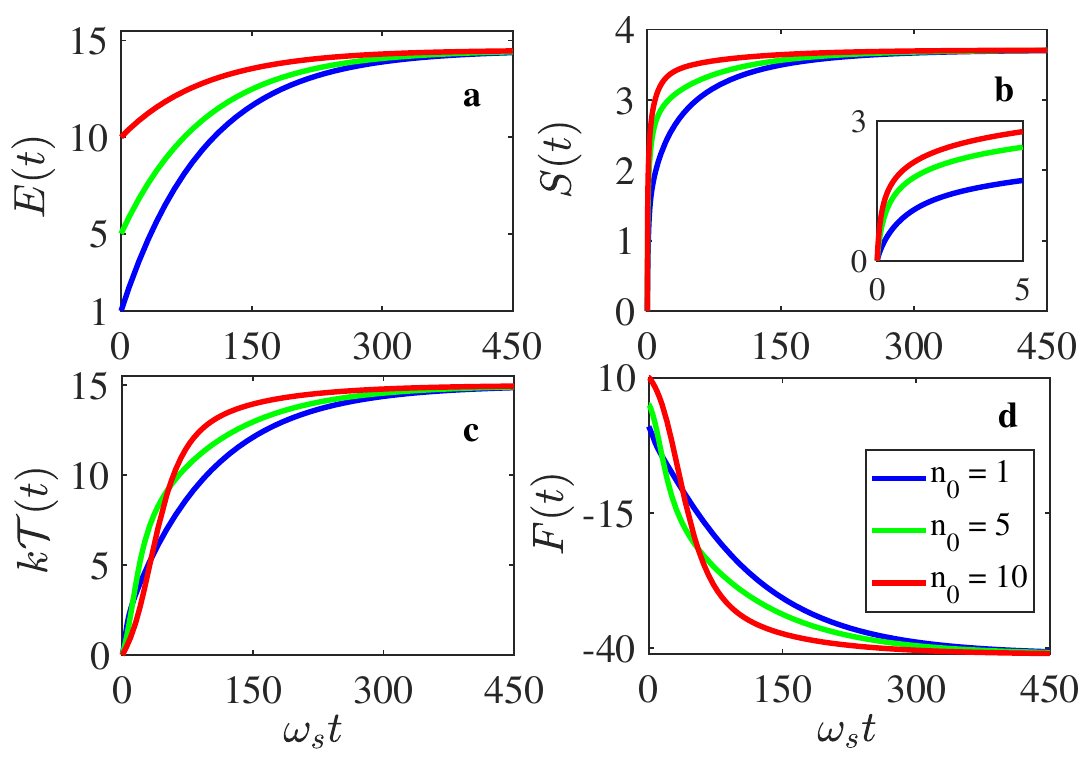}
  \caption
  {
   The non-equilibrium dynamics of the quantum system in terms of various thermodynamic quantities in the weak system-reservoir coupling regime with different initial states.
   Different colors represent initial states with system energy $E_0=n_0\omega_s, n_0=1,5,10$.
   (a) Internal energy; (b) Entropy; (c) Local temperature; (d) Free energy.
   Reproduced by permission of the authors of \cite{Ali18}.
   }
  \label{fig-mechanical}
 \end{figure}
 The above thermodynamic properties and the local temperature $T^*(t)$
 have been applied to a single-mode photonic cavity in contact with a heat bath \cite{Ali18}.
 The frequency of the photon field in the cavity is denoted as $\omega_s$.
 Initially, the system is assumed to be in the pure state with zero entropy at zero temperature,
 and the bath is at thermal equilibrium with a relative high temperature $k_{\rm B} T_0 = 15\omega_s$.
 \Figure{fig-mechanical} shows the evolution of the above thermodynamic quantities in the weak-coupling regime
 for different initial states of the system with different energies.
 The entropy increases during the non-equilibrium processes.
 The increase of internal energy and the decrease of free energy amount to heat transfer into the system,
 resulting in the rise of the local temperature.
 \Figure{fig-mechanical}(c) clearly shows that the local temperature of the system becomes the same as the temperature of the reservoir.
 Therefore, the local temperature defined by \Eq{def-direct-generlize-temp} is physically meaningful because it coincides with the thermal equilibrium temperature when the system approaches the thermal equilibrium state with a reservoir in the weak-coupling regime.
 In the long-time limit, the free energy approaches a minimum value with maximum entropy in the equilibrium state.

 The above analysis suggests that classical thermodynamic behaviors may emerge from quantum dynamics of a single-particle system in the weak system-reservoir coupling regime, without introducing any hypothesis on an equilibrium state.
 The main advantage of this approach is that a time-dependent local temperature is applicable to the real-time dynamics of nanosystems
 before the system approaches the steady state or thermal equilibrium.
 However, the thermodynamic relation-based definitions still face some challenges.
 For instance, from the theoretical point of view, the internal energy is a linear function of the reduced density matrix,
 while the entropy is a non-linear one \cite{Ven0940004}.
 Equation (\ref{def-direct-generlize-temp}) may thus yield multiple values of local temperature at a specific time.
%
 In general, from the experimental perspective,
 the practical implementation of, e.g., \Eq{def-direct-generlize-temp} requires the simultaneous measurement of the change in energy and entropy of the system.
 Although some efforts have been made to measure entropy in a nanosystem which consists of a few particles \cite{Cam16240502,Isl1577,Isl1577,Li11752},
 experimental measurement of energy change is a challenging problem, especially in a many-body system \cite{Aba12020504}.
 Therefore, it is not easy to realize the definition in practice, at least with the existing experimental techniques.

\subsubsection{Definitions based on statistical relations} \label{subsubsec:DF&FDR}

 Since local temperature is closely related to a local Bose--Einstein distribution
 for bosonic systems (such as phonons) or a Fermi--Dirac distribution for fermionic systems (such as electrons),
 one can employ definitions based on local distribution functions \cite{Pek04056804,Gia07184518,Hei10100408,Gal0710320}.
 For an electronic system described by the Fermi--Dirac distribution,
 Pekola \emph{et al.} have proposed a definition of $T^*$ for spinless electrons in the normal-metal region of
 normal-metal-insulator-superconductor tunnel nanojunctions \cite{Pek04056804}.
 By using the Sommerfeld expansion, the local temperature is related to the thermal energy density of the electrons.
 In consideration of electron spin,
 Giazotto \emph{et al.} \cite{Gia07184518} and Heikkil\"{a} \emph{et al.} \cite{Hei10100408}
 have extended this definition to the
 spin-polarized tunneling junctions and the spin valve setups, respectively.
 In these works, a spin-dependent local temperature is defined as follows:
 \be
  T^*_{s}=\frac{\sqrt{6}}{k_B\pi}\sqrt{\int^\infty_{-\infty}\{f_{{\rm neq},s}(\epsilon) -{\lbrack}1-\theta(\epsilon-\mu_s){\rbrack}\}{\epsilon}d\epsilon}.
 \label{def-Sommerfeld}
 \ee
 Here, $\mu_s$ is the electrochemical potential for the electrons in the normal-metal region,
 with $s=\uparrow$ ($\downarrow$) labeling the spin-up (spin-down) electrons,
 and $\theta(\epsilon-\mu_s)$ is the Heaviside step function.
 The spin-dependent non-equilibrium distribution function is expressed in terms of $f_{{\rm neq},s}=n_{L,s}f_L+n_{R,s}f_R$,
 with $f_{\{L,R\}}$ the equilibrium Fermi-Dirac distribution
 and $n_{\{L,R\},s}$ the model-dependent weight coefficient of left/right leads \cite{Gia06217,Pot973490}.
 The term $1-\theta(\epsilon-\mu_a)$ in \Eq{def-Sommerfeld} is equivalent to the equilibrium Fermi-Dirac distribution in the zero-temperature limit \cite{Gia07184518}.

 For a phonon system described by the Bose--Einstein distribution,
 Galperin \emph{et al.} have proposed a definition of local temperature
 which directly compares the non-equilibrium local distribution function,
 $f_{neq}$, with an equilibrium one, $f_{eq}(E_a,T_a)$, with a given temperature $T_a$ \cite{Gal0710320}.
 $T^*$ is determined by varying $T_a$ to give the best match.
 In short, the definition is expressed as
 \be
 f_{\rm neq}(\ep) \simeq f_{\rm eq}(\ep,T^*).
 \label{def-distribution}
 \ee
The idea is to approximate $f_{\rm neq}$ by $f_{\rm eq}$  with a tunable local temperature $T^\ast$.

\begin{figure}[htbp]
  \centering
  \includegraphics[width=0.6\columnwidth]{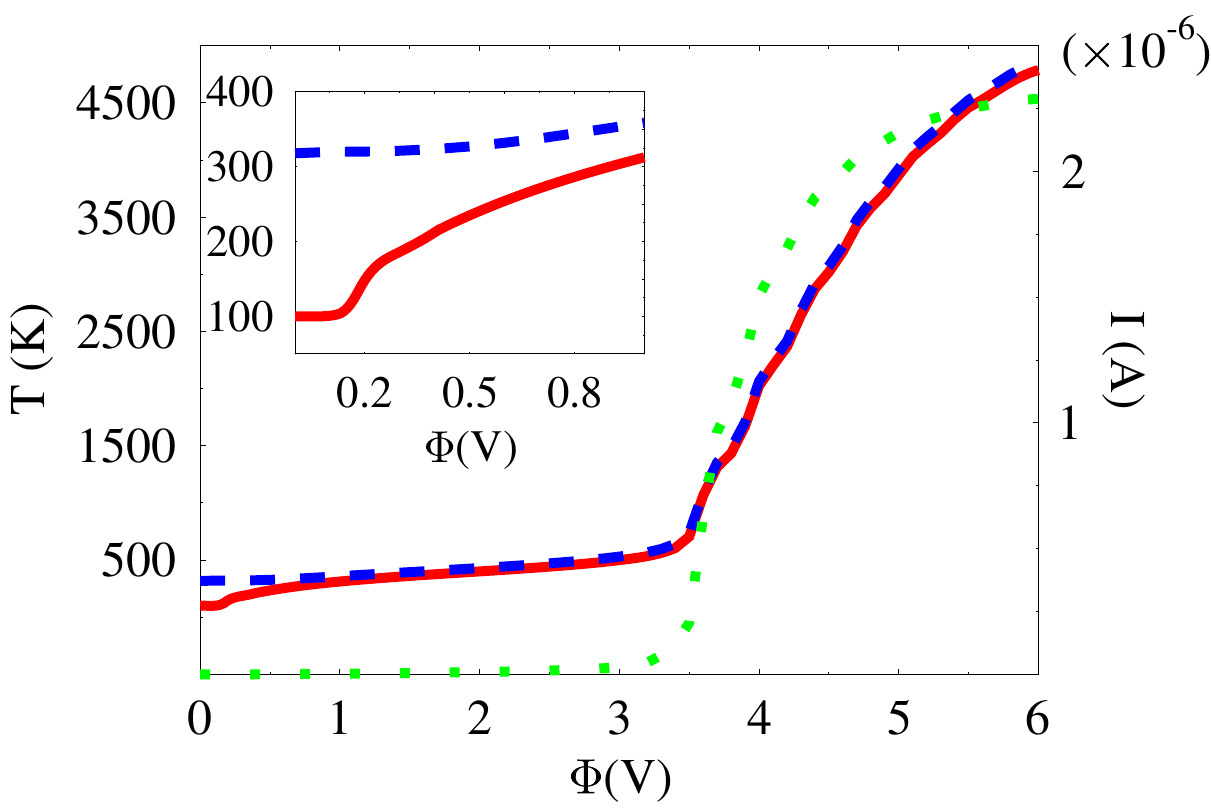}
  \caption
  {
   The local temperature $T^{*,BE}$ defined by the equilibrium distribution assumption, \Eq{def-distribution}, (dashed line, blue)
   and $T^{*,ZCC}$ defined by the \gls{zcc} (solid line, red) plotted against the applied bias $\Phi$, for a single-level model junction coupled to one vibrational \gls{dof}.
   The lead temperature is 100K.
   The inset shows the low bias region.
   The dotted line (green) shows the junction current (right vertical axis).
   $T^{*,ZCC}$ and $T^{*,BE}$ are the $T_{th}$ and $T_{BE}$ in the figure.
   Reprinted with permission from \cite{Gal07155312}. Copyright 2007 American Physical Society.
   }
  \label{fig-distribution}
 \end{figure}

 Galperin \emph{et al.} have estimated the local phonon temperature as a function of voltage \cite{Gal07155312,Gal0710320}
 in a single-level model system coupled to one vibrational \gls{dof}.
 The system contacts with two leads which have the same temperature as that of the system (100 K) in the zero-bias limit.
 These authors have then compared the value of the local temperature defined by \Eq{def-distribution} with that defined by the \gls{zcc}.
 These two temperatures are termed $T^{*,BE}$ and $T^{*,ZCC}$, respectively,
 and are plotted in \Figure{fig-distribution}.

 The two definitions give similar values of local temperature at a large bias voltage,
 but $T^{*,BE}$ is substantially higher than $T^{*,ZCC}$ as the bias approaches zero.
 The \gls{zcc} gives a good estimation,
 while the equilibrium distribution-based definition produces an erroneous result.
 The reason for the deviation is that the equilibrium distribution function in \Eq{def-distribution} ignores
 the possible interaction between the phonon mode and the electrons in the molecule.
 The \gls{zcc} considers this contribution by coupling the whole molecule to an external probe, then adjusting the heat flux to zero.
 The agreement at a high voltage bias is partially an artifact of the single resonance level and/or the single vibrational mode model,
 whereby the results depend on properties of the phonon distribution at a relatively narrow frequency range and not on the full non-equilibrium distribution \cite{Gal07155312}.

 The above distribution-function-based definition of \Eqs{def-Sommerfeld} and (\ref{def-distribution}) have the advantage
 that the local temperature is uniquely determined once the non-equilibrium distribution function is given.
 However, in principle, the validity of the two definitions requires that
 there is no significant difference between the non-equilibrium distribution function and an equilibrium one,
 namely, the nanosystem stays in a near-equilibrium state \cite{Koc06155306}.
 In addition, before the local temperature can be determined by using such definitions,
 one needs to know the non-equilibrium distribution function,
 while the values of $T^*$ determined from different distribution functions may not agree with each other.
 Particularly, it is challenging for the definition of \Eq{def-Sommerfeld} to be applied to systems at a high temperature.
 This is because the Sommerfeld expansion is only valid for systems at sufficiently low temperatures.

 For the definitions of \Eqs{def-Sommerfeld} and (\ref{def-distribution}),
 the key quantity, the non-equilibrium distribution function, is experimentally accessible
 at low temperature using the scanning tunneling spectroscopy \cite{Pot973490,Pie00437,Pie011078,Hua04599}.
 When a biased tunneling probe (such as the tip of an \gls{stm}) approaches towards the surface of a sample,
 the electrons tunneling through the energy barrier between the sample and the probe produces a tunneling current.
 Since the tunneling current depends on the local electron distribution
 in the probe and at the coupled site of the sample,
 the local distribution function can be obtained by measuring the tunneling current \cite{Che09036804}
 or the differential conductance \cite{Pot973490} of the tunneling junction.
 This scanning tunneling technique has been applied to measure $f_{\rm neq}$ of electrons or quasi-particles
 which obey the Fermi--Dirac distribution
 in carbon nanotube junctions \cite{Bro13161409} and mesoscopic metallic wire junctions \cite{Ant03076806}.
 However, the technique is limited to nanoscale conductors because the tunneling barrier for an insulator
 is generally so high that electron tunneling is forbidden.
 Non-Fermi energy distributions could not be measured either \cite{Alt1034}.

 Temperature not only determines the distribution function of a nanosystem, but also provides a measure of the magnitude of thermal fluctuations/excitations.
 The \gls{fdt} connects the response of a system from a prepared non-equilibrium state to its statistical fluctuations at equilibrium \cite{Cha87}.
 However, when the system is far from equilibrium, for instance, driven by an external field, the detailed-balance relation may break down
 and the \gls{fdt} does not hold exactly \cite{And06046124,Bat16604,Gne18066601,Zha1620004,Spe06391}.
 Nevertheless, an \gls{fdr} between the response of a system in a non-equilibrium state to an external perturbation and
 the correlation functions computed in the unperturbed system,
 can still be derived for a large number of systems \cite{Pug171}, from which a local temperature can be extracted.

 Arrachea and Cugliandolo have developed a local \gls{fdr} involving single-particle Green's functions for a mesoscopic metallic ring driven by a time-dependent magnetic field and coupled to an electron reservoir \cite{Arr05642}.
 %
 Performing the Fourier transform for the Green's function $G_{ij}^{K}(t,t^{\prime})$ with respect to $\tau\equiv(t-t^{\prime})$,
 the \gls{fdt} for electrons at equilibrium reads \cite{Cug11483001}
 \be
 G^K_{ij}(t,\omega)=\tanh{(\frac{\beta\omega}{2})}{\lbrack}G^r_{ij}(t,\omega)-G^a_{ij}(t,\omega){\rbrack},
 \label{single-FDR-equilibrium}
 \ee
 where $G_{ij}^{K,a,r}$ are the causal, advanced and retarded Green's functions;
 $i$ and $j$ label the site of the nanosystem ($G^K$ was named Keldysh Green's function in \cite{Arr05642}).
 To extend \Eq{single-FDR-equilibrium} to a nanosystem driven out of equilibrium,
 the background temperature $T$ (or $\beta=1/T$) in the r.h.s. of \Eq{single-FDR-equilibrium} has been replaced by another parameter $T^{\prime}$.
 Comparing the left and right hand sides of the inverse Fourier transform of \Eq{single-FDR-equilibrium} averaged over time $t$,
 these authors have adjusted $T^{\prime}$ such that the two sides of \Eq{single-FDR-equilibrium} fit with each other.
 This leads to a modification of the \gls{fdr} for a non-equilibrium nanosystem by defining the local temperature $T^*=T^{\prime}$ for such a system.

 Caso \emph{et al.} have formulated a similar modified \gls{fdr} for a nanosystem driven by electrical fields, already shown in \Figure{figure2-zcc} \cite{Cas10041301,Cas11165419}.
 The local temperature can be extracted from the following equation \cite{Cas10041301,Cas11165419}
 \be
 iG^K_{ll}(0,\omega)-iG^K_{ll}(0,\mu)=\tanh\bigg{\lbrack}{\beta_l^*(\frac{\omega-\mu}{2})}\bigg{\rbrack}\bar{\varphi}_l(\omega),
 \label{def-single-FDR}
 \ee
 for the site $l$ of the system.
 Here, $\bar{\varphi}^*_l(\omega)=-2\textup{Im}{G^r_{ll}(0,\omega)}$ and $G^{K,r}_{ll}(0,\omega)$ are the causal and retarded Green's functions, respectively, for the site $l$ at time $t=0$.
 The zero-time condition arises from the assumption that in the weak-driving adiabatic regime
 the driving frequency $\Omega_0$ is much smaller than the inverse of the dwell time of the electrons within the central system,
 and the external field is treated as a perturbation \cite{Cas10041301,Cas11165419}.
 However, the local temperatures defined by \Eqs{single-FDR-equilibrium} and (\ref{def-single-FDR}) are not directly accessible experimentally,
 because the Green's function is difficult to be directly measured in practice \cite{Arr05642}.

 To find an experimentally feasible definition,
 an \gls{fdr} has been developed in the framework of current-current correlation functions \cite{Cas12266}
 \be
 iC^K_{\alpha\alpha}(0,\omega)=\coth\bigg{\lbrack}{\beta_l^*(\frac{\omega}{2})}\bigg{\rbrack}\bar{\varphi}^*_\alpha(\omega),
 \label{def-current-FDR}
 \ee
 where ${\bar\varphi}^*_\alpha=-2\textup{Im}{C^R_{\alpha,\alpha}(0,\omega)}$
 and $C^{K,R}_{\alpha\alpha}(0,\omega)$ are causal and retarded current-current correlation functions for currents through the reservoir $\alpha$ at time $t=0$.
 Compared with the Green's function involved in the definitions of \Eqs{single-FDR-equilibrium} and (\ref{def-single-FDR}),
 current-current correlation functions are related to noise which is measurable in practice \cite{Bla001}.
 Therefore, \Eq{def-current-FDR} provides, in principle, an experimentally-accessible definition \cite{Cha952363}.

 \begin{figure}[htbp]
  \centering
  \includegraphics[width=0.9\columnwidth]{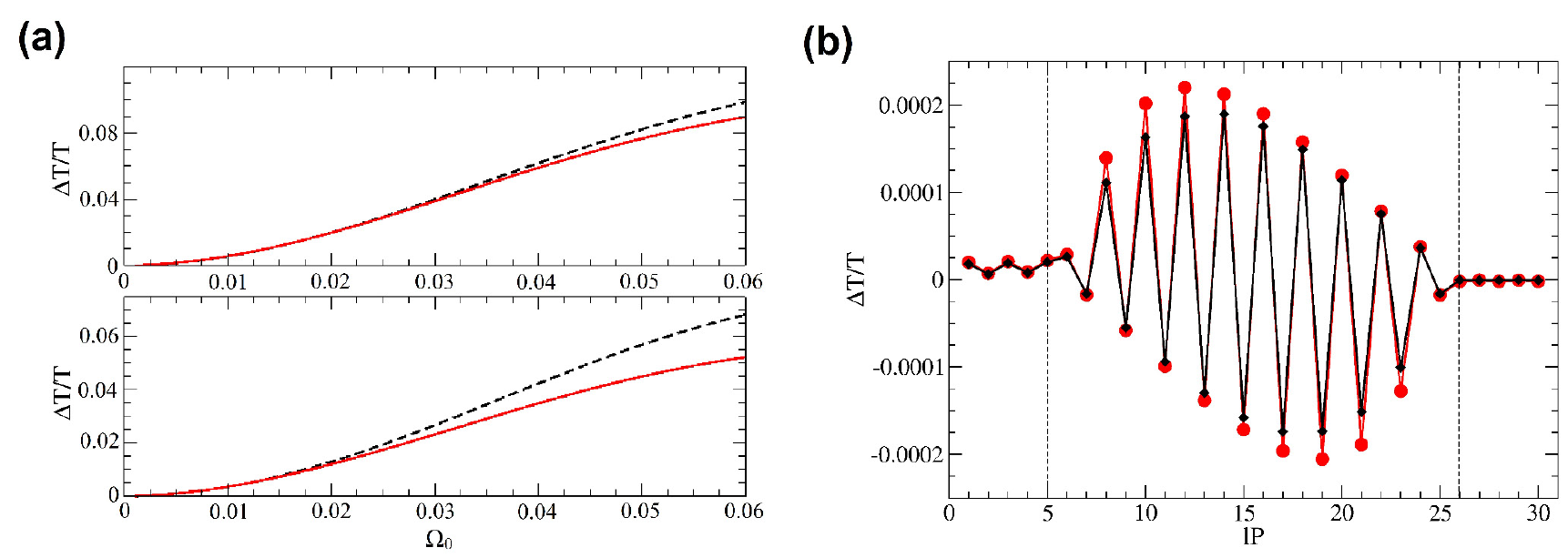}
  \caption
  {
   (a) Local temperature differences $\Delta{T^{ZCC}}$ (dashed black) and $\Delta{T^{FDR}}$ (solid red),
   relative to the temperature $T$ of the reservoirs, $\Delta{T}/T$,
   for the site $l = 0$ connected to the left reservoir as a function of driving frequency $\Omega_0$, for the system depicted in \Figure{figure2-zcc}(a).
   The upper and lower panels correspond to $T=0.016$ and $T=0.005$, respectively.
   (b) Local temperature differences $\Delta{T^{ZCC}}$ (black diamonds) and $\Delta{T^{FDR}}$ (red circles)
   along a one-dimensional nanowire model system of $N = 30$ sites
   with two ac fields operating with a phase lag of $\delta=\pi/2$ at the positions indicated by dotted lines.
   Reprinted with permission from \cite{Cas12266}. Copyright 2012 Springer Nature.
   }
  \label{fig-fluctuation}
 \end{figure}
 Caso \emph{et al.} have compared the value of the \gls{zcc}-defined local temperature $T^{*,ZCC}$
 and that of the local temperature $T^{*,FDR}$ defined by \Eq{def-current-FDR}
 in a one-dimensional model nanosystem \cite{Cas12266}.
 The system is shown in \Figure{figure2-zcc}(a).
 %
 \Figure{fig-fluctuation}(a) shows $\Delta{T_l^{FDR}}$ and $\Delta{T_l^{ZCC}}$
 calculated for the site $l = 0$ connected to the left reservoir,
 as a function of the driving frequency $\Omega_0$.
 Here, $\Delta{T^{FDR}}=T^{*,FDR}_l-T$ and $\Delta{T^{ZCC}}=T^{*,ZCC}_{l}-T$
 are the differences between the temperature $T$ of the reservoir and two local temperatures, respectively.
 The two local temperatures $T^{*,FDR}_l$ and $T^{*,ZCC}_l$ coincide with each other at a low frequency $\Omega_0$.
 As $T$ increases, the region in which the two definitions agree quantitatively becomes broader;
 see the upper panel of \Figure{fig-fluctuation}(a).
 \Figure{fig-fluctuation}(b) exhibits a good agreement between the two local temperatures along the whole nanostructure,
 and an almost perfect agreement within the left and right regions.
 The Friedel-like oscillations appear in the central region.
 The local temperature oscillations here have the same microscopic origin as that of the temperature oscillations discussed in \Sec{subsubsec:zcc}.

 Similar to the equilibrium distribution definition, the \gls{fdr}-based definition generalizes a statistical property established at equilibrium to the non-equilibrium situation by defining a local temperature.
 However, limited by the weak- driving adiabatic assumption, this definition is valid for near-equilibrium systems with slow dynamics. Nonetheless, in view of the experimental developments of measuring current-current correlations and low-frequency noise (see, e.g., \cite{Bla001}), these definitions are experimentally conducive to the study of local temperatures in realistic nanosystems.

\subsubsection{Definitions based on minimal perturbation condition} \label{subsubsec:mpc}

 Dubi and Di~Ventra have proposed an operational method to determine $T^*$ by mimicking a probe-based thermometer technique \cite{Dub0997}.
 The method is described as follows.
 As shown in \Figure{fig-model}, one couples a thermal probe with temperature $T_p$ to a local site of a nanosystem.
 Due to the interaction between the probe and the coupled site, the system dynamics is generally modified.
 Consequently, the \gls{qme} of the system reduced density matrix $\rho_s$ reads
\be
 \dot{\rho_{\rm s}}=-i\lbrack{H_{\rm s},\rho_s}\rbrack+\mathcal{R}_{\rm env}{\rho_{\rm s}}+\mathcal{R}_{\rm probe}(T_p){\rho_{\rm s}},
 \label{MPC-master-equation}
\ee
 where $H_{\rm s}$ is the system Hamiltonian,
 $\mathcal{R}_{\rm env}$ is a superoperator that represents the dissipative interactions between the system and its environment,
 and the temperature-dependent superoperator $\mathcal{R}_{\rm probe}(T_p)$ corresponds to the influence of the thermal probe, with temperature $T_p$, on the microscopic states of the system.
 Then the temperature $T_p$ is varied (or ``floated" as discussed in their work) so that the perturbation on some local or global properties of the nanosystem is minimized (ideally to zero),
 and a local temperature is determined as $T^*=T_p$.
 This minimal perturbation condition (\gls{mpc}), also named as the ``temperature floating-probe method'', reads
\be
 \mathcal{R}_{\rm probe}(T^*)\rho_{\rm s}=0.
 \label{MPC-density}
\ee

\begin{figure}[htbp]
  \centering
  \includegraphics[width=0.7\columnwidth]{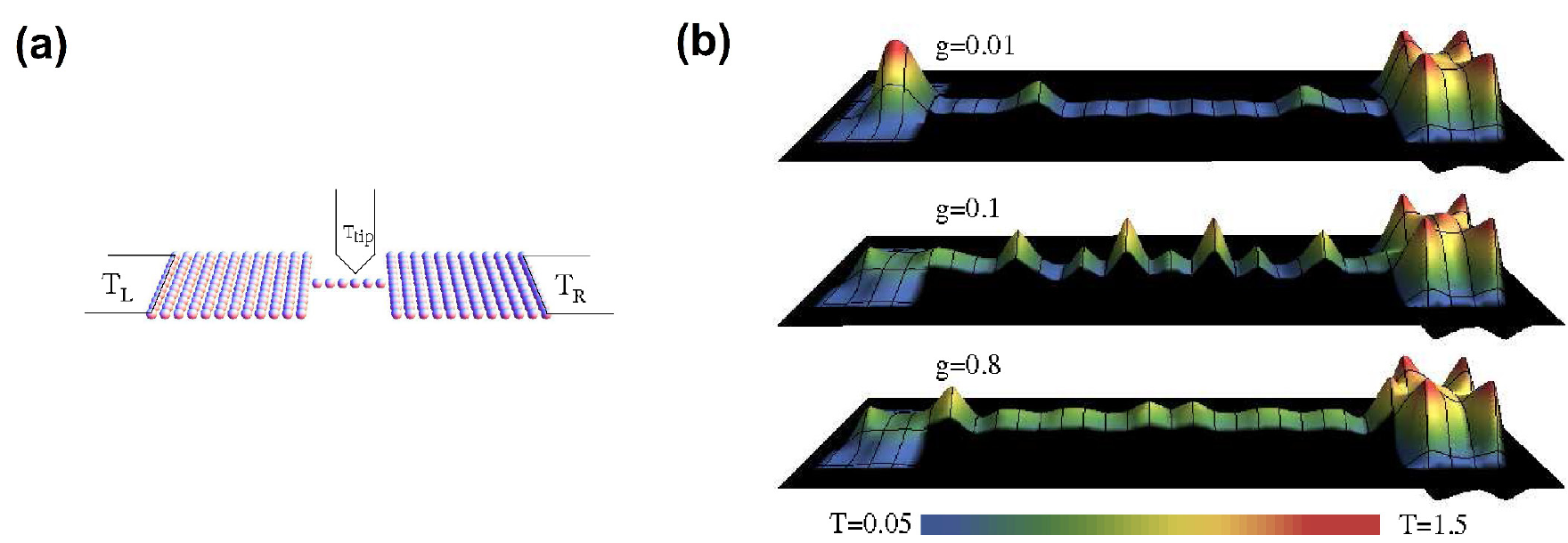}
  \caption
  {(a) Schematic representation of the calculation to determine the local temperature via the addition of a probe.
   A nanowire is coupled to the left and right leads with coupling strength $g$.
   The solid lines at the edges correspond to the contact area between the leads and the thermal baths at temperatures $T_L$ and $T_R$, respectively.
   The $T_{\rm tip}$ represents the temperature of the probe.
   The left and right leads have the same chemical potential.
   (b) The local temperature along the nanostructure for three different values of the lead-wire coupling, $g=0.01, 0.1, 0.8$.
   Three effects are observed: a hot spot in the cold lead at small coupling, temperature oscillations in the wire at intermediate coupling, and uniform temperature along the wire at large coupling.
   Reprinted with permission from \cite{Dub0997}. Copyright 2009 American Chemical Society.
   }
  \label{fig-MPC-density}
 \end{figure}

 Based on the \gls{mpc}-based definition of local temperature,
 the temperature profile has been calculated for a one-dimensional model nanosystem \cite{Dub0997}; see \Figure{fig-MPC-density}(a).
 \Figure{fig-MPC-density}(b) plots the local temperature at steady state for three different values of the lead-wire coupling strength $g$.
 At a weak coupling of $g = 0.01$, the temperature inside the wire is very low, but a ``hot spot" develops in the cold lead.
 As $g$ increases the hot spot vanishes and temperature oscillations emerge in the wire.
 At a strong coupling of $g = 0.8$, the temperature distribution along the system is uniform,
 and the wire equilibrates at a temperature that is roughly the average of $T_L$ and $T_R$.
 %

 In its original formulation, the \gls{mpc} approach of \Eq{MPC-density} only considered
 the energy exchange between probe and nanosystem,
 while particle transfer was excluded.
 This is a strong limitation, particularly for the studies of current-carrying nanoelectronic devices subjected to bias voltages.
 In order to overcome this limitation,
 Ye \emph{et al.} have modified the original formulation of the \gls{mpc} \cite{Ye15205106}.
 Another key quantity, the local electrochemical potential $\mu_p$ of the probe, was introduced;
 and the \gls{mpc} was imposed on the expectation value of a specific system observable $\hat{O}$.

 The modified \gls{mpc} definition is described as follows.
 One varies $T_p$ and $\mu_p$, until the electric current $I_p$ through the probe vanishes and simultaneously the probe-induced perturbation $\delta{O_p(T_p,\mu_p)}$ on the observable $\hat{O}$ gets minimized:
 \begin{equation}
 \left\{
        \begin{array}{lr}
            \left. I_p \right|_{T_p=T^*,{\mu}_p={\mu}^*}=0, & \\
            T^*=\textup{arg} \min\limits_{T_p}\left\{\delta{O_p}(T_p,\mu^*)\right\}.
        \end{array}
 \right.
 \label{def-MPC-observable}
 \end{equation}
 When \Eq{def-MPC-observable} is satisfied, there is no particle exchange between the probe and the nanosystem,
 and the probe influence on the system is minimized (ideally to zero).
 This suggests that the system locally equilibrates with the probe to some degree.
 Therefore, this definition can be referred to as a ``local-equilibrium condition'' and be somewhat understood from the zeroth law of thermodynamics.

 Particularly, when a nanosystem is connected to two leads, $L$ and $R$, and under a voltage bias $V=\mu_R-\mu_L$,
 a convenient approximation for the local electrochemical potential $\mu_p$ is determined by a linear combination
 \begin{align}
 {\mu^*\simeq\zeta_L\mu_L+\zeta_R\mu_R}.
 \label{MPC-mu}
 \end{align}
 Then the local temperature $T^*$ is measured by varying $T_p$, with $\mu_p$ fixed at the value given by \Eq{MPC-mu}
 \begin{align}
 {\delta{O_p(T_p,\mu_p)} = [\zeta_L{O_p(T_L,\mu_L)+\zeta_R{O_p(T_R,\mu_R)}]-{O_p(T_p,\mu_p)}}}.
 \label{MPC-T}
 \end{align}
 Here, $O_p(T_p = T_\alpha,\mu_p = \mu_\alpha)$ denotes the local observable $\langle \hat{O} \rangle$ measured
 by setting the electrochemical potential and temperature of the probe to be identical with those of the leads $\alpha=L,R$.
 The subscript $P$ represents the presence of the probe.
 The weight coefficients $\zeta_L$ and $\zeta_R$ in \Eqs{MPC-mu} and (\ref{MPC-T}) are determined by
 \begin{align}
 {\zeta_\alpha=1-\frac{I_p(T_\alpha,\mu_\alpha)}{I_p(T_L,\mu_L)+I_p(T_R,\mu_R)}}.
 \label{MPC-weight}
 \end{align}
 Equations (\ref{MPC-mu}) and (\ref{MPC-weight}) are the result of the conservation of particle number in the system \cite{Ye15205106}.
 \begin{figure}[htbp]
  \centering
  \includegraphics[width=0.6\columnwidth]{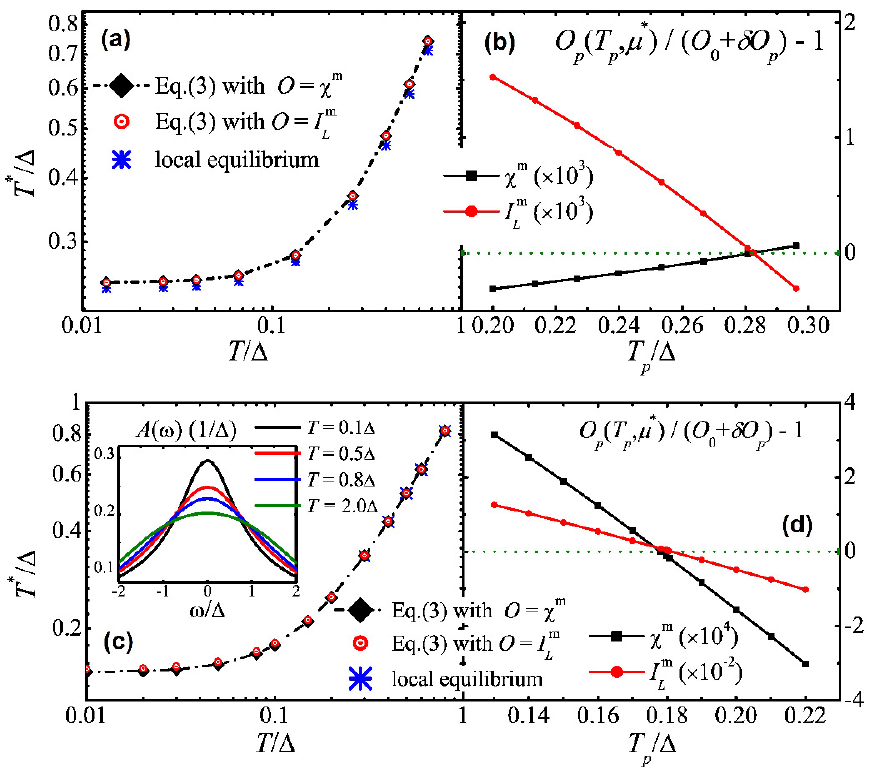}
  \caption
  {
    Local temperature $T^*$ of a single-level \gls{qd} determined by different protocols versus background $T$.
   (a) A noninteracting \gls{qd} without electron-electron Coulomb interactions;
   (b) An interacting \gls{qd} with electron-electron Coulomb interactions;
   The inset shows the Kondo spectral peak of the dot at various $T$.
   (c) and (d) are variations $\delta{O_p}$ of system observables $\chi^m$ (local magnetic susceptibility) and $I^m_\alpha$ (global spin-polarized current) with $\mu^*$ determined for a noninteracting/interacting \gls{qd}, respectively.
   Reprinted with permission from \cite{Ye15205106}. Copyright 2015 American Physical Society.
   }
  \label{fig-MPC-local}
 \end{figure}

 Ye \emph{et al.} have utilized the \gls{mpc} approach to a \gls{qd} system described by a single impurity Anderson model \cite{And6141},
 and have compared the value of the \gls{mpc}-defined $T^{*,MPC}$ with the \gls{zcc}-defined $T^{*,ZCC}$ under the same situation \cite{Ye15205106}.
 To test the robustness of $T^{*,MPC}$ with respect to the choice of observables,
 two properties that are particularly sensitive to temperature variations have been chosen to impose the \gls{mpc}:
 the local magnetic susceptibility $\chi^m= \frac{\partial \langle\hat{m}_z\rangle}{\partial H_z}|_{H_z \to 0}$
 and the global spin-polarized current $I^m_\alpha=\langle\hat{I}^m_{\alpha\uparrow}\rangle-\langle\hat{I}^m_{\alpha\downarrow}\rangle$ through the lead $\alpha$.
 Here, $\hat{m}_z$ is the dot magnetization operator due to the dot spin polarization, $H_z$ is the magnetic field,
 and $\hat{I}^m_{\alpha{s}}$ is the electric current operator for spin-$s$.

 The results in \Figure{fig-MPC-local} show a remarkable agreement with the local temperature $T^*$ determined by the minimal perturbation of either $I^m_\alpha$ or $\chi^m$, thus confirming its robustness with respect to the choice of observables.
 \Figure{fig-MPC-local}(a) and (b) illustrate that the \gls{mpc}-defined $T^{*,MPC}$ also agrees closely and consistently with that obtained with $T^{*,ZCC}$ in a large range of background temperatures.
 This is because the electric and heat currents flowing into the probe are both close to zero when the minimal perturbation is satisfied.
 This result numerically confirms the notion that the \gls{mpc} can be referred to as a local equilibrium condition to some degree.

 Although the Kondo effect is evident from the inset of \Figure{fig-MPC-local}(b),
 which shows that the \gls{qd} exhibits more prominent Kondo features when $T$ is low (as confirmed by the higher and sharper Kondo peak centered at $\omega=\mu$),
 the \gls{mpc} definition applies equally well for a \gls{qd} system from non-interacting to the Kondo-correlated regime.
 These results highlight the generality of the \gls{mpc}.

 In contrast to the \gls{zcc}-based definitions,
 the \gls{mpc} method relies on the measurement of local (or semilocal) observables,
 instead of the heat current through the probe.
 Therefore, one of the advantages of the \gls{mpc} method is that
 it offers an operational protocol for local temperature measurement that
 is feasible for experimental implementations.

\subsection{Uniqueness/non-uniqueness of local temperature out of equilibrium} \label{subsec:uni}

 In principle, for systems at a global equilibrium state,
 the value of temperature should be uniquely determined by any thermometer or
 by any theoretical definition that is physically sound.
 Let us instead discuss the uniqueness of the values of local temperature for systems
 out of equilibrium. Ideally, the various definitions should produce
 the same (or at least similar) values of local temperature for a given system under a given non-equilibrium condition.
 For macroscopic systems out of equilibrium, it has been shown that
 different definitions give the same temperature as long as the local equilibrium condition is satisfied \cite{Cas031937}.
 However, the situation is more complicated for nanosystems out of equilibrium.

 Summarizing the various definitions of local temperature that have been introduced in this section,
 we indeed find that they yield similar quantitative results in many cases.
 For instance, it has been shown analytically in \cite{Ven0940004} that for a two-level system,
 the $T^{*,MPC}$ determined by imposing the \gls{mpc} condition on the reduced density matrix 
 is exactly equivalent to $T^*$ defined based on the thermodynamic-type relation of \Eq{information-compressibility}.
 By performing numerical calculations on \gls{qd}s under low temperatures,
 Ye \emph{et al.} \cite{Ye15205106} have shown that the $T^{*,MPC}$ extracted from the local magnetic susceptibility or spin-polarized current agrees closely with $T^{*,ZCC}$ in a wide range of temperatures, even in the presence of strong electron correlations; see \Figure{fig-MPC-local}.
 As for $T^{*,BE}$ reviewed in \Sec{subsubsec:DF&FDR}, a numerical analysis has shown that $T^{*,ZCC}$ coincides with $T^{*,BE}$ in a single-level nanosystem subjected to a high bias voltage.

 \begin{figure}[htbp]
  \centering
  \includegraphics[width=0.5\columnwidth]{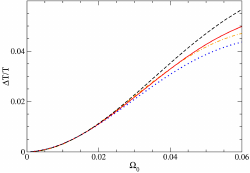}
  \caption
  {
   Local temperature differences $\Delta{T}^{*,ZCC}$ (dashed black) and $\Delta{T}^{*,FDR}$ for Case I (solid red), Case II (dotted blue) and Case III (dashed and dotted orange),
   relative to the temperature $T$ of the reservoirs,
   for the site connected to the left reservoir as a function of driving frequency $\Omega_0$
   (see the main text for a detailed explanation of the three cases).
   Reprinted with permission from \cite{Cas12266}. Copyright 2012 Springer Nature.
   }
  \label{fig-uniqueness1}
 \end{figure}

 Caso \emph{et al.} have compared the difference between the $T^{*,FDR}$ and $T^{*,ZCC}$.
 It has been shown analytically that the $T^{*,FDR}$ determined by the single-particle Green's functions is in good agreement with the $T^{*,ZCC}$
 at a low background temperature and in the weak-driving regime \cite{Cas10041301,Cas11165419}.
 For the $T^{*,FDR}$ derived from the current-current correlation function,
 the correlation functions involve the equilibrium Fermi--Dirac function $f_p$,
 and thus depend on the temperature $T_p$ and the electrochemical potential $\mu_p$ of the probe.
 To clarify the influence of such dependency on the value of local temperature defined by \gls{fdr},
 three different choices of $T_p$ and $\mu_p$ have been considered \cite{Cas12266}.
 In the first choice (Case I), $\mu_p$ is considered to be equal to the reservoir electrochemical potential $\mu$,
 and $T_p$ is equal to the local temperature $T^{*,ZCC}_{l}$.
 The second choice (Case II) also considers $\mu_p=\mu$, but $T_p = T^{*,FDR}_l$.
 In the third choice (Case III), $T_p = T^{*,FDR}_l$ is adopted, while
 $\mu_p=\mu^{*,ZCC}_{l}$ (the \gls{zcc}-defined local electrochemical potential for site $l$).
 \Figure{fig-uniqueness1} shows that the local temperatures $T^{*,FDR}$ and $T^{*,ZCC}$
 calculated for the above three cases agree well with each other in the slow-driving regime.
 When the driving frequency $\Omega_0$ is further reduced,
 the differences among the values of local temperatures determined
 by different definitions become vanishingly small.
 Conspicuous deviations only emerge in the high-frequency region.

 The above results confirm that in near-equilibrium nanosystems,
 such as the systems subjected to slow driving fields or low biases,
 the local temperatures determined by different definitions agree quantitatively with each other.
 This is consistent with the intuition that the local temperature
 should inherit all the physical implications of the conventional concept of temperature,
 if the local system can be properly described by a local equilibrium approximation.
 Moreover, it is remarkable that for certain systems subjected to
 rather large biases or involving strongly correlated states,
 such as the systems studied in \Figure{fig-distribution} and \Figure{fig-MPC-local},
 the different definitions of local temperature still yield similar values.
 This suggests that the physical considerations behind
 these definitions are somewhat universal and may remain valid
 even in the far-from-equilibrium or quantum regime.

 Of course, in some circumstances the local temperatures resulted from different definitions
 may differ distinctly from one another.
 The reason is mainly twofold. First, in a general non-equilibrium situation,
 the various physical implications of the concept of temperature
 (as presented in \Sec{subsec:mot}) may become independent of each other.
 For instance, it is difficult to define a unique local temperature
 if nonlocal excitations emerge in the system \cite{Ye16245105}.
 Second, the practical implementation of a theoretical definition
 may involve non-universal approximation or parametrization procedure.
 For example, as shown in \Figure{fig-uniqueness1}, the value of $T^{*,FDR}$ relies heavily
 on the specific choice of the temperature or electrochemical potential of the probe
 in the fast-driving regime \cite{Cas12266}.

 \begin{figure}[htbp]
  \centering
  \includegraphics[width=0.6\columnwidth]{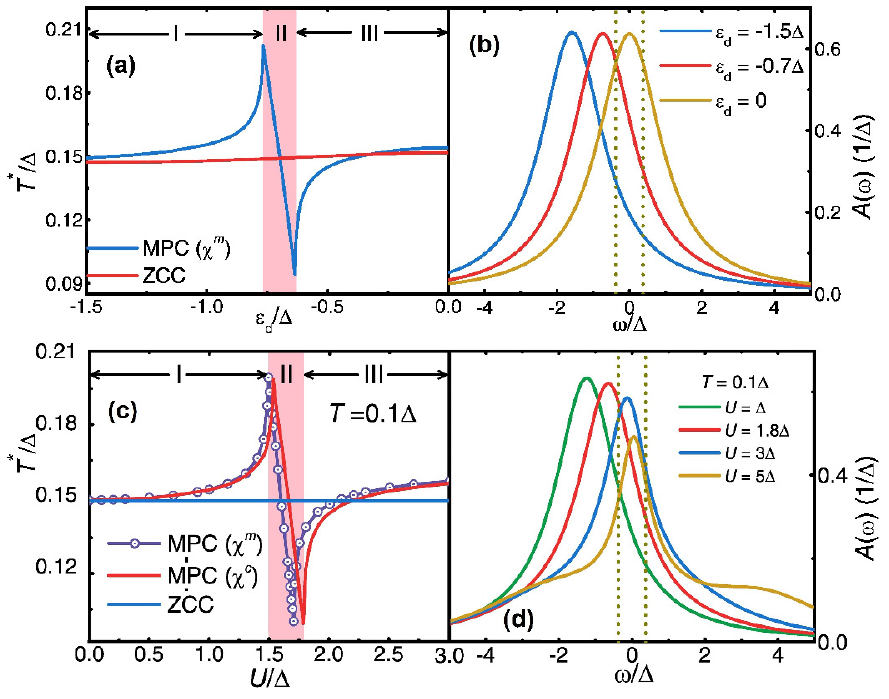}
  \caption
  {
   (a) Calculated $T^*$ versus single-particle energy level, $\epsilon_d$, for a noninteracting \gls{qd} under a bias voltage $V$;
   (b) Dot spectral function $A(\omega)$ for different $\epsilon_d$;
   (c) Calculated $T^*$ versus interaction strength, $U$, for an interacting \gls{qd} under a bias voltage $V $;
   (d) Dot spectral function $A(\omega)$ of a non-equilibrium \gls{qd} with varying $U$.
   The region between the two vertical dot lines in (c) is the excitation window in which the dot level is strongly resonant with the lead states.
   Reprinted with permission from \cite{Ye16245105}. Copyright 2016 American Physical Society.
   }
  \label{fig-uniqueness2}
 \end{figure}

 Ye \emph{et al.} have compared the calculated $T^{*,MPC}$ and $T^{*,ZCC}$ of \gls{qd}s \cite{Ye16245105}.
 Under a low background temperature, the local excitations on the \gls{qd} are
 dominated by the scattering events and correlation effects among electrons,
 while the phonon modes are not promoted.
 \Figure{fig-uniqueness2}(a) shows the $T^{*,MPC}$ as a function of energy level $\epsilon_d$ for a non-interacting \gls{qd}
 and \Figure{fig-uniqueness2}(c) shows $T^{*,ZCC}$ as a function of Coulomb interaction energy $U$ for an interacting \gls{qd}.
 In a non-interacting \gls{qd},
 the \gls{zcc} predicts an almost constant $T^{*,ZCC}$ over a large range of $\epsilon_d$.
 In contrast, the \gls{mpc} results in a conspicuous fluctuation of $T^{*,MPC}$ around $\epsilon_d = - 0.7 \Delta$,
 where the magnitude of $T^{*,MPC}$ differs significantly from $T^{*,ZCC}$.
 The vertical lines mark the three distinct regions.
 The three regions of $T^{*,MPC}$ have also been observed in an interacting \gls{qd},
 while the change in $T^{*,ZCC}$ is negligibly small with increasing $U$.

 The existence of the three distinct regions can be explained by the changes in peak positions of the spectral function $A(\omega)$, as shown in \Figure{fig-uniqueness2}(b) and (d).
 With increasing $\epsilon_d$ or $U$, the spectral peak gradually approaches the thermal activation window.
 As a result, the dot-lead resonances become stronger, and long-range nonlocal excitations emerge in the system.
 For a dot in the region I of \Figure{fig-uniqueness2}
 the dot level is off-resonant with the lead states.
 The electronic excitations are largely local on the dot,
 and $T^{*,ZCC}$ precisely captures the magnitude of these local excitations.
 In contrast, for a dot in the regions II and III,
 the spectral peak approaches towards, and finally enters into the thermal activation window.
 Electronic excitations can then occur inside the leads to
 create hot electrons (holes) above (below) the Fermi energy. Such excitations are hence nonlocal.
 In particular, \Figure{fig-uniqueness2}(a) and (c) illustrate that,
 for a dot in the region II where both local and nonlocal excitations take place,
 the $T^{*,MPC}$ effectively characterizes the magnitude of the nonlocal excitations, while $T^{*,ZCC}$ does not.


\section{Experimental measurement of non-equilibrium local temperatures} \label{sec:Experiment}

\subsection{Experimental strategies to measure the local temperature} \label{subsec:str}

A variety of experimental strategies for measuring the local temperatures
of non-equilibrium nanosystems have been proposed.
Like the theoretical definitions of local temperature reviewed in \Sec{sec:Def},
the existing experimental efforts on local temperature measurement
can also be ascribed to three general categories.

First, from the thermodynamical point of view,
the local temperature can be determined by monitoring
local equilibration of energy and particles within a non-equilibrium nanosystem.
According to the \gls{zcc}-based definition reviewed in \Sec{subsubsec:zcc},
the fulfillment of the local equilibrium condition can be examined by
probing the particle and heat currents flowing
through a probe which is weakly coupled to the nanosystem; see \Eq{def-zcc}.
However, it is worth pointing out that, while the particle current (such as the electric current)
through a nanostructure
can be measured directly with a high-precision instrument (such as an ammeter) \cite{Rip10112001},
the measurement of heat current at the nanoscale
is much less straightforward and more challenging.
Because of its importance for characterizing the thermal properties of nanomaterials,
as well as its close connection to the measurement of local temperature,
we shall briefly review the development of nanocalorimetry
in \Sec{subsec:mueasurement-heat-current},
and discuss its relevance to the local temperature measurements.
We will then review the experimental progress
on local temperature measurements on current-carrying nanosystems
with the use of \gls{mcbj} setups in \Sec{subsubsec:break-junction}.

 Second, from the statistical perspective, the local temperature determines
 the distribution of particles among the microscopic states of a nanosystem.
 For instance, as discussed in \Sec{subsubsec:DF&FDR},
 the local distribution of electrons in a nanowire has been obtained
 by measuring the tunneling current or differential conductance
 between the nanowire and a biased probe \cite{Che09036804,Pot973490}.
 Alternatively, the local distribution function can be extracted from
 the fluctuation-dissipation relations
 by exploring the statistical features of measured local physical observables
 (such as the shot noise spectra) \cite{Dut81497,Bla001,Cle101155}.
 Moreover, the deviation of local temperature from the background temperature
 reveals how the magnitude of local excitations/fluctuations in a
 non-equilibrium system differs from the thermal equilibrium situation.
 Experimentally, the local excitations/fluctuations can be probed by spectroscopic techniques
 which monitor the scattering processes between the nanosystem and
 incident particles (such as photons or electrons).
 In spectroscopy experiments,
 the local distribution of particles and the magnitude of local excitations
 can be characterized and quantified by the line shapes of the measured spectra.
 Particularly, with a spectroscopy measurement,
 it is possible to extract both the local phonon temperature ($T^\ast_{\rm ph}$)
 and local electron temperature ($T^\ast_e$)
 which are distinctly different in a non-equilibrium nanosystem \cite{War1133}.
 In \Sec{subsubsec:spectroscopy} we will review the spectroscopic
 techniques for local temperature measurement.

 Third, the temperature of a local region of a nanostructure or material can be measured by
 nano-sized thermometers which are fabricated through the miniaturization of conventional thermometers.
 The nano-sized thermometers include thermal sensors or probes as their key components.
 Thus, they are capable of detecting thermally sensitive properties or behaviors
 within the studied local region,
 such as thermovoltage due to the Seebeck effect \cite{Mil982900,Kim118700},
 change of electric resistance \cite{Ham964268,Tov10114317},
 thermal expansion due to local heating \cite{Har0689}, etc.
 In practice, a nano-sized thermometer can be mounted on or integrated into
 the atomically sharp tip of a \gls{stm} or \gls{afm} to form a \gls{sthm}.
 Nanoscopic thermal imaging techniques can then be realized based on
 the \gls{sthm} apparatus \cite{Men12596,Men1610874},
 which enables a non-invasive measurement of the temperature field
 of nanostructures under applied bias voltage or thermal gradient.
 In \Sec{subsubsec:SThM} we will review the \gls{sthm} techniques.

 The following subsections will cover the working principles and
 practical implementations of the experimental techniques mentioned above.
 The advantages and limitations of these techniques will be analyzed.
 At the end of this section, we will relate the existing experimental
 strategies for local temperature measurement
 to the theoretical definitions presented in \Sec{sec:Def}.

 Before going to the next subsection, we shall mention that
 the term ``effective temperature'' ($T_{\rm eff}$) is often adopted
 in the literature (experimental works particularly).
 In this section, any experimentally measured $T_{\rm eff}$
 is regarded as a ``physical'' temperature
 as long as it has certain characteristics
 of the thermodynamical temperature.

\subsection{Measurement of heat current at the nanoscale} \label{subsec:mueasurement-heat-current}

\subsubsection{General principle of nanocalorimetric techniques} \label{subsec:nanocalorimetry}

Calorimetry at the nanoscale has developed into to a crucial and useful technique
for quantifying the generation, dissipation and transport of heat in a nanosystem.
Recent advancement in the experimental instruments and operational protocols
has enabled the measurement of heat flow through nano-devices with a resolution of picowatts \cite{Gim94589,Var97306,Sad12084902,Sad13163110,Dec14094903,Son15253}.
Presently, the calorimeters have been applied to
monitor cellular activities \cite{Lee0915225,Wan16105005};
and to study radiative heat transfer \cite{Zhu19239,Tom18216,Fio18806,Cui1814479,Kim15387,Son16509},
thermal conductivity of nanostructures \cite{Lee17371,Lee158573,Shi03881,Can13105002,Kim14203107},
and phase transitions of nanomaterials \cite{Gao1953}.

To explain the working principle of a nanocalorimeter and
elucidate its relation to the local temperature measurement,
we consider the nano-contact between a non-equilibrium nanosystem
(with a local temperature $T^\ast$) and a probe
(with a background temperature $T_p$).
As discussed in the paragraph after \Eqs{I-linear-response} and \eqref{J-linear-response},
in the absence of electric current and considering only the linear response,
the heat current flowing into the probe through the nano-contact ($J_p$)
follows the Fourier's law \cite{Dub11131,Cha06085901} of
\be
  J_p = - \kappa \Delta T.  \label{fourier-law-1}
\ee
Here, $\kappa$ is the thermal conductance of the nano-contact.
From \Eq{fourier-law-1},
the magnitude of the heat current $J_p$ can be deduced from
the preset or measured thermal gradient $\Delta T = T_p - T^\ast$
(or vice versa) provided that $\kappa$ is known.

In the following, we will focus on two types of nanocalorimeters
that are widely used in current experimental studies:
the bimaterial cantilever-based calorimeters and the resistance-based calorimeters.
The readers may refer to \cite{Efr04176,Wil17125108,Gao1953}
for other types of nanocalorimeters.

\subsubsection{Bimaterial cantilever-based calorimeters} \label{subsec:bimaterial-cantilever-based-calorimeter}

 In this subsection, the voltage bias across a sample is left out of consideration,
 and temperature gradient is the only force to drive the heat flow from the hot side of the sample to the cold side.
 The heat flow $J$ leads to heating in the cold side,
 and in experiments it is generally quantified by the thermal power $P$,
 the amount of incident thermal energy per unit time, \emph{i.e.}, $J=P$.
 The Fourier's law of \Eq{fourier-law-1} is always considered valid.

\begin{figure}[htbp]
  \centering
  \includegraphics[width=0.75\columnwidth]{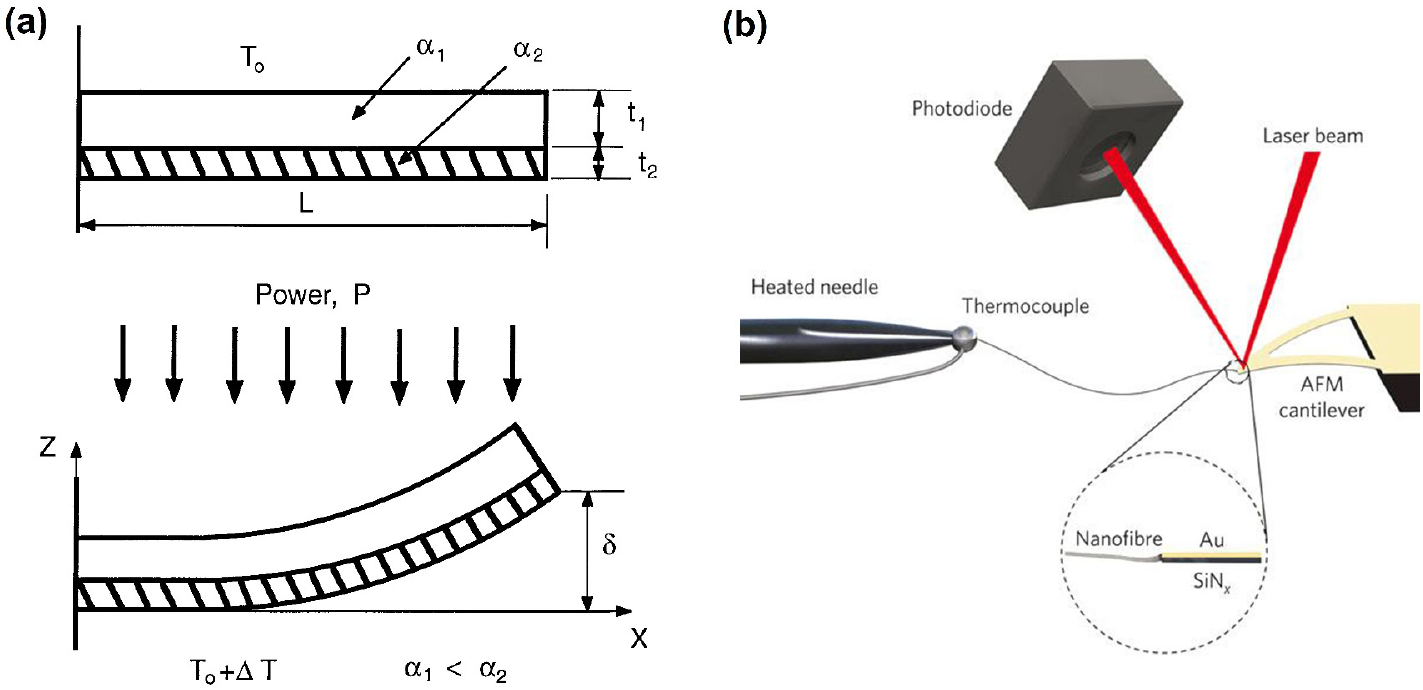}
  \caption
  {(a) Schematic illustration of a bimaterial cantilever beam heated uniformly
   by an incident thermal flux $P$.
   One end of the cantilever is connected to a bulk reservoir of ambient temperature $T_0$,
   while the other end experiences a deflection due to the unbalanced thermal expansion of
   the two beam materials.
   The meanings of the symbols are explained in the main text.
   Reprinted with permission from \cite{Lai97113}. Copyright 1997 Elsevier.
   (b) An \gls{afm} setup which includes a bimaterial cantilever-based calorimeter
   for the measurement of heat transport through one-dimensional nanosystems.
   In the depicted setup, a nanofibre connects the bimaterial cantilever to a thermocouple.
   The deflection of the cantilever is measured by focusing a laser beam onto the cantilever
   and receiving the reflected beam with a photodetector.
   Reprinted with permission from \cite{She10251}. Copyright 2010 Springer Nature.
   }
  \label{figure2-principle-cantilever-calorimeter}
 \end{figure}

 The design and implementation of the bimaterial cantilever-based calorimeter
 was pioneered by Gimzewski and co-workers \cite{Gim94589,Gimzewski95}.
 They have demonstrated a calorimetric technique with a sensitivity to heat flux of 1 nW,
 and measured the heat generation of the catalytic conversion
 of hydrogen and oxygen to form water over a platinum surface \cite{Gim94589,Gimzewski95}.
 The same device has been used for measurements of photothermal power with a resolution of 100 pW \cite{Bar9479,Bar943793}.
 Figure \ref{figure2-principle-cantilever-calorimeter}(a) illustrates
 the working principle of the cantilever-based calorimeter.
 The incident heat current (or thermal power) results in a deflection of the cantilever
 due to the mismatch in thermal expansion of the two beam materials.
 The cantilever deflection can be detected and measured by optical means,
 from which the temperature profile along the cantilever can be deduced.

 Assuming that temperature is constant at any cross section of the cantilever,
 the relation between the cantilever deflection 
 and temperature profile is described by \cite{You01}
\be
 \frac{d^2 \delta}{dx^2} = 6 (\alpha_1 - \alpha_2) \frac{t_1 + t_2}{t_2^2 K} \Delta T(x).
 \label{general-formula-bimaterial-cantilever}
\ee
 Here, $\delta(x)$ is the vertical deflection of the cantilever at position $x$ along its length $L$,
 $\Delta{T}(x)$ is the change of local temperature at $x$ with respect to the ambient temperature $T_0$,
 $t_i$ ($\alpha_i$) is the thickness (thermal expansion coefficient) of the $i$th layer ($i=1,2$) of the beam,
 and $K$ is a constant determined by the thicknesses and Young's moduli of the two beam materials.

If the thermal power $P$ is fed to the loose end of the cantilever,
such as by a laser beam, the temperature profile along the cantilever $\Delta T(x)$
can be evaluated via the Fourier's law as
\be
 \Delta T(x) = \frac{ P } { w (\kappa_1 t_1 + \kappa_2 t_2)} x,
 \label{temperature-power-one-end-bimaterial-cantilever}
\ee
where $\kappa_i$ is the thermal conductivity of the $i$th layer and $w$ is the cantilever width.
It is assumed that most of the heat absorbed by the cantilever
is finally conducted into the bulk reservoir at temperature $T_0$
through the fixed end,
and the heat dissipation through air is negligible.
The cantilever deflection at the free end of the cantilever can be obtained by
solving \Eq{general-formula-bimaterial-cantilever} as \cite{Bar943793}
\be
 \delta(x = L) = 2 (\alpha_1 - \alpha_2) \frac{t_1 + t_2}{t_2^2 K} \frac{L^3 P} {w (\kappa_1 t_1 + \kappa_2 t_2)}.
 \label{one-end-bimaterial-cantilever}
\ee
Based on \Eq{temperature-power-one-end-bimaterial-cantilever} and \Eq{one-end-bimaterial-cantilever},
the temperature profile $\Delta T(x)$ and the incident heat current $J=P$
are determined simultaneously once $\delta(x=L)$ is measured.

 Figure \ref{figure2-principle-cantilever-calorimeter}(b) depicts an
 experimental setup in which the bimaterial cantilever-based calorimeter
 is integrated into an \gls{afm}.
 Such a setup has been employed to study thermal transport in one-dimensional nanosystems \cite{She10251}.
 As shown in \Fig{figure2-principle-cantilever-calorimeter}(b),
 a nanofibre connects the bimaterial cantilever to a micro thermocouple.
 The temperature of the thermocouple can be adjusted to create a thermal gradient across the cantilever,
 and the deflection of the cantilever is measured by focusing a laser beam onto the cantilever
 and receiving the reflected beam with a photodetector.
 Note that the heat current reaching the cantilever ($J_c$) consists of two components:
 the heat conducted through the nanofibre ($J_n$) and
 the heat generated by the laser beam ($J_l$).
 Since $J_l$ is controllable (by tuning the laser intensity)
 and $J_c$ is measurable from the cantilever deflection via \Eq{one-end-bimaterial-cantilever},
 the heat current through the nanofibre can be determined as $J_n = J_c - J_l$.

 Using the setup of \Fig{figure2-principle-cantilever-calorimeter}(b),
 Shen \emph{et al.} have measured the thermal conductivity
 of polyethylene nanofibres \cite{She10251}.
 They kept the laser power $J_l$ constant and changed the temperature of the thermocouple from $T_a$ to $T_b$
 (here $a$ and $b$ label two different temperatures).
 As a result, the heat conducted through the nanofibre varies from $J_{n,a}$ to $J_{n,b}$,
 and $J_{n,a}-J_{n,b}=J_{c,a}-J_{c,b}$.
 Thus, by measuring $J_{c,a}$ and $J_{c,b}$ with the bimaterial (Au/Si$_3$N$_4$)
 cantilever-based calorimeter and using the Fourier's law of \Eq{fourier-law-1},
 the thermal conductance of the nanofibre is
 determined by $\kappa=({J_{c,a}-J_{c,b}})/({T_{a}-T_{b}})$.
 Shen \emph{et al.}'s experimental findings indicate that
 nanofibres with diameters of $50\sim500$ nm and lengths up to tens of millimeters
 may possess thermal conductivities larger than many pure bulk metals.

 It is interesting to note that the \gls{zcc}-based definition of local temperature
 discussed in \Sec{subsubsec:zcc} is, in principle, realizable
 with the use of a bimaterial cantilever-based calorimeter.
 For instance, to measure the local temperature $T^\ast$ of a nanosystem,
 the fixed end of the cantilever (the probe) can be connected to
 a metal lead with a tunable $T_p$ and $\mu_p$,
 while the loose end of the cantilever is placed close to
 the examined nanosystem to achieve local equilibration.
 Then, one tunes $T_p$ and $\mu_p$ until the \gls{zcc} is satisfied,
 which means $I_p = 0$ (no electric current through the metal lead,
 measured by an ammeter)
 and $J_p = 0$ (no deflection on the bimaterial cantilever),
 and the local temperature is determined as $T^\ast = T_p$.
 Of course, there are surely many practical issues to address before
 such an experiment can be actually conducted.

\subsubsection{Resistance thermometry-based calorimeters} \label{subsec:resistance-thermometry-based-calorimeter}

\begin{figure}[htbp]
  \centering
  \includegraphics[width=0.85\columnwidth]{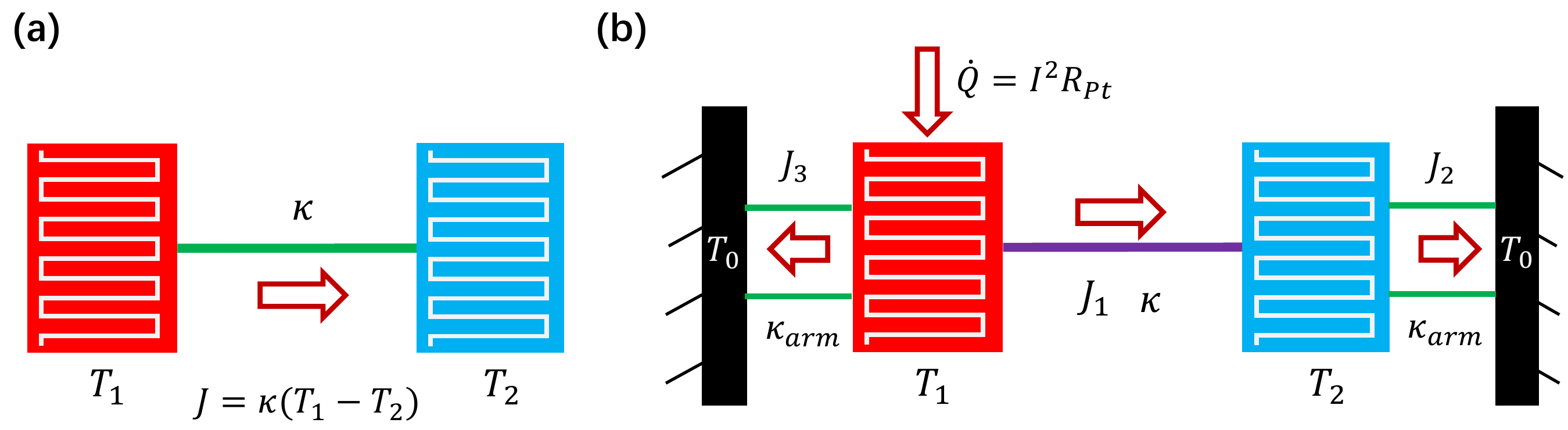}
  \caption
  {
   (a) Schematic illustration of the setup and working principle of a resistance thermometry-based calorimeter.
   The nanopads 1 (red) and 2 (blue) are connected by a conductor with a thermal conductance $\kappa$.
      The temperature of each nanopad is measured by a resistance thermometer (the serpentine white line),
      and the heat current through the conductor is determined as $J = \kappa (T_1 - T_2)$.
   (b) Schematic illustration of measurement of the
   thermal conductance $\kappa$ of a one-dimensional nanobeam (purple).
   Each of the nanopads is suspended to a bulk material (black) of temperature $T_0$
   by an arm (green) with a thermal conductance $\kappa_{\rm arm}$.
   A Pt resistance line (serpentine white line) with an electric resistance $R_{\rm Pt}$
   is integrated on each nanopad, which serves as both a resistance heater and a thermometer.
   The detailed procedure of measuring $\kappa$ is explained in the main text.
   }
  \label{figure4-priciple-resistance-thermometer}
 \end{figure}

 Another widely used type of calorimetric device is the resistance thermometry-based calorimeter.
 Figure \ref{figure4-priciple-resistance-thermometer}(a)
 illustrates the setup and working principle of a resistance thermometry-based calorimeter.
 Two nanopads of different temperatures ($T_1$ and $T_2$) are thermally connected
 by a bridging conductor with a thermal conductance $\kappa$.
 Except the bridging conductor, the nanopads are considered to be
 thermally isolated from the surrounding environment,
 with other means of heat exchange (such as heat radiation)
 presumed to be negligible.
 The temperature of each nanopad is measured by a resistance thermometer
 which includes a temperature-dependent resistive element
 (the serpentine white lines in \Fig{figure4-priciple-resistance-thermometer}).
 The heat current flow from one nanopad to another through the bridging conductor
 is thus determined by \Eq{fourier-law-1}, provided the
 thermal conductance of the conductor $\kappa$ is known.

 The setup displayed in \Fig{figure4-priciple-resistance-thermometer}(a)
 can be utilized to study thermal transport in one-dimensional nanostructures.
 For instance, \Fig{figure4-priciple-resistance-thermometer}(b) illustrates
 how the unknown $\kappa$ of a one-dimensional nanobeam (purple) can be determined
 by using a resistance thermometry-based calorimeter.
 To carry out the measurement, each of the two nanopads is suspended to
 a bulk material of ambient temperature $T_0$ by an arm of
 thermal conductance $\kappa_{\rm arm}$.
 A serpentine Pt resistance line is integrated on each pad
 and is used as both a resistance heater and a thermometer.
 Measurement of $\kappa$ of the central nanobeam proceeds as follows \cite{Shi03881,Lee17371}.

 As illustrated in \Fig{figure4-priciple-resistance-thermometer}(b),
 an electric current $I$ flows through the serpentine Pt resistance line
 integrated on the pad 1, which generates a Joule heat with the rate $\dot{Q} = I^2 R_{\rm Pt}$,
 where $R_{\rm Pt}$ is the electric resistance of the Pt line.
 The Joule heat raises the temperature of the pad 1 to $T_1 \equiv T_0 + \Delta T_1$,
 with $T_0$ being the environmental temperature.
 Some of the Joule heat transfers across the central nanobeam (denoted as $J_1$),
 and raises the temperature of the pad 2 to $T_2 \equiv T_0 + \Delta T_2$
 (measured by the Pt resistance line in the pad 2),
 while the rest of the generated heat dissipates into the environment
 through the suspended arms in contact with the pad 1 ($J_3$) and pad 2 ($J_2$).
 Again, other means of heat dissipation (such as by air or by radiation)
 are considered to be negligible.
 In the steady state, the heat dissipation into the environment
 through the suspended arms should balance
 the Joule heat generated by the incident electric current.
 This implies $\dot{Q} = J_1 + J_3$ and $J_1 = J_2$.
 Assuming the Fourier's law of \Eq{fourier-law-1} is valid for $J_i (i = 1,2,3)$,
 one has the following relations \cite{Lee17371,Shi03881}:
 \begin{align}
 J_1 & = \kappa (T_1 - T_2) = \kappa (\Delta T_1 - \Delta T_2),
 \label{heat-current-J1}
 \\
 J_2 & = \kappa_{\rm arm} (T_2 - T_0) = \kappa_{\rm arm} \Delta T_2,
 \label{heat-current-J2}
 \\
 J_3 & = \kappa_{\rm arm} (T_1 - T_0) = \kappa_{\rm arm} \Delta T_1.
 \label{heat-current-J3}
 \end{align}
The thermal conductance of the nanobeam can then be easily deduced as
\be
 \kappa= \frac{I^2 R_{\rm Pt}\, \Delta T_2}{\Delta T^2_2-\Delta T^2_1}.
 \label{kappa-nanobeam-1}
\ee

Using the above technique, Shi and co-workers have measured
the thermal conductances of nanowires
and carbon nanotubes, and observed a
dependence of $\kappa$ on temperature \cite{Shi03881,Kim01215502,Yu051842,Li032934}.
Lee \emph{et al.} have measured the electronic contribution
to the thermal conductance of a VO$_2$ nanowire
and found the breakdown of the Wiedemann--Franz law \cite{Kit04}
at high temperatures ranging from 240 to 340 K
in the vicinity of its metal-insulator transition \cite{Lee17371}.
The violation of the Wiedemann--Franz law is
attributed to the formation of a strongly correlated,
incoherent non-Fermi liquid, in which charge and heat are independently transported.

 \begin{figure}[htbp]
  \centering
  \includegraphics[width=0.9\columnwidth]{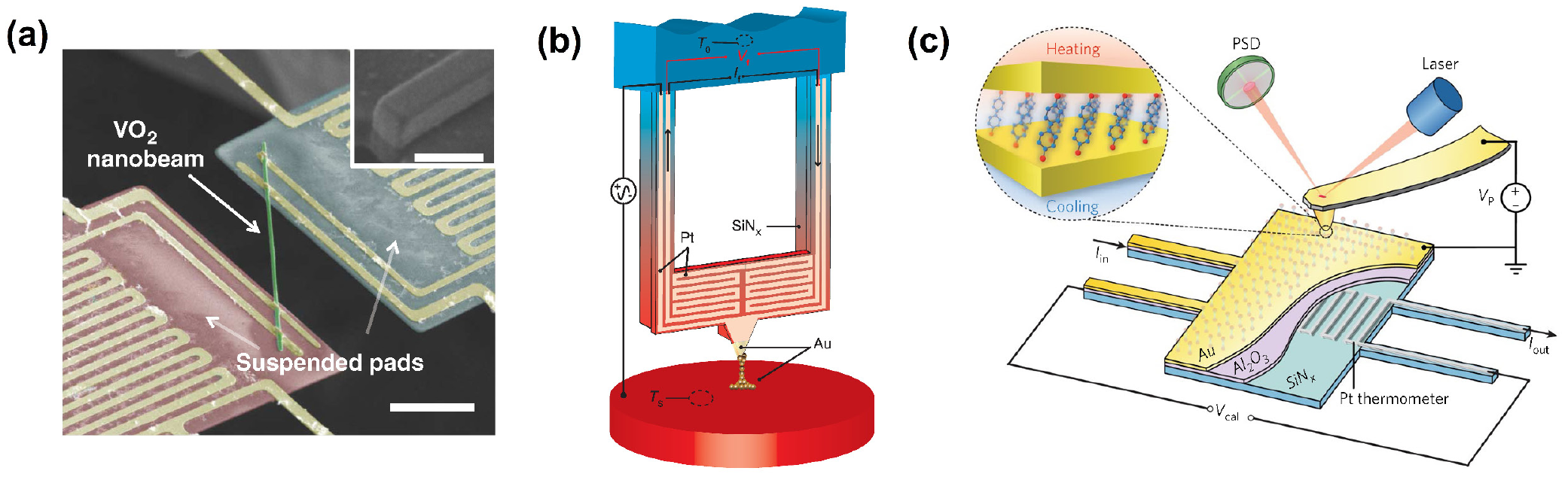}
  \caption
  {
   Several experimental setups using resistance thermometry-based calorimeter.
   (a) False-color \gls{sem} image of a nanojunction setup.
   A VO$_2$ nanobeam bridges two suspended pads, and heat transports
   from the heated pad (red) to the sensing pad (blue) through the nanobeam (green).
   The Pt resistance line integrated on each pad acts as both a micro-heater and a thermometer.
   The inset shows the \gls{sem} image of the rectangular cross section of the nanobeam.
   Reprinted with permission from \cite{Lee17371}. Copyright 2017 AAAS.
   (b) A custom-fabricated calorimetric \gls{sthm} probe
   for thermal conductance measurement of atomic contact junctions.
   The serpentine Pt thermometer is integrated on the probe of the \gls{sthm}.
   Reprinted with permission from \cite{Cui171192}. Copyright 2017 AAAS.
   An \gls{afm}-based calorimetric device for investigating the
   Peltier cooling effect in molecular junctions.
   A serpentine Pt thermometer is embedded into the SiN$_x$ substrate.
   Reprinted with permission from \cite{Cui18122}. Copyright 2018 Springer Nature.
  }
  \label{figure5-resistance thermometry-based-calorimeter}
 \end{figure}

 Cui \emph{et al.} have investigated the
 thermal transport in single atomic junctions \cite{Cui171192,Seg171125}.
 Figure \ref{figure5-resistance thermometry-based-calorimeter}(b) shows the experimental setup,
 in which a custom-fabricated high-resolution Pt resistance thermometer is integrated on the
 probe of an \gls{sthm}.
 An atomic junction is formed when the \gls{sthm} probe is in contact with a heated metallic substrate.
 Driven by the temperature difference between the substrate and the probe,
 the heat current across the atomic junction
 gives rise to a change of the probe temperature,
 which can be quantified by the resistance change of the Pt thermometer.
 Consequently, the incident heat current can be determined from
 \Eq{fourier-law-1} with the known thermal conductance of the probe.

 Cui \emph{et al.} have also investigated local heating and
  Peltier cooling in molecular junctions by integrating
 a Pt thermometer-based calorimeter into a conducting-probe \gls{afm};
 see Figure \ref{figure5-resistance thermometry-based-calorimeter} \cite{Cui18122}.
 Molecular junctions are created by placing the tip of an \gls{afm} probe
  in gentle contact with a self-assembled monolayer of molecules formed on the substrate.
 Current flow across the molecular junctions gives rise to heat generation or cooling
 due to Joule heating or Peltier cooling effect,
 which in turn results in a temperature change $\Delta T$ of
 the calorimeter on the substrate that can be quantified by the integrated Pt resistance thermometer.
 The total cooling or heating power $P$ can be directly determined from by $P = \kappa \Delta T$,
 where $\kappa$ is the predetermined thermal conductance of the device \cite{Cui171192,Cui18122}.

\subsubsection{Advantages and limitations of nanocalorimetric techniques}
\label{subsubsec:strengthes-and-limitations-calorimetry}

 The present nanocalorimetric techniques allow for measuring heat flow
 through nano-devices with a picowatt resolution.
 For the bimaterial cantilever-based calorimeter,
 Lai \emph{et al.} have optimized the beam materials and
 the thickness ratio of two beam layers
 to achieve measurement of thermal power within 76 pW \cite{Lai97113}.
 Varesi \emph{et al.} have
 modulated the incident laser at sufficiently high frequency,
 which has enabled detection of thermal power within 40 pW \cite{Var97306}.
 Canetta and Narayanaswamy have used two laser beams with different wavelengths
 to heat up the cantilever and detect its deflection.
 They have also optimized the geometry of the cantilever.
 The resulting calorimeter is capable of measuring thermal power
 with a resolution of less than 1 pW \cite{Can13103112}.
 Regarding the resistance thermometry-based calorimeter,
 Sadata \emph{et al.} have used a high-resolution resistance thermometer
 with a sensitivity to temperature in the range of 20-100 $\mu$K \cite{Sad12084902},
 and reported measurement of heat currents with a resolution of $\sim$5 pW \cite{Sad13163110}.

 In practical applications, the main difficulty for reaching higher resolutions is the noise.
 For a cantilever-based calorimeter, noise may arise from external sources
 such as electronic components, mechanical vibrations of the cantilever beam \cite{Sal972480},
 and intensity fluctuations in the laser source \cite{Bar9479}.
 For a resistance thermometry-based calorimeter, the noise can be either intrinsic
 (such as Johnson--Nyquist noise and shot noise) or non-intrinsic
 (such as low frequency noise and electronic components noise) \cite{Sad12084902}.

 We now relate the nanocalorimetric techniques to local temperature measurements.
  As discussed in \Sec{subsec:bimaterial-cantilever-based-calorimeter},
 in principle, with a bimaterial cantilever-based-calorimeter,
 the incident heat current and the local temperature profile of the cantilever
 are determined simultaneously by measuring the cantilever deflection; see
 \Eqs{general-formula-bimaterial-cantilever}--\eqref{one-end-bimaterial-cantilever}.
 However, with most of the existing experimental instruments and setups,
 the measurement of heat current usually requires the knowledge of temperature gradient
 (and thus the local temperature) as a prerequisite condition,
 \emph{i.e.}, the heat current is determined by measuring the temperature,
 rather than the other way around.

There are also some practical limitations concerning the application of nanocalorimeters.
One of them is spatial resolution.
While a local temperature has been defined for a single molecular junction
and atom-sized regions of nanostructures \cite{Cas11165419,Ber14235438,Ber15125407},
experimental measurements of heat current were done for
nanojunctions consisting of a large number of molecules \cite{Cui18122}
and one-dimensional nanowires with diameters of hundreds of nanometers \cite{Lee17371,Lee158573,Shi03881,Can13105002,Kim01215502,Yu051842,Li032934,Can14104901}.
Moreover, as discussed in \Sec{subsec:nanocalorimetry},
the present nanocalorimetric techniques presume that
the Fourier's law of \Eq{fourier-law-1} generally holds for nano-sized systems.
However, it is well-known that the Fourier's law does not properly
describe ballistic phonon transport when the material dimension
is similar to or even smaller than the phonon mean free path
\cite{Cha08075903,Roy08062102,Dub09115415,Dub09042101,Yan1085,Joh13025901}.

\subsection{Measurement of local temperatures of non-equilibrium nanosystems} \label{subsec:mueasurement-local-temp}

\subsubsection{Mechanically-controlled break junction techniques} \label{subsubsec:break-junction}

\begin{figure}[htbp]
  \centering
  \includegraphics[width=0.85\columnwidth]{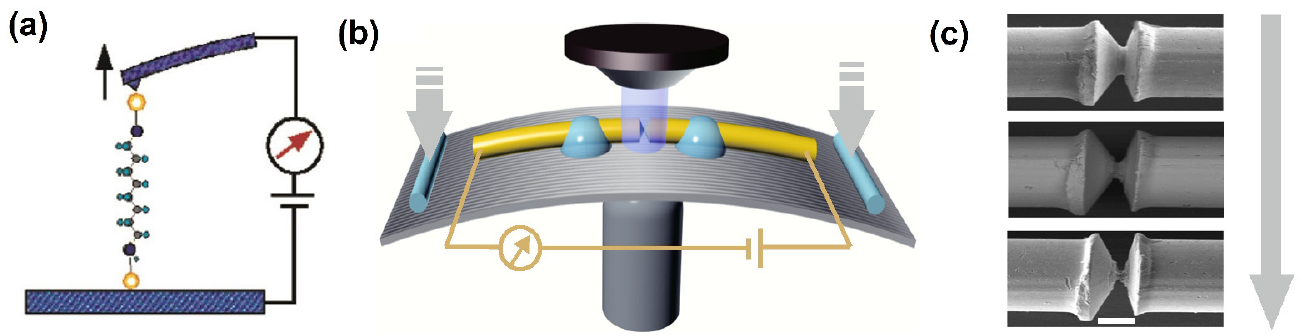}
  \caption
  {
   (a) A schematic of an AFM-based \gls{mcbj} setup.
   The cantilever (probe) is gently brought into contact with a
   molecule adsorbed on the bottom substrate.
   Continuously pulling the tip of the probe away from the
   substrate finally breaks the molecular junction.
   The breaking force is recorded.
   Reprinted with permission from \cite{Hua061240}. Copyright 2006 American Chemical Society.
   (b) A schematic of \gls{mcbj} for metallic point contacts.
   A metallic nanowire (yellow) is fixed to two electrodes (blue) at a flexible substrate.
   One makes a notch near the middle of the nanowire to reduce the cross section of the line.
   A push rod is integrated under the substrate.
   A vertical movement of the rod bends the substrate.
   As the beam is bent, the metal wire starts to elongate,
   which results in the reduction of the cross section at the notch and finally results in a complete fracture of the metallic nanowire.
   (c) \gls{sem} images of the notched Au nanowire during the stretching process.
   The scale bar marks the length of $50 \mu$m.
   (b) and (c) reprinted with permission from \cite{Zha1934}. Copyright 2019 Springer Nature.
  }
  \label{figure8-MCBJ-bed}
 \end{figure}

 \gls{mcbj} has become a versatile and useful platform for the study of
 electron transport through single molecules,
 with which many important measurements/observations have been made
 \cite{Ree97252,Liu119048,Per14830,Bee06026801,Hui093909,Bru1235,Par101370}.
 Figure \ref{figure8-MCBJ-bed} gives schematic representations of
 two typical \gls{mcbj} setups.
 Specifically, Figure \ref{figure8-MCBJ-bed}(a) depicts an
 \gls{afm}-based \gls{mcbj} setup, in which
 the cantilever (or probe)  is gently brought into contact with
 a molecule adsorbed on the bottom substrate.
 Likewise, \gls{stm}-based \gls{mcbj}s have also been constructed,
 in which the probe also acts as an electrode.
 Figure \ref{figure8-MCBJ-bed}(b) demonstrates an \gls{mcbj} setup
 with a metallic point contact, in which the mechanical force
 at the contact is controlled by adjusting a push rod
 integrated under the substrate.
 More details about the design and working mechanisms of \gls{mcbj}
 setups can be found in Refs. \cite{Agr03377,Xia134845,Hua15889,Wan17375}.

\begin{figure}[htbp]
  \centering
  \includegraphics[width=0.65\columnwidth]{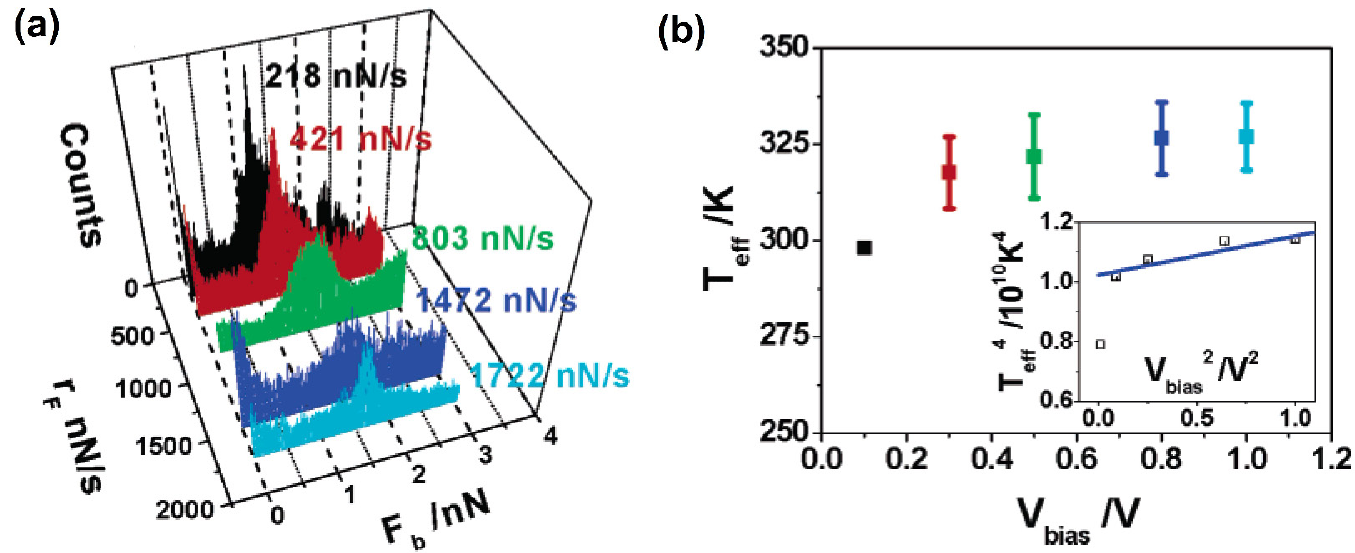}
  \caption
  {
   (a) Breakdown force histograms are constructed from $\sim500$ measurements
   for individual force curves at each loading rate.
   (b) Local temperature of an octanedithiol molecule,
   extracted from $F^*$ as a function of bias voltage.
   The inset shows that the local temperature has a fourth-power dependence on the square of the bias.
   The blue line in the inset is the best fit of the data using \Eq{local-temp-of-ep}.
   Reprinted with permission from \cite{Hua061240}. Copyright 2006 American Chemical Society.
  }
  \label{figure9-MCBJ-octanedithiol-junctions}
 \end{figure}

 Based on an \gls{mcbj} setup,
 one method for measuring a local temperature is to study the force at which an molecular junction breaks with
 the increasing current-induced local heating.
 The idea behind the method is that the breakdown process is thermally activated.
 Consequently, the higher the local temperature is in the molecule-electrode contact,
 the smaller the average breakdown force it requires \cite{Hua061240,Hua07698}.
 A large number of breaking experiments were carried out repeatedly
 to obtain breakdown force histograms for subsequent statistical analyses; see \Fig{figure9-MCBJ-octanedithiol-junctions}(a).
 The broad distribution of the breakdown force reflects different molecule-electrode contact configurations,
 and the peak position gives the most probable force $F^*$ required to break a molecular junction.
 Local temperatures $T^*_{ph}$ can be extracted from the most probable breakdown force $F^*$ as \cite{Eva01105}
\be
 F^*=\frac{E_b}{x_\beta} + \frac{k_BT^*_{ph}}{x_\beta}\, \ln\bigg(\frac{r_Ft_{\rm off}x_\beta}{k_BT^*_{ph}}\bigg).
 \label{local-temp-breakdown-force}
\ee
 Here, $E_b$ is the dissociation energy of the junction, $x_\beta$ is the average thermal bond length along the pulling direction until breaking,
 $r_F$ is the force loading rate,
 and $t_{\rm off}$ is the lifetime of the bond before it breaks down.
 Considering current-induced forces which may reduce the dissociation energy barrier and thus yield a smaller breakdown force,
 $t_{\rm off}$ can be expressed by $t_{\rm off}=t_D{\rm exp}\lbrack -{(E_b-\alpha V_{\rm bias})}/{k_BT^*_{ph}} \rbrack$ with $t_D$ the diffusion relaxation time \cite{Smi04S472}.
 The empirical coefficient $\alpha$ describes the influence of current-induced forces.

 The temperature $T^*$  in \Eq{local-temp-breakdown-force} was originally described as an effective temperature of the molecular junction
 in Ref.\ \cite{Hua061240}.
 Nevertheless, as it will be discussed below, the agreement between the experimental results and theoretical estimates of \Eq{local-temp-of-ep}
 indicates that the obtained $T^*$ can provide a quantitative measure of the thermal excitation of local phonon modes,
 We thus assign $T^*$ to the local temperature of local phonon modes ($T^*_{ph}$) in the following text.

 Huang \emph{et al.} have investigated current-induced local heating in single molecular junctions \cite{Hua061240}.
 They used an \gls{stm}-based \gls{mcbj} to measure the variation of local temperature versus bias voltage
 $V_{\rm bias}$ for an octanedithiol molecule connected in the junction.
 The results show that local heating raises the local temperature $T^*_{ph}$ $\sim$30 K above the ambient temperature $T_0$ at a bias voltage of $\sim$1 V.
 Higher bias makes the single molecule junction unstable \cite{Yas971069,Ita9911163}.
 The experimental data shown in the inset of \Fig{figure9-MCBJ-octanedithiol-junctions}(b)
 roughly agree with theoretical predictions from \Eq{local-temp-of-ep} at high bias.
 This agreement suggests that the electron-phonon scattering
 dominates the local heating in the octanedithiol junction, while
 the effect of electron-electron scattering in the junction is relatively weak \cite{Hua061240}.
 A possible reason for the deviation in the low-bias region
 is that a low bias voltage cannot provide sufficient energy
 to excite the local phonon modes in the molecule.

\begin{figure}[htbp]
  \centering
  \includegraphics[width=0.6\columnwidth]{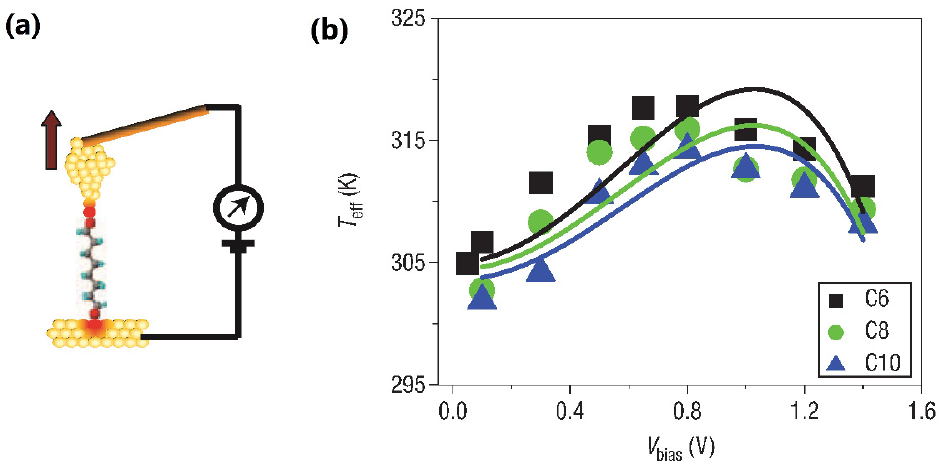}
  \caption
  {
   (a) A schematic of \gls{stm}-based \gls{mcbj} setup.
   Molecules may bridge between the \gls{stm} probe and the substrate via covalent bonds and form a molecule junction.
   Pulling the probe away from the substrate results in a breakdown of individual molecules from contacting the electrodes.
   (b) Local ionic temperature of a molecular junction, for three different types of molecule n-alkanedithiol, with $n=6$ (squares), $n=8$ (circles), and $n=10$ (triangles).
   The solid lines are best fit to the theoretical estimation of \Eq{local-temp-of-ep-ee}.
   Reprinted with permission from \cite{Hua07698}. Copyright 2007 Springer Nature.
  }
  \label{figure11-MCBJ-heating-effect}
 \end{figure}

 Huang \emph{et al.} have also measured local phonon temperatures of
 alkanedithiols of different lengths covalently bonded to two gold electrodes \cite{Hua07698}.
 Their experimental results shown in \Fig{figure11-MCBJ-heating-effect}(b)
 indicate that $T^*_{ph}$ first increases with voltage bias, and then decreases after reaching a peak value.
 The measured data are in general agreement with the theoretical prediction of \Eq{local-temp-of-ep-ee} (solid lines)
 regardless of the length of the molecule.
 This suggests that both electron-phonon and electron-electron scattering occur in the alkanedithiols junctions.
 In the same experiment, \Fig{figure11-MCBJ-heating-effect}(b) also suggests
 that the lengths of alkanedithiols have a non-negligible impact on local temperature.
 Under a given bias voltage the local temperature increases with decreasing molecular length.
 Based on \gls{dft} calculations, Chen \emph{et al.} have attributed such a behavior
 to the insulating character of alkanedithiols \cite{Che05621}.
 In other words, the increasing electronic resistance of longer alkanedithiols
 weakens the Joule heating effect in nanojunctions.

 For atom-sized contacts, an alternative method to evaluate local temperatures
 is to measure the change of electric conductance when the junction heats up.
 The principle is described as follows.
 The atomic configuration of the atom-sized contact corresponds to
 a local potential minimum.
 When the contact is stretched by an external force,
 the configuration will become unstable and then transform to
 a different configuration corresponding to a new local potential minimum \cite{Kra96228}.
 Since the electric conductance of atom-sized metal contacts depends sensitively on the atomic configuration of the junction,
 a rearrangement in the contact geometry will lead to a sudden jump in electric conductance.
 For example, when there are two stable atomic configurations,
 a two-level conductance fluctuation emerges \cite{Rai8911561,Nee113593}.
 The atomic rearrangement is achieved by atom jump between two distinct positions.
 The atom jump is a thermally activated process.
 Thus, the rate of the conductance fluctuation depends on local temperature as follows
 \be
 \nu=\nu_0 \exp\bigg({-\frac{E_B-\alpha V_{\rm bias}}{k_BT^*_{ph}}}\bigg),
 \label{TLF-frequency}
 \ee
 where $\nu_0$ is the transition rate between two stable atomic configurations,
 $E_B$ is an energy gap between two configurations,
 and $\alpha$ characterizes the influence of current-induced force \cite{Tod013606}.

\begin{figure}[htbp]
  \centering
  \includegraphics[width=\columnwidth]{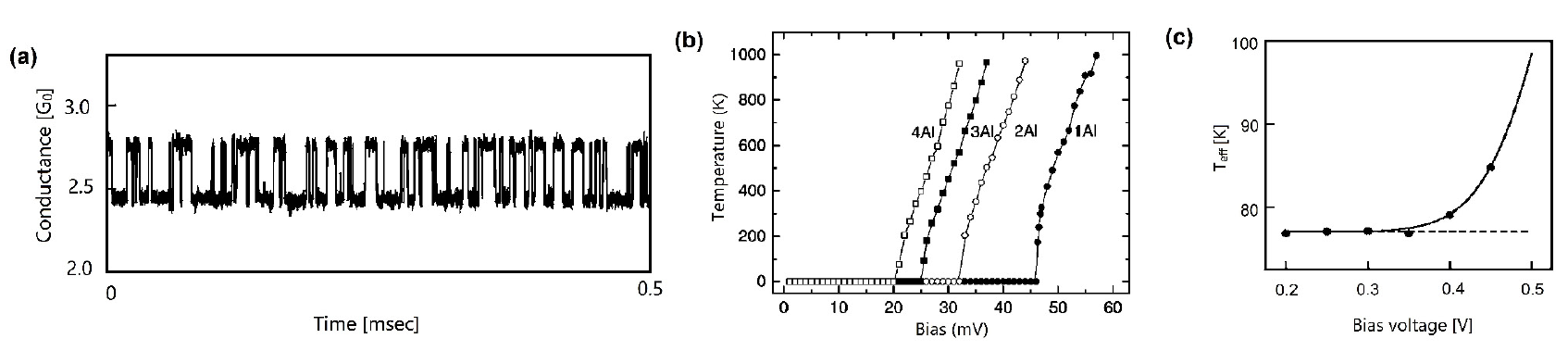}
  \caption
  {
   (a) A typical two-level fluctuation trace of electric conductance obtained at 0.35 V for atom-sized metal contacts.
   (b) Local temperature calculated from \Eq{def-thermal-power}
   as a function of bias for Al nanowires of different lengths.
   The calculation assumes there is no heat dissipation into the bulk electrodes.
   (a) and (c) reprinted with permission from \cite{Tsu07133121}. Copyright 2007 American Institute of Physics.
   (c) Local ionic temperature of an atomic-size Zn contact with the ambient temperature $T_0=77$ K.
   The solid curve is a fit to the relation $T^*_{ph}=aV_{\rm bias}^3$.
   The dash line is the initial slope of the curve.
   (b) reprinted with permission from \cite{Yan05041402}. Copyright 2005 American Physical Society.
  }
  \label{figure10-MCBJ-atomic-junctions}
 \end{figure}

 Using the \gls{mcbj} setup, Tsutsui \emph{et al.} have studied local heating in
 atom-sized Au contacts \cite{Tsu065334}.
 Figure \ref{figure10-MCBJ-atomic-junctions}(a) demonstrates
 an entire trace of two-level fluctuations of electric conductance under a given bias voltage.
 The frequency $\nu$ is derived by counting the number of these conductance switching cycles in the entire trace.
 A large number of repeated measurements were performed to
 generate a sufficient amount of data for the subsequent statistical analysis.
 This gives rise to the most probable rate of the conductance fluctuation,
 with which the local temperature is determined from \Eq{TLF-frequency}.
 In Figure (5) of Ref. \cite{Tsu065334}, the local temperature of the Au contact is plotted as a function of bias.
 The measured data exhibit a $\sqrt{V_{\rm bias}}$ dependence of $T^*$,
 which agrees well with the theoretical prediction of \Eq{local-temp-of-ep},
 though the measured data are distributed within a limited range of bias voltage.
 This also suggests that electron-phonon interactions contribute significantly to
 the current-induced local heating at the atom-sized Au contact,
 while the contribution of electron-electron interaction may be small at a high bias.

 Yang \emph{et al.} have investigated local heating in Al nanowires in a junction using DFT calculations.
 They calculated the local temperature of the nanowire with \Eq{def-thermal-power}
 by assuming that there is no heat dissipation into the bulk electrodes \cite{Yan05041402}.
 They have found that, a bias larger than the lowest vibrational frequency of the junction
 is necessary to generate heat.
 Once above that threshold, a severe overheating occurs in the nanowire,
 which results in a drastic elevation of local temperature
 with a slightly increased bias by only a few mV; see \Fig{figure10-MCBJ-atomic-junctions}(b).
 Tsutsui \emph{et al.} have studied local heating in atom-sized Zn contacts
 at the ambient temperature of $T_0=77$ K \cite{Tsu07133121}.
 Figure \ref{figure10-MCBJ-atomic-junctions}(c) exhibits a similar dependence of local temperature on bias.
 The local temperature remains unchanged when the bias is below 0.35 V,
 while it rises rapidly as the bias increases further.
 The experimental data fit well to the relation $T^*_{ph}=aV_{\rm bias}^3$
 with $a$ being a fitting parameter.
 The sharp increase of local temperature has been attributed to contact overheating \cite{Tsu07133121},
 which implies the heat generated is not efficiently dissipated into the electrodes
 and severely accumulates at the atom-sized Zn contact.

 The \gls{mcbj} techniques have the following advantages:
 They allow for the measurement of local temperature
 in single-molecule and atom-sized junctions under bias voltages.
 Besides measuring local electric properties,
 \gls{mcbj} can also be combined with other techniques in practice,
 such as noise spectroscopy and surface enhanced Raman spectroscopy \cite{Xia134845},
 to characterize voltage noise and Raman scattering.

 Since the present \gls{mcbj} setups can measure local properties of nanojunctions,
 this platform is promising for the practical implementation of \gls{mpc}-based definitions
 such as that described by \Eq{def-MPC-observable}.
 A possible scheme is described as follows.
 By using a specialized detector, one can monitor the variation of
 a certain local property of the nanosystem as the probe approaches the system.
 When the probe is weakly coupled to the system,
 one can tune the temperature and the applied bias on the probe
 until the electric current through the probe vanishes
 and the local property is minimally perturbed by the probe.
 Thus, the local temperature and local electrochemical potential are determined
 to be equal to the probe counterparts.
 Such a scheme would require a high-precision measurement of
 the local property.

 Nevertheless, there are still some limitations to the \gls{mcbj} techniques.
 For instance, it is difficult to precisely control the microscopic details of
 the break process, such as the atomic configuration of the contacts.
 To minimize the uncertainty, a large number of repeated measurements are required to obtain sufficient data for statistical analysis.

\subsubsection{Spectroscopic techniques} \label{subsubsec:spectroscopy}

 An alternative method makes use of spectroscopic techniques, including
 optical interferometry \cite{Asheghi05,Hau14024001}, Raman spectroscopy \cite{Lee11081419} and luminescence \cite{Liu1514879,Oka12705,Eva12328,Kuc1354,Tan1622071}.
 The technical details of the above spectroscopic techniques can be found in relevant reviews \cite{Bri124799,Jaq124301,Kim15129,Far09143001}.
 For spectroscopic techniques, the height, width, and peak position of measured spectral lines reflect local excitations in nanosystems and the distribution of relevant microstates of the particles.
 The principle behind spectroscopic techniques is that the change of local temperature has an impact on the distribution of particles and local excitations in nanosystems,
 which changes the shape and peak position of spectral lines.

\begin{figure}[htbp]
  \centering
  \includegraphics[width=0.9\columnwidth]{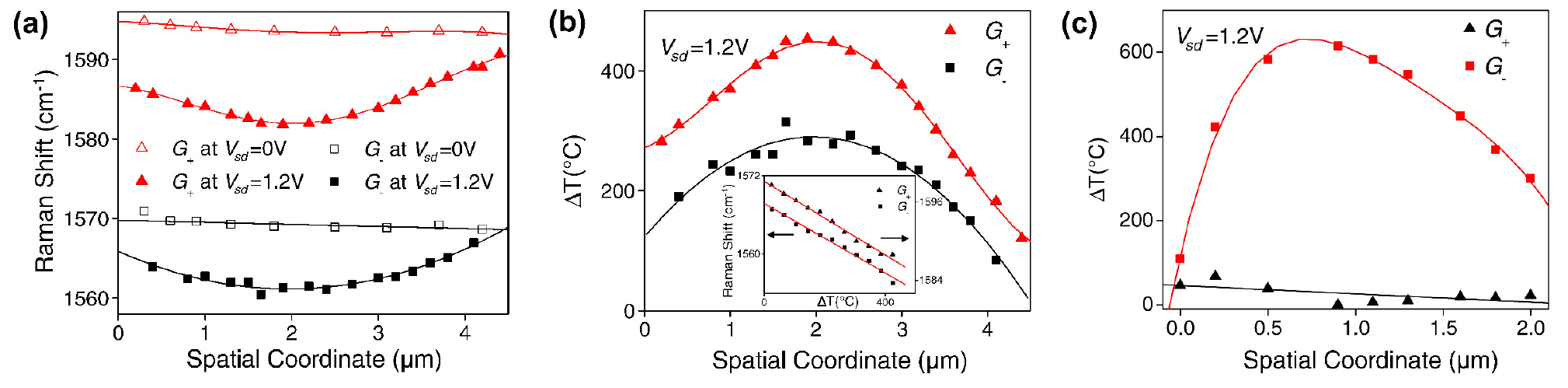}
  \caption
  {
   (a) Spatial profile of Raman shifts of $G_+$ and $G_-$ bands for a 5$\mu$m carbon nanotube at $V_{sd} = 0$ V and 1.2 V.
   (b) Temperature profiles of $G_+$ and $G_-$ bands for 5$\mu$m carbon nanotube at $V_{sd} = $ 1.2 V.
   (c) Temperature profiles of $G_+$ and $G_-$ bands for 2$\mu$m carbon nanotube at $V_{sd} = $ 1.2 V.
   $\Delta T = T^* - T_0$ with $T_0$ being the room temperature.
   Reprinted with permission from \cite{Des09105501}. Copyright 2009 American Physical Society.
  }
  \label{figure17-Raman-Gband-temperature}
 \end{figure}

 Raman spectroscopy has been applied to the study of electrically-heated suspended carbon nanotubes \cite{Bus073618,Des09105501}.
 and to measure local temperature of high energy optical phonons of $G_+$ and $G_-$ bands.
 Generally the $G$ band is related to the different C-C vibrational modes in a carbon nanotube \cite{Mal0951}.
 In carbon nanotubes, the $G$ band is split into many spectra lines around 1580 cm$^{-1}$ \cite{Dre0547}.
 Here, the $G_+$ and $G_-$ bands are characteristic of the transverse and longitudinal optical phonon modes at the $\Gamma$ point, respectively.
 \Figure{figure17-Raman-Gband-temperature}(a) shows the Raman shifts of the $G_+$ and $G_-$ bands, along the length of the nanotube with 5$\mu$m length, for $V_{sd} = 0$ V and 1.2 V.
 The $G$ bands downshift significantly in response to the voltage bias.
 To image the temperature profiles along the nanotubes,
 Deshpande \emph{et al.} have measured the $G$ band shifts with a varying temperature; see the inset of \Figure{figure17-Raman-Gband-temperature}(b) \cite{Des09105501}.
 The local temperature at each point is obtained by dividing the voltage-induced downshift of $G$ bands in \Figure{figure17-Raman-Gband-temperature}(a)
 by the slope of the contour line in the \Figure{figure17-Raman-Gband-temperature}(b)'s inset.
 \Figure{figure17-Raman-Gband-temperature}(b) and (c) show the profiles of phonon temperature along 2 $\mu$m and 5 $\mu$m carbon nanotubes under the bias $V_{sd} = $ 1.2 V, respectively.
 The voltage-induced local heating is prominent in the middle of the nanotubes.
 Particularly,
 the local temperature of the $G_-$ mode significantly varies along the 2 $\mu$m nanotube,
 while the local temperature of the $G_+$ mode is around the ambient temperature.
 This is because the electrons are preferentially coupled with the $G_-$ mode phonons \cite{Wu07027402,Bus073618},
 and induce critical local ionic heating.

 Iof \emph{et al.} have directly monitored the local phonon temperature in a current-heating molecular junction,
 using a surface-enhanced Raman spectroscopy technique that can simultaneously measure both the Stokes, $I_S$, and anti-Stokes, $I_{AS}$, components of the Raman scattering \cite{Iof08727}.
 The idea is that the ratio between the Stokes and anti-Stokes intensities is directly related to their non-equilibrium populations,
 from which a local phonon temperature $T^*_{ph}$ can be extracted by the following equation \cite{Oro08127401}
 \begin{align}
 {\frac{I_{AS}}{I_S}=A_\nu\bigg\lbrack\frac{\omega_L+\omega_v}{\omega_L-\omega_v}\bigg\rbrack^4{\rm exp}\bigg\lbrack\frac{-\hbar\omega_v}{k_BT^*_{ph}}\bigg\rbrack},
 \label{Raman-phonon-temperature}
 \end{align}
 where $I_{AS(S)}$ is the intensity of the anti-Stokes (Stokes) Raman mode,
 $\omega_{L(v)}$ is the frequency of the laser (Raman mode) and $\hbar\omega_v$ is the Raman shift.
 $A_\nu$ is a bias-dependent correction factor related to the ratio of the anti-Stokes and Stokes cross sections.

\begin{figure}[htbp]
  \centering
  \includegraphics[width=0.45\columnwidth]{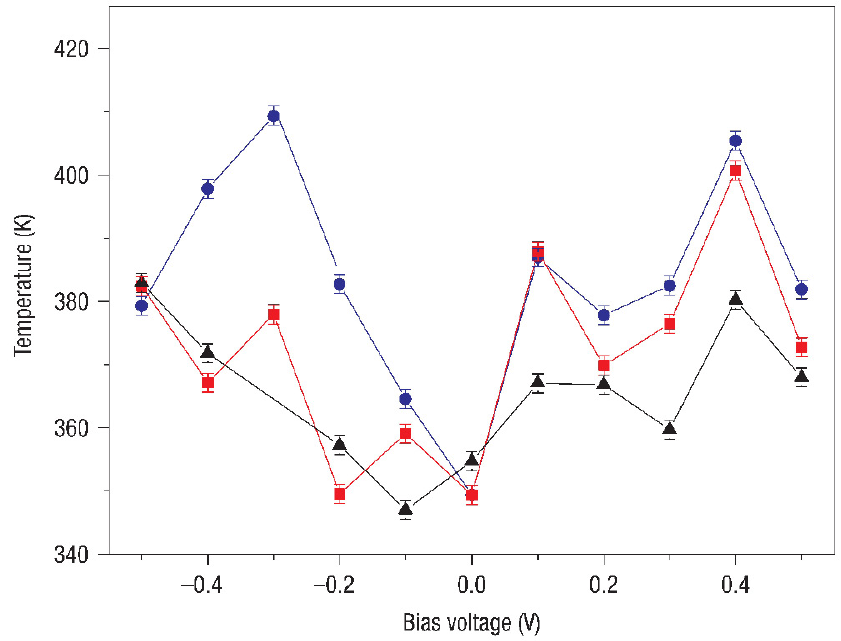}
  \caption
  {
   Plot of $T^*_{ph}$ as a function of bias voltage $V$ for the 1,585 cm$^{-1}$ (black triangles), 1,280 cm$^{-1}$ (blue circles) and 1,083 cm$^{-1}$ (red squares) modes in biphenyldithiol nanojunction.
   While the temperature is slightly different, the overall voltage dependence shows roughly similar features.
   Reprinted with permission from \cite{Iof08727}. Copyright 2008 Springer Nature.
  }
  \label{figure18-Raman-temperature}
 \end{figure}

 In Figure \ref{figure18-Raman-temperature} the local temperature is plotted as a function of voltage bias
 for different phonon modes of a biphenyldithiol nanojunction.
 Although the temperature is slightly different between different modes (due to the different electron-phonon coupling strengths and symmetries),
 the overall voltage dependence shows roughly similar features.
 An apparent decrease in local temperature has also been observed at high bias,
 which roughly agrees with results in \Figure{figure11-MCBJ-heating-effect}(b).
 This agreement suggests that both electron-electron and electron-phonon interaction contribute to local heating in the junction.

 Ward \emph{et al.} have shown that the anti-Stokes component of surface-enhanced Raman scattering could be used to determined the local electron temperature \cite{War1133}.
 The idea is that the intensity of the anti-Stokes scattering $I_{\rm AS}(\epsilon)$ at a specific shift $\epsilon$
 is proportional to the joint \gls{dos} $J(\epsilon) = \int{ g(E + \epsilon) f(E + \epsilon, T^*_e) g(E) \lbrack{ 1- f(E , T^*_e) }\rbrack}dE$ for electrons at energy $E + \epsilon$ and holes at energy $E$, with $g(E)$ being the density of electrons and holes \cite{Ott06307}.
 The integral over the conduction band is related to the local electron temperature \cite{War1133},
 \begin{align}
 {I_{AS}(\epsilon) \propto \int{ f(E + \epsilon, T^*_e) \lbrack{ 1- f(E, T^*_e) }\rbrack}dE \simeq \frac{\epsilon} {{\rm exp} \lbrack \epsilon/(k_BT^*_e) \rbrack - 1}},
 \label{Raman-electron-temperature}
 \end{align}
 when the band densities of states vary slightly over the anti-Stokes shift range of interest.
 Here $f(E,T)$ is the Fermi-Dirac function.

\begin{figure}[htbp]
  \centering
  \includegraphics[width=\columnwidth]{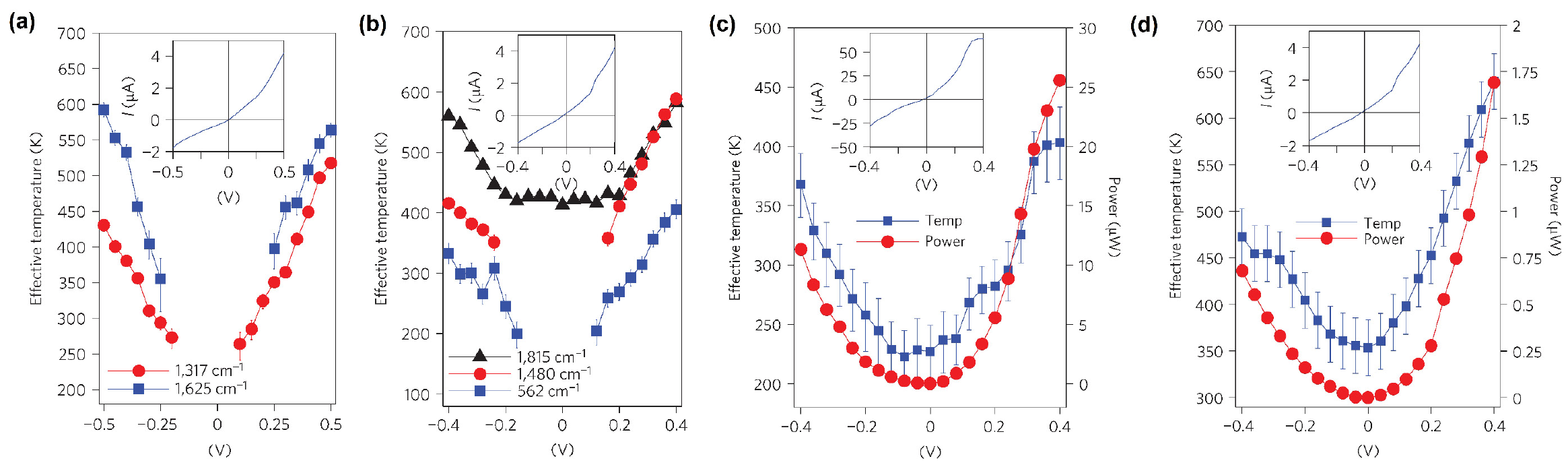}
  \caption
  {
   Local phonon (a,b) and electron (c,d) temperature as a function of $V$.
   Inset: $I-V$ curve for each device.
   For OPV$_3$ molecule junction, (a) shows local temperatures for two vibrational modes: 1,317 cm$^{-1}$ (red) and 1,625 cm$^{-1}$ (blue),
   and (b) exhibits another three modes: 1,815 cm$^{-1}$ (black), 1,480 cm$^{-1}$ (red) and 562 cm$^{-1}$ (blue).
   The vibrational modes in (a) are pumped by charge flow through the molecule and the vibrational pumping for modes in (b) results from both current and optical excitation.
   The curve is limited below $\left|{V}\right|\leq0.2V$ by the detector noise floor for anti-Stokes signal, except for the mode with 1,815 cm$^{-1}$.
   (c) and (d) show local electron temperature (blue; left axis) and dissipated electrical power (red; right axis) for the bare nanojunction without any molecule and OPV$_3$ molecule junction, respectively.
   The nanogap between the two electrodes allows a relatively large tunneling current through the bare junction.
   Reprinted with permission from \cite{War1133}. Copyright 2011 Springer Nature.
  }
  \label{figure19-Raman-temperature}
 \end{figure}

 \Figure{figure19-Raman-temperature}(a) and (b) depict the local phonon temperature as a function of $V$ in an OPV$_3$ molecule junction
 (OPV$_3$, three-ring oligophenylene vinylene terminating in amine functional groups).
 However, local ionic cooling at a high bias does not appear.
 An approximately linear increase of local phonon temperature $T^*_{ph}$ for measured Raman modes is observed with increasing $\left|{V}\right|$,
 while $T^*_{ph}$ does not act as $T^*_{ph}\propto\sqrt{V}$ predicted by \Eq{local-temp-of-ep} at low biases,
 as would be expected for heat dissipation to environment (bulk substrates and electrodes) via elastic scattering between phonons of the nanojunction and phonons of the bulk system.
 This suggests that local vibrational relaxation to environment takes place by other means, such as coupling to phonons with reduced \gls{dof} \cite{War1133}.

 \Figure{figure19-Raman-temperature}(c) and (d) plot the local temperature of electrons as a function of $V_{\rm bias}$ for the bare nanojunction without any molecule and OPV$_3$ molecule junction, respectively.
 The nanogap between the two electrodes allows a relatively large tunneling current through the bare junction,
 while the presence of the OPV$_3$ molecule limits the electric current through the junction.
 A linearly increasing $T^*_e$ with respect to $V$ is consistent with the prediction for local electron heating in nanoscale conductors when no electron-phonon interaction is present \cite{Dag062935}.

 Mecklenburg \emph{et al.} have developed a non-invasive thermometric technique, plasmon energy expansion thermometry,
 which can measure local electron temperatures of bulk systems with nanometer-scale spatial resolution \cite{Mec15629,Col15611}.
 Incorporating \gls{eels} with \gls{stem},
 the thermometry can measure local temperatures from the minimally required energy $E$ to excite a bulk plasmon,
 which represents the oscillations of electron density around ion cores within the bulk material containing an electron gas.
 The principle behind is that the temperature-induced change of free electron density $\rho_e$ of bulk plasmon affects the plasmon energy $E$.
 When the temperature of a bulk material rises by $\Delta T$ above room temperature $T_0$,
 the material expands.
 The thermal expansion of a bulk material changes its density,
 as well as the density of free electrons inside the material $\rho_e(T) \simeq \rho_e(T_0) \lbrack 1-\alpha_1 \Delta T - \alpha_2 (\Delta T)^2 \rbrack$.
 Here, $\alpha_i (i = 1,2)$ are the first-order and second-order coefficients of thermal expansion.
 Since the energy $E$ of free electrons depends on $\rho_e$, given by $E = \sqrt { {4 \pi \rho_e e^2} / {m} }$,
 the temperature rise $\Delta T$ is related to the change in plasmon energy $\Delta E(T) = E(T)-E(T_0)$ by the following equation
\be
 {\Delta{T} = \frac{\alpha_1} {2\alpha_2} \bigg( \sqrt{1 - \frac{8 \alpha_2 \Delta E (T^*)} {3\alpha_1^2 E(T_0)}} -1 \bigg)}.
 \label{EELS-temperature}
\ee
 By focusing the \gls{stem} electron beam into a nanometer-sized probe, placing it over the sample,
 and analyzing the shift of the plasmon peak in \gls{eels} according to \Eq{EELS-temperature},
 one obtains an image for the temperature map of the sample; see Figure~1 in \cite{Mec15629}.

\begin{figure}[htbp]
  \centering
  \includegraphics[width=0.9\columnwidth]{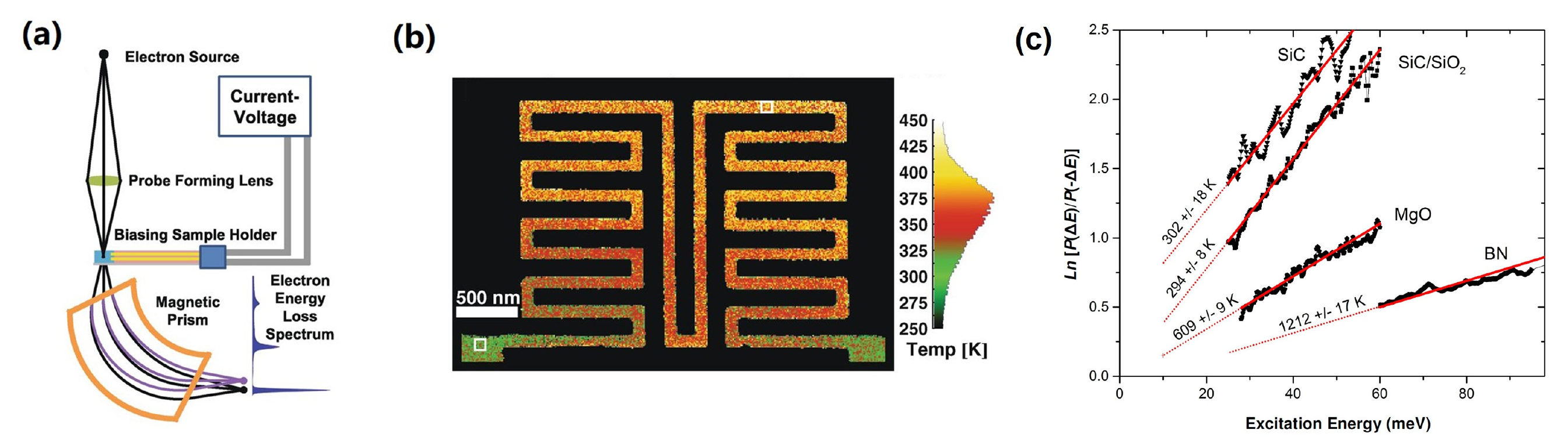}
  \caption
  {
   (a) Schematic of the plasmon energy expansion thermometer.
   It consists of a \gls{stem}, a biasing sample holder, a power source for Joule-heating the sample, and an \gls{eels} spectrometer;
   (b) A false-color temperature map of a serpentine aluminum wire heated by the electrical current.
   The histogram indicates the color scale and bins each pixel according to its temperature.
   (a) and (b) reprinted with permission from \cite{Mec15629}. Copyright 2015 AAAS.
   (c) The result of the bulk and surface phonon scattering in different nanostructures:
   a single SiC nanoparticle and a SiO$_2$/SiC interface sample at room temperature and for a single MgO nanosphere and a BN flake at high temperatures.
   The plots follow the linear behavior as predicted by the detailed balance relation.
   The measured temperature values for each nanosystem are shown over the dotted lines.
   Reprinted with permission from \cite{Lag184556}. Copyright 2018 American Chemical Society.
  }
  \label{figure20-EELS-temperature}
 \end{figure}

 In this way, Mecklenburg \emph{et al.} have imaged the local temperature distribution of a serpentine Al nano-device subjected to a bias voltage \cite{Mec15629}.
 \Figure{figure20-EELS-temperature}(b) reveals the spatial resolution of the temperature field of the nano-device is, technically, less than 2 nm.
 However, as stated by Mecklenburg \emph{et al.}, temperature differences do not exist
 on the length scales smaller than the mean free path of a particle.
 Since the electron transport is ballistic over distances less than the electron mean free path, $l_e$,
 the setup cannot measure a temperature gradient for separations smaller than $l_e$.
 Thus $l_e$ provides the minimal spatial resolution of local electron temperature, estimated as $l_e \le 4-15$ nm in their experiment.
 The precision of the temperature measurement is limited to several Kelvins (standard error of $\sim$3 to $\sim$5 K).

 Another application of \gls{stem}-\gls{eels} to the temperature measurement has been reported by Lagos \emph{et al.} \cite{Lag184556,Idr18095901}.
 When an electron probe of the \gls{stem} setup is coupled to a nanostructure,
 the incident high energy electron from the probe will be scattered elastically and inelastically, producing a large variety of vibrational/phonon excitations.
 For an inelastic process involving the transfer of the energy $\Delta{E}$ and momentum $\Delta{q}$ between the electrons and phonons in materials,
 the detailed balance relation associates the scattering probability of a loss energy event with that of a gain energy process
 \begin{align}
 {\frac{P(\Delta{E},\Delta{q})}{P(-\Delta{E},-\Delta{q})}=e^{-{\beta^*_{ph}}\Delta{E}}},
 \label{STEM-detailed-balance}
 \end{align}
 thus providing a straightforward access to temperature information on objects through scattering measurements.
 Here, $\beta^*_{ph} \equiv 1/T^*_{ph}$ and $P$ is the probability of energy loss or gain in scattering.
 Although the detailed balance relation is only valid for equilibrium systems,
 Lagos \emph{et al.} have held a view that this technique is still applicable for nonequilibrium nanosystems, which violate the \gls{fdt} \cite{Lag184556}.
 The energy gain and loss by phonons in inelastic scattering have been measured for different nanostructures \cite{Lag184556,Lag17529}.
 The logarithm of the ratio of the loss and gain scattering is fitted as a function of the excitation energy.
 \Figure{figure20-EELS-temperature}(c) shows that all of the curves exhibit the typical linear behavior, with slope scaling like $T^{-1}$, as described by \Eq{STEM-detailed-balance}.
 Using an energy gain and loss spectroscopy,
 one can measure the local temperature of a nanosystem with precision down to about 1 K.
 The spatial resolution depends on the type of scattering signal used to determine the local temperature:
 short wavelength bulk phonons technically provide a highly localized scattering signal down to the atom-level ($\sim2 \AA$).

 Some efforts have been made to incorporate an \gls{afm} setup with spectroscopic techniques.
 For instance, Aigouy \emph{et al.} have developed a scanning thermal imaging method based on a temperature-sensitive phenomenon: fluorescence \cite{Aig05184105}.
 Fluorescent nanoparticles have been glued on the apex of an \gls{afm} probe, as a temperature sensor.
 When nanoparticles are in thermal equilibrium with the measured sample,
 the population of energy levels in fluorescent nanoparticles is ruled by the Boltzmann distribution \cite{Ale044753,Sam08023101}.
 The relative intensity of the two fluorescent lines is linked by the following relation \cite{Ber901100,Mau958019,Ale044753}
\be
 \frac{I_{\nu_1}}{I_{\nu_2}} = A \exp\left(-\frac{\Delta \hbar \omega}{k_B T^*} \right).
 \label{intensity-fluorescence-peak}
\ee
 Here, $I_\nu$ is the intensity of each peak located at specific frequency $\nu$,
 $\Delta\omega$ is the energy separation between these two groups of lines,
 and $A$ is a parameter that depends on the fluorescent material.
 By scanning the surface of the sample and calculating the ratio of the intensity of the two fluorescence lines,
 one can obtain information on the temperature distribution over the sample.

 The spectroscopic techniques described above have the following advantages: some of the spectroscopic techniques enable temperature measurements with atom-level spatial resolution;
 since spectroscopic techniques are mainly based on statistical properties of nanosystems, they have the potential to realize the theoretical definitions in \Sec{subsubsec:DF&FDR}.
 In addition, spectroscopic techniques are non-invasive in the sense that the measurement has a minor effect on the measured system's temperature \cite{Mec15629}. This is different from the \gls{mcbj} techniques which require breaking the nanojunction by applying mechanical forces.
On the other hand, the spatial resolution of some spectroscopic techniques,
 such as Raman spectroscopy and infrared thermography may be restricted by the diffraction limit (around half wavelength) \cite{Bri124799}.
 Moreover, the techniques with a high spatial resolution, such as plasmon energy expansion thermometry,
 are only valid for metals and semiconductors which have sufficiently sharp plasmon resonances \cite{Mec15629}.

 Finally, it should be pointed out that, since some spectroscopic techniques probe locally the energy space rather than the real space,
 the measured non-equilibrium temperature is spatially nonlocal.
 Notable examples are the mode-specific local phonon temperature $T^*_{ph}$ exhibited in Figures \ref{figure18-Raman-temperature} and \ref{figure19-Raman-temperature},
 and the mode-specific temperature of magnons (quasi-particles representing the collective excitations of electronic spins)
 defined by an energy repartition principle in spin systems at a non-equilibrium steady-state \cite{Yan17024417}.
 Such non-equilibrium temperatures should be distinguished from the local temperatures that are truly local in real space.

\subsubsection{Scanning thermal microscope} \label{subsubsec:SThM}

\begin{figure}[htbp]
  \centering
  \includegraphics[width=0.45\columnwidth]{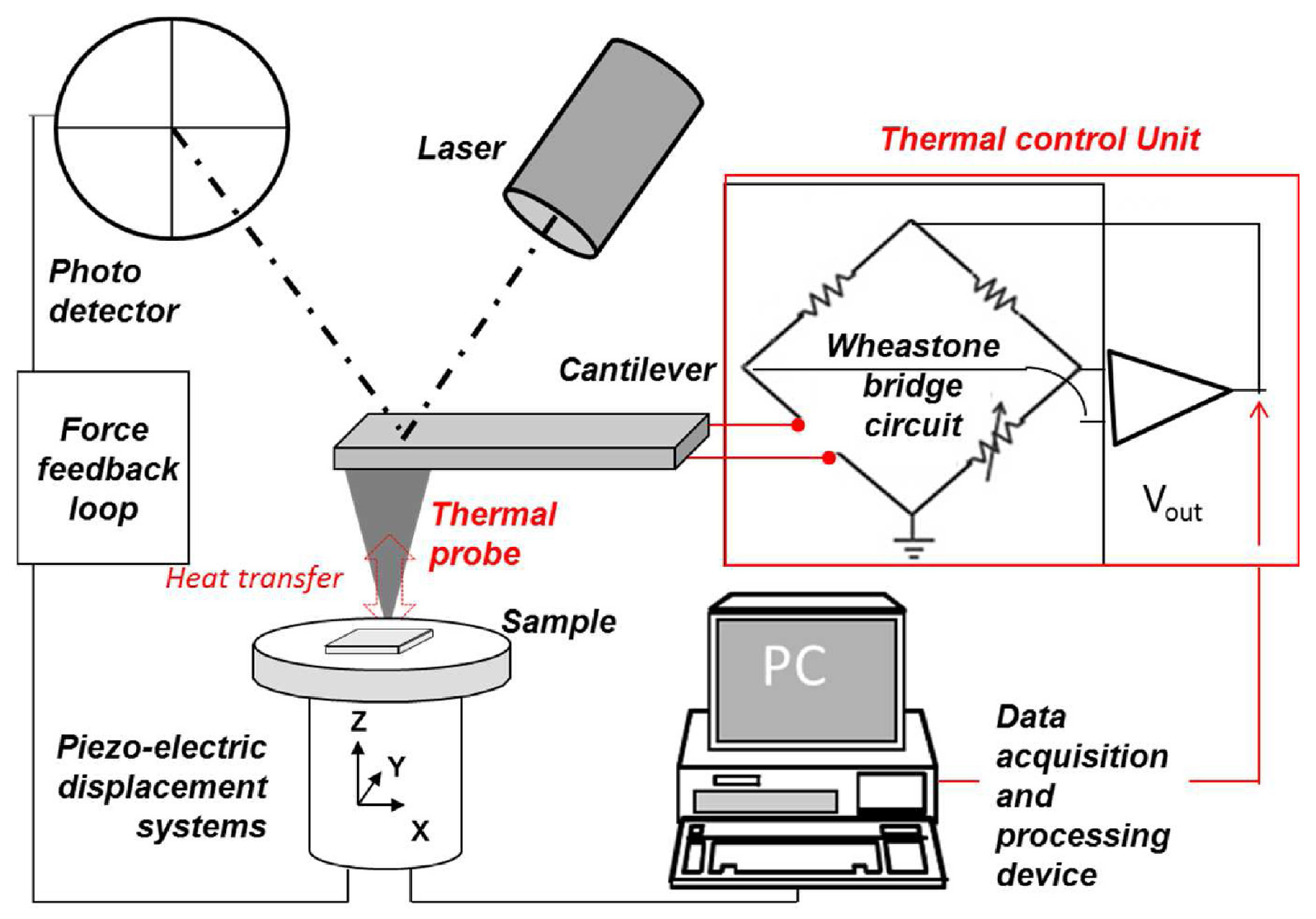}
  \caption
  {
   The set-up of an \gls{afm}-based \gls{sthm} system.
   Here, the output signal is the voltage delivered by a ``thermal control unit" and a balanced Wheastone bridge can be used to maintain the probe mean temperature at constant value.
   Reprinted with permission from \cite{Gom15477}. Copyright 2015 John Wiley and Sons.
  }
  \label{figure12-SThM-steup}
 \end{figure}

 The \gls{sthm}, designed for measuring the temperature distribution of a large nanostructure,
 is widely used to investigate thermal transport of nanoscale/mesoscopic structures \cite{Pao14025506,Yu11183105,Men13205901}.
 Figure \ref{figure12-SThM-steup} shows the setup of an \gls{afm}-based \gls{sthm}.
 The principle of an \gls{sthm} is that the local temperature is obtained from the value of a temperature-dependent property, measured by the thermal sensors integrated onto the \gls{afm} probe.
 In practice, the probe closely approaches a sample and scanning its surface to obtain information of its topography and temperature simultaneously but independently.
 Various thermal sensors have been developed based on different thermally-sensitive properties or phenomena, such as thermoelectricity \cite{Mil982990,Luo96325}, change in electrical resistance \cite{Zha112435,Ptl943785}, thermal expansion \cite{Nak95694}, or fluorescence \cite{Sai09115703}.
 Here, we first review two types of widely-used thermal probes.

 The first type of probe is based on a resistance thermometer.
 The temperature dependence of the resistance $R_p$ can be expressed as \cite{Gom15477}
 \begin{align}
 {R_p(T) = R_p(T_0) \lbrack 1 - \alpha (T - T_0) \rbrack},
 \label{SThM-resistance}
 \end{align}
 where $T_0$ is a reference temperature and $\alpha$ is the temperature coefficient of its electrical resistivity.
 The probe determines the local temperature by measuring the variation of the electrical resistance $R_p$ of the probe in response to a temperature change.
 Although the resistance thermometer has been used in the nanocalorimeters we previously reviewed,
 the experimental utilization of a resistance thermometer in \gls{sthm} is different that in calorimetry.
 Here, a resistance thermometer is integrated on the \gls{sthm} probe, and can measure the temperature of a local region of the sample.
 In nanocalorimeters, a resistance thermometer is embedded into the pad, and can only measure the temperature for one end of the nanostructures.

\begin{figure}[htbp]
  \centering
  \includegraphics[width=0.75\columnwidth]{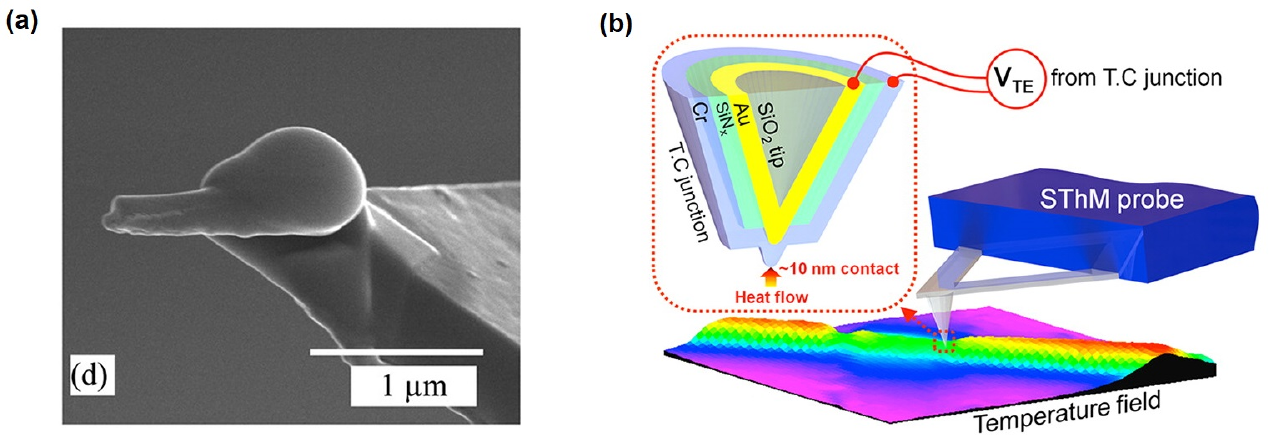}
  \caption
  {
   (a) \gls{sem} images of the \gls{sthm} probe assembling carbon nanotube.
   High thermal conductivity and small tip radius of the carbon nanotube improve the thermal resolution of \gls{sthm}.
   Reprinted with permission from \cite{Tov141174}. Copyright 2014 Royal Society of Chemistry.
   (b) Schematic illustrating the \gls{uhv}-\gls{sthm} technique.
   The device resides in a \gls{uhv} chamber.
   The nano-thermocouple junction fabricated by Au and Cr layers is assembled outside the end of the tip.
   The thermoelectric voltage $V_{\rm TE}$ generated from the Au-Cr junction is directly proportional to local temperature of the sample at the contact region.
   Reprinted with permission from \cite{Kim124248}. Copyright 2012 American Chemical Society.
  }
  \label{figure13-SThM-thermometer}
 \end{figure}

 To improve the spatial resolution of a resistance thermometer-based \gls{sthm} a
 carbon nanotube has been proposed as a thermal probe because of its small tip radius and high thermal conductivity \cite{Tov12114317,Hir131}.
 Tovee \emph{et al.} have developed a carbon nanotube-based \gls{sthm} probe, as shown in \Figure{figure13-SThM-thermometer}(a).
 Compared with the conventional Pt tip with the $\sim$100 nm spatial resolution \cite{Bri124799},
 this probe could achieve a spatial resolution of $\sim30$ nm \cite{Tov141174}.

 The second type of probe is the thermocouple-based \gls{sthm} probe.
 Based on the Seebeck effect between two different types of metals,
 a thermocouple uses the thermoelectric voltage $V_{\rm TE}$ to carry out the temperature measurement \cite{Lee13209}
 \begin{align}
 {S_{\rm Sb} = \frac{V_{\rm TE}}{T_1 - T_2}},
 \label{SThM-thermalcouple}
 \end{align}
 where $S_{\rm Sb}$ is the Seebeck coefficient of the thermocouple, $T_1$ and $T_2$ are temperatures of the two different metals.
 Given the excellent thermal contact between the thermocouple and the sample,
 $T_1$ can be assumed to be the same as the local temperature $T^*$ of the measured region of the sample.
 Given excellent thermal contact between the thermocouple and the tip, $T_2$ is equivalent to the probe temperature $T_p$.
 Since the thermoelectric voltage $V_{\rm TE}$ across the thermocouple is directly proportional to $T^*$ of the sample,
 one can quantitatively determine $T^*$ by measuring $V_{\rm TE}$ with a known $S_{\rm Sb}$.

 A point-contact thermocouple has been developed by Sadat \emph{et al.} \cite{Sad102613}.
 The thermocouple forms when a Pt-coated \gls{afm} probe at the ambient temperature is brought into physical contact with the surface made by another type of metal.
 This method can map the temperature field of metallic surfaces with a $\sim0.01$ K temperature resolution and $<100$ nm spatial resolution \cite{Sad102613}.

\begin{figure}[htbp]
  \centering
  \includegraphics[width=0.80\columnwidth]{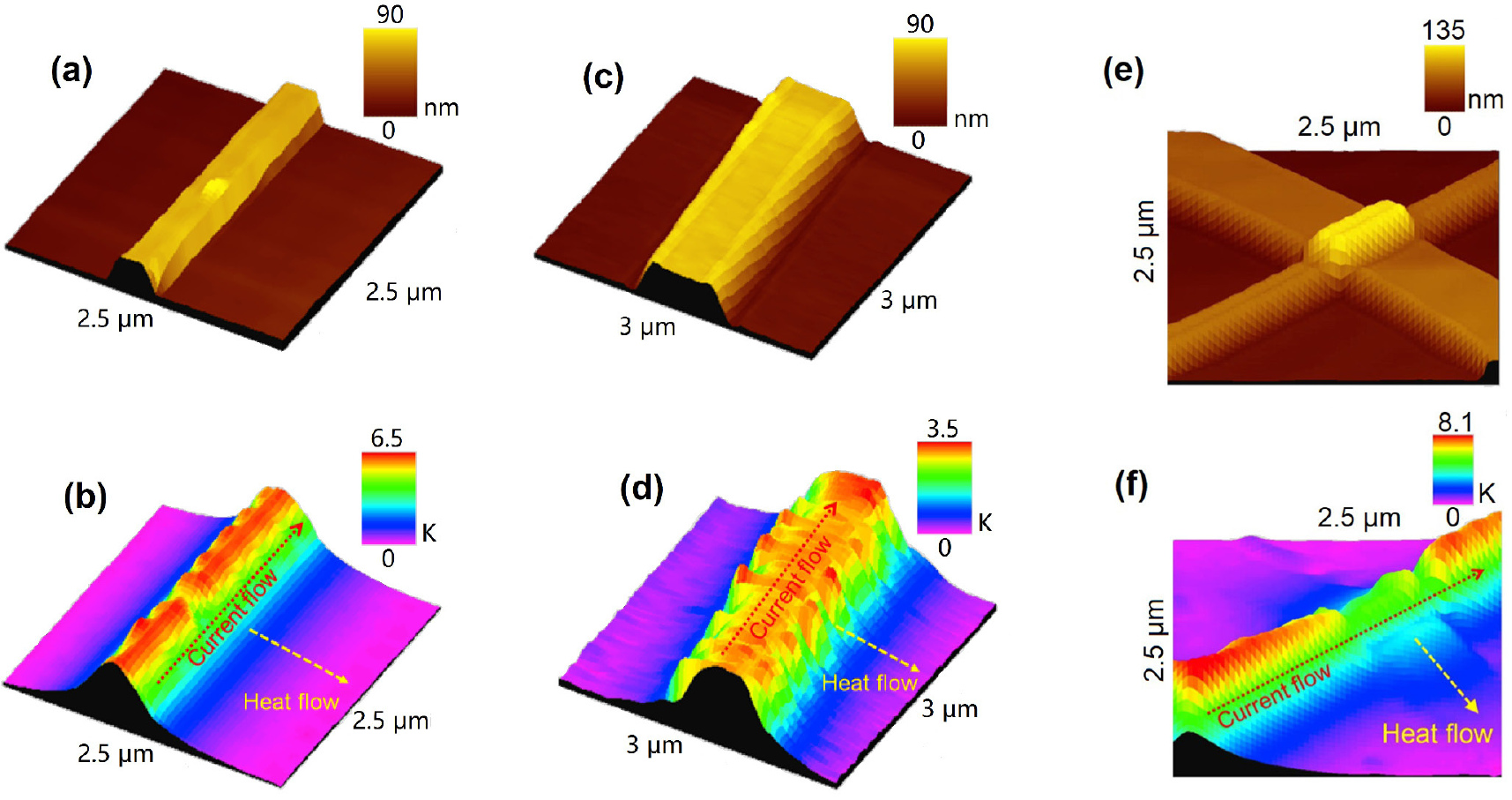}
  \caption
  {
   Simultaneously-measured topographical and temperature images on (a), (b) 200 nm and (c), (d) 1 $\mu$m wide Pt lines, respectively.
   The directions of current and heat flow are represented on temperature images.
   The measured temperatures decay rapidly away from the Pt lines.
   Simultaneous-measured topographical and temperature (e), (f) images, respectively, on a 2.5 $\mu$m - 2.5 $\mu$m region where the 200 nm wide Pt line lies on top of a 1 $\mu$m wide Pt line.
   The 1 $\mu$m wide Pt line acts as a fin for the 200 nm Pt line in the cross sections.
   Reprinted with permission from \cite{Kim124248}. Copyright 2012 American Chemical Society.
  }
  \label{figure14-SThM-Pt-line}
 \end{figure}

 Kim \emph{et al.} have reported an \gls{uhv}-based \gls{sthm} technique \cite{Kim124248}.
 A nanoscale Au-Cr thermocouple coats the SiO$_2$ tip of the probe, as shown in \Figure {figure13-SThM-thermometer}(b).
 Since an \gls{uhv} environment eliminates heat transport between the tip and the sample through air,
 the \gls{uhv}-\gls{sthm} is capable of quantitatively mapping temperature fields with the $\sim15$ mK temperature resolution and $\sim10$ nm spatial resolution on metal and dielectric surfaces.
 \Figure{figure14-SThM-Pt-line}(a) and (c) show the topographical image of 200 nm (narrow) and 1 $\mu$m (wide) Pt lines, respectively.
 The local temperature decays rapidly in the direction perpendicular to the Pt lines; see yellow lines in \Figure{figure14-SThM-Pt-line}(b) and (d).
 When the narrow Pt line (through which an electrical current flows) is connected to the wide Pt line (through which no electrical current passes),
 the temperature rise of the 200 nm line is significantly lower in the region where it intersects the 1 $\mu$m line; see \Figure{figure14-SThM-Pt-line}(e) and (f).
 This is because heat current dissipates from the 200 nm Pt line by the 1 $\mu$m Pt line in the cross section.

 Varesi and Majumdar have developed a \gls{sjem}, an \gls{afm}-based setup that can image both surface topography and Joule heating-induced thermal expansion of a sample \cite{Var9837,Maj98297}.
 The thermal expansion $\Delta L_z$ of the sample thickness is related to the rise of local temperature $T^* - T_0$ via the following equation \cite{Var9837,Maj98297}
 \begin{align}
 {\Delta L_z = \alpha L_z (T^* - T_0) },
 \label{SJEM-thermal-expansion}
 \end{align}
 where $T_0$ is the background temperature,
 $\alpha$ is the linear thermal expansion coefficient,
 and $L_z$ is the local thickness of the sample, which is obtained from the surface topography.
 According to \Eq{SJEM-thermal-expansion}, the change in local temperature can be determined by detecting the thermal expansion $\Delta L_z$ of the sample.
 It is reported that the spatial resolution of \gls{sjem} has reached $\sim$ 20 nm \cite{Can01S67}.

\begin{figure}[htbp]
  \centering
  \includegraphics[width=0.95\columnwidth]{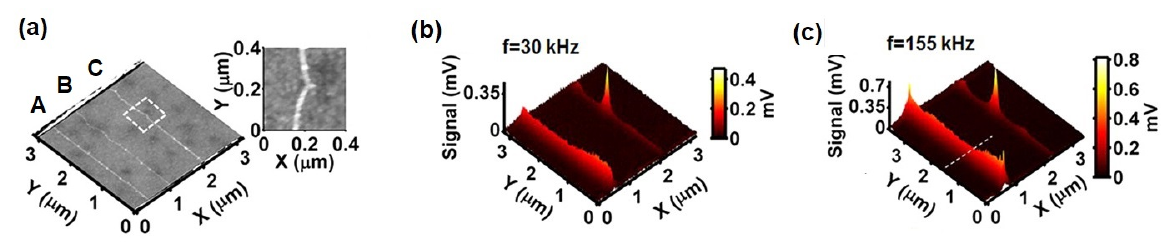}
  \caption
  {
   (a) Topographical \gls{afm} image of three single-wall carbon nanotubes.
   The inset provides a magnified view of a ``kink" in one of the single-wall carbon nanotubes.
   (b) \gls{sjem} image of the same nanotubes at low frequency.
   A strong peak appears at the position of the kink, illustrating significant heat generation at this location.
   (c) \gls{sjem} image of the carbon nanotubes at high frequency.
   $f$ is the frequency of the AC voltage bias.
   The overall shape of (c) is similar to that of (b).
   Reprinted with permission from \cite{Xie1210267}. Copyright 2012 American Chemical Society.
  }
  \label{figure15-SThM-thermal-expansion}
 \end{figure}

 \gls{sjem} has also been applied to investigate temperature fields of various nanosystems,
 such as nanowires \cite{Can01S67}, two-dimensional metallic films \cite{Gur05809}, and graphene-metal contacts \cite{Gro11287}.
 Using \gls{sjem}, Xie \emph{et al.} have investigated the temperature profile of three single-wall carbon nanotubes subjected to a bias voltage \cite{Xie1210267}.
 \Figure{figure15-SThM-thermal-expansion}(a) shows the topographical image of three single-wall carbon nanotubes.
 Compared with nanotube A and C, nanotube B has high combined electrical resistance in the electrode-nanotube contacts.
 There is a prominent ``kink" in the middle of nanotube C.
 The "kink" arises from the growth process of the carbon nanotube and can be considered as a topographic defect.
 AC voltage bias is applied to three nanotubes and provides Joule heating.
 \Figure{figure15-SThM-thermal-expansion}(b) illustrates the \gls{sjem} image of the nanotubes under AC voltage with a frequency $f = 30$ kHz.
 Nanotube A exhibits nearly uniform thermal expansion along its length.
 Nearly no topographical signal is obtained from nanotube B due to its high resistance.
 Nanotube C shows a sharp peak corresponding to the site of the ``kink'' defect, indicating substantial local heating.
 Under a high frequency voltage with $f = 155$ kHz,
 the overall image of \Figure{figure15-SThM-thermal-expansion}(c) is similar to that of \Figure{figure15-SThM-thermal-expansion}(b).
 This defect-induced local heating can be used to examine whether a topographic deflect exists in carbon nanotubes \cite{Xie1210267}.

 Combining \gls{afm} with spectroscopic techniques,
 Aigouy \emph{et al.} have proposed a \gls{sthm} setup based on fluorescence \cite{Sam08023101,Aig09074301,Sai09115703}.
 Rare-earth ion-based fluorescent particles are attached to the apex of the \gls{afm} probe.
 Scanning a Ni nanoscale (Joule) heater, they have imaged the topography and the temperature profile of the whole nanostructure; see Figure 8(a) and (b) in \cite{Sai09115703}.
 The temperature cross section exhibits a homogenous temperature distribution along the axis direction of the Ni heater; see Figure 8(c) in \cite{Sai09115703}.
 The spatial resolution is in the range of the fluorescent particle size (smaller than 500 nm), and the temperature sensitivity is smaller than 5 K \cite{Aig09074301}.

 A large advantage of the \gls{sthm} is simultaneously imaging surface topography and temperature field.
 Under an \gls{uhv} environment, a custom-fabricated thermocouple can reach a high spatial resolution with an accuracy of $\sim$10 nm \cite{Kim124248}.
 However, a limitation is that the \gls{sthm} generally demands physical contact with nanosystems.
 The thermal probe inevitably perturbs the local properties of nanosystems to some extent,
 and therefore disturbs the value of the local temperature \cite{Mec15629}.
 More strengths and limitations of \gls{sthm} have been analyzed by
 Brites \emph{et al.} \cite{Bri124799}, Gom\.{e}s \emph{et al.} \cite{Gom15477} and Kim \emph{et al.} \cite{Kim15129}.

\subsubsection{Relation between experiments and theories} \label{subsubsec:expermental-theorey-gap}

 In \Sec{subsec:mueasurement-local-temp}
 we have reviewed the present thermometric techniques designed
 for a variety of non-equilibrium nanosystems,
 ranging from one-dimensional molecular junctions and  nanowires to
 two-dimensional nanosheets and layered nanostructures.
 Besides these works, there are still many other experimental efforts on
 thermometers at the nanoscale which are not covered in the above discussion,
 such as \cite{Bri124799,Jaq124301,Gom15477,Kim15129}.
 In the following, we establish the connections between the
 existing experimental efforts presented in this Section and
 the theoretical definitions of local temperature reviewed in \Sec{sec:Def}.

 First, it is important to recognize that some theoretically proposed
 schemes for local temperature measurement have been realized in real experiments.
 For instance, the extraction of a local temperature from
 the local distribution function of particles has been
 realized by spectroscopic techniques.
 As reviewed in \Sec{subsubsec:spectroscopy},
 both the local phonon temperature and local electron temperature
 of a molecular junction have been obtained by analyzing the
 surface enhanced Raman spectroscopy \cite{War1133}.
 Moreover, experimental observations have confirmed some theoretical predictions.
 For instance, as reviewed in \Sec{subsubsec:break-junction}
 the local temperatures measured by \gls{mcbj}-based techniques
 agree well with the theoretical prediction of \Eq{local-temp-of-ep-ee}
 which quantifies the contribution of electron-phonon
 and electron-electron couplings to current-induced local ionic heating.
 Furthermore, the advancement of experimental instruments and techniques
 has encouraged novel theoretical proposals.
 For instance, the latest \gls{sthm} has achieved high-precision
 measurement of local physical properties
 which are highly sensitive to the local temperature.
 This would make some newly proposed protocols,
 such as the \gls{mpc}-based definitions reviewed in \Sec{subsubsec:mpc},
 practically feasible and potentially promising.

 Second, there is a still noticeable gap between the experiments and the theories.
 On the experimental side, many experimental measurement approaches
 are based on the miniaturization of conventional macroscopic thermometers \cite{Bri124799}.
 For instance, as reviewed in \Sec{subsubsec:SThM},
 the application of \gls{sthm} has presumed that the local thermal properties
 or phenomena of nano-sized materials,
 such as thermoelectricity, thermal expansion, and temperature-dependent
 electrical resistance, preserve their conventional relations or forms
 as in bulk materials.
 The validity of such a presumption requires
 the nano-sized material should include a sufficiently large number of \gls{dof}.
 As discussed in \Sec{subsubsec:eff},
 the minimal length scale for the existence of a local Boltzmann distribution is estimated
 to be $\sim 100$ nm for real materials \cite{Har0689}.
 Such a length scale coincides with the typical spatial resolution of the temperature field
 measured by a resistance thermometer-based \gls{sthm} \cite{Bri124799}.
 In contrast, on the theoretical side, the proposed
 local temperature definitions are, in principle, suitable for general nanosystems of all sizes.
 However, numerical simulations of real nano-devices or materials for practical purposes
 are often limited by the computational capabilities of modern computers.
 Consequently, most of the reported theoretical studies were carried out
 for nano-sized devices or materials with a relatively small number of \gls{dof},
 such as nanojunctions consisting of a single molecule and
 \gls{qd}s with a few energy levels.

 Third, a number of theoretical definitions have not been put to experimental test.
 This is because some key quantities involved in these definitions
 cannot be measured directly with the existing instruments/apparati.
 For instance, despite the possible scheme presented at the end of
 \Sec{subsec:bimaterial-cantilever-based-calorimeter},
 the \gls{zcc}-based definitions have not been realized,
 because of the difficulty in measuring the heat current without
 the prerequisite knowledge of the temperature gradient.
 As for the thermodynamic relations-based definitions,
 the measurement of thermodynamic entropy for nanosystems
 which is at the heart of many definitions, has remained
 a formidable challenge for experiments \cite{Mer1720}.
 In addition, some dynamic response properties required for
 the statistical relations-based definitions,
 such as the \gls{negf}s, are also hard to access directly
 in experiments.

 The above discussion suggests that there is plenty of room
 for improvement for both experimental and theoretical studies
 regarding the determination of local temperatures
 for non-equilibrium nanosystems.
 Particularly, it is important for the experimental techniques
 to go beyond the conventional strategies for classical systems
 and be able to account for the quantum effects at the nanoscale.
 On the other hand, it is also crucial for theoretical proposals
 to pay more attention to their practical feasibility,
 so that they can be realized by the existing experimental setups.
 For instance, the \gls{mpc}-based definitions discussed in \Sec{subsubsec:mpc}
 were designed with unambiguous operational protocols,
 and their implementation does not require the potentially difficult procedure
 of measuring the heat current or thermodynamic properties.
 Thus, these operational definitions are expected to be
 promising candidates for future practical applications.


\section{Implications and applications of non-equilibrium local temperatures} \label{sec:Imp&App}

\subsection{Physical Implications} \label{subsec:imp}

\subsubsection{Generalization of the thermodynamic laws} \label{subsec:law}

 For a nanosystem in a non-equilibrium steady state,
 Stafford \textsl{et al.} have shown that the \gls{zcc}-defined local temperature is consistent with the thermodynamic laws,
 if a broadband probe is weakly coupled to the system.

 \noindent {\textsl{The Zeroth Law} --}
 The zeroth law has been discussed in two scenarios.
 The first one is as follows \cite{Mea14035407,Sha16245403}:
 If the local temperatures and the local electrochemical potentials of two non-interacting fermionic systems,
 as measured by a probe, are equal,
 the two systems will be in thermal and electrical equilibrium with each other
 when connected by a transmission line coupled locally to the same two positions.
 %
 %
 \begin{figure}[htbp]
  \centering
  \includegraphics[width=0.9\columnwidth]{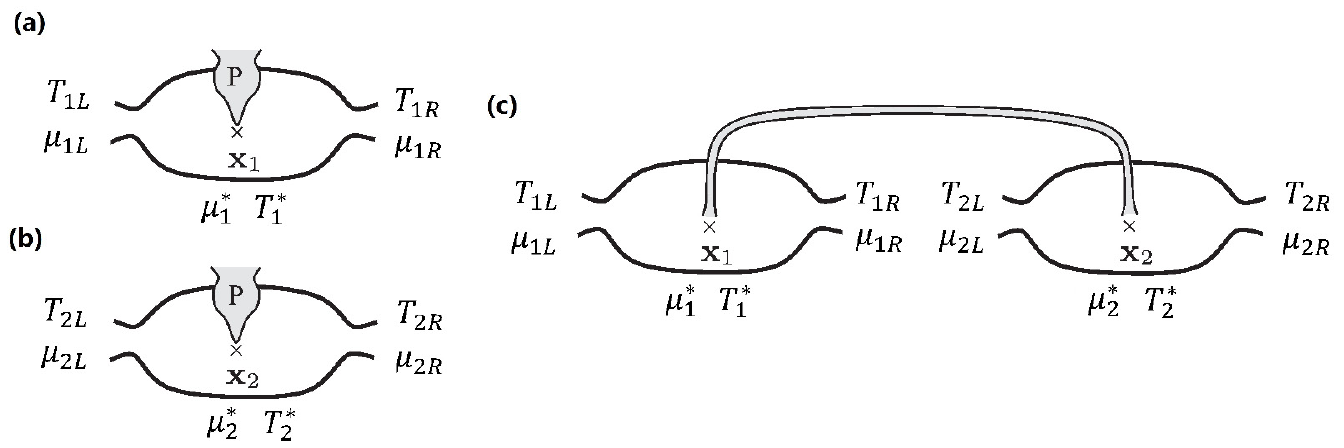}
  \caption
  {
   (a) and (b): Two independent systems sequentially probed by the same probe at local positions $x_1$ and $x_2$.
   (c) A transmission line is locally coupled to the positions at $x_1$ of system 1 and $x_2$ of system 2.
   Here, $T^*_1$ ($T^*_2$) and $\mu^*_1$ ($\mu^*_2$) are the probe-measured local temperature and local electrochemical potential of system 1 (2).
   Reprinted with permission from \cite{Mea14035407}. Copyright 2014 American Physical Society.
   }
  \label{fig1-zeroth-law}
 \end{figure}

 Figure \ref{fig1-zeroth-law}(a) and (b) show two independent two-terminal nanojunctions.
 The electron reservoirs in contact with the systems are in an equilibrium state
 in the absence of the thermal and electrical biases across the junction.
 The values of the equilibrium temperature and electrochemical potential are assumed to be $T_0$ and $\mu_0$ for each reservoir \cite{But904869}.
 Once the thermal and electrical biases are imposed on the junctions,
 there is a pair of positions $x_1$ of system 1, and $x_2$ of system 2,
 in which a probe measures the identical $\mu^*$ and $T^*$.
 %
 A transmission line then connects the position $x_1$ of system 1 to the position $x_2$ of system 2, with the same coupling strength as that of the probe; see \Figure{fig1-zeroth-law}(c).
 When the bias voltage is sufficiently small,
 the net electric currents $I$ and thermal current $J$ between the two systems can be evaluated as \cite{Mea14035407}:
 \begin{equation}
 \left[
 \begin{array}{ccc}
 I\\
 J
 \end{array}
 \right ]
  =
 \left[
 \begin{array}{ccc}
 \mathcal{M}(E)                   &      a \, \partial_E \mathcal{M}(E)\\
 a \, \partial_E \mathcal{M}(E)   &      \mathcal{M}(E)
 \end{array}
 \right ]_{E = \mu_0}
 \left[
 \begin{array}{ccc}
 \mu^*_2 - \mu^*_1 \\
 (T^*_2 - T^*_1)/T_0
 \end{array}
 \right ],
 \label{zeroth-law-linear}
 \end{equation}
 in the linear response regime.
 Here the coefficient $a = (\pi k_B T_0)^2/3$, and $\mathcal{M}(E)$ is the transmission coefficient between the two connected systems.
 Equation (\ref{zeroth-law-linear}) implies that the net currents through the line cancel
 when the local temperature and the local electrochemical potential of the connected positions are identical.
 In this case, the two systems will be in equilibrium with each other,
 when connected by a transmission line coupled locally to the connected positions.
 This equilibrium condition has been extended to a non-interacting electron system beyond the linear response regime \cite{Sha16245403}:
 If the $T^*$ and $\mu^*$ of the two systems under small bias are equal,
 there will be no energy/particle flow between the systems
 when connected by a transmission line coupled locally to the connected positions.

 Another scenario has also been discussed:
 Under what conditions two different thermometers yield the same value of local temperature for the same non-equilibrium system \cite{Sha16245403}?
 A sufficient condition has been proposed for the zeroth law as follows.
 Consider an electron system consisting of many orbitals, denoted by $i, j, \dots$, weakly coupled to a probe with a broadband width.
 If the probe is only coupled to a single orbital $n$ of the system,
 the probe-system tunneling matrix $\bm \Gamma^p(\omega)$ has only one non-zero element $\Gamma^p_{ij}(\omega) = \bar{\Gamma}^p(\mu_0) \delta_{ i,n} \delta_{ n,j}$.
 In the case of such system-probe coupling,
 the values of local temperature and local electrochemical potential depend only on the non-equilibrium state of the system to which the probe is coupled.
 These values are thus independent of the properties of the probe.
 The weak system-probe coupling ensures that a temperature measurement does not strongly perturb the microstates of the system.

 \noindent {\textsl{The First Law} --}
 Consider a fermionic non-interacting system subjected to bias.
 A broadband probe is coupled to a single orbital of the system and obtains the local temperature $T_1^*$ and local electrochemical potential $\mu^*$.
 A small change is imposed on the bias,
 This change finally leads the system to a new non-equilibrium steady state
 which is assumed to be characterized by the same value of the local electrochemical potential $\mu^*$,
 but by a different local temperature $T^*_2$.
 In absence of the work done by the environment on the system,
 the heat flowing into the system can be evaluated by the first law \cite{Sha16245403}: $\Delta Q \equiv \Delta \la E - \mu^* N \ra$,
 where $\la N \ra$ and $\la E \ra$ are the mean occupation number and the energy of the orbital coupled to the probe, respectively.
 It has been found that if the local specific heat $C_s$ of the system is properly defined,
 the heat $\Delta Q$ will satisfy \cite{Sha16245403}
\be
 \Delta Q = C_s (T^*_1 - T^*_2).
 \label{heat-flow-first-law}
\ee
 The local specific heat $C_s$ depends on both the local occupation number and the local \gls{dos} of the system sampled by the probe.
 Equation (\ref{heat-flow-first-law}) shows that the first law
 associates the absolute local temperature difference $T^*_1 - T^*_2$
 with the heat $\Delta Q$ into the system \cite{Sha16245403}.

 \noindent {\textsl{The Second Law} --}
 For an interacting fermionic system in a steady state arbitrarily far from equilibrium,
 it has been shown that the local temperature is consistent with the Clausius's statement of the second law \cite{Sha16245403},
 that heat spontaneously flows from a hot body to a cold body.
 \begin{figure}[htbp]
  \centering
  \includegraphics[width=0.9\columnwidth]{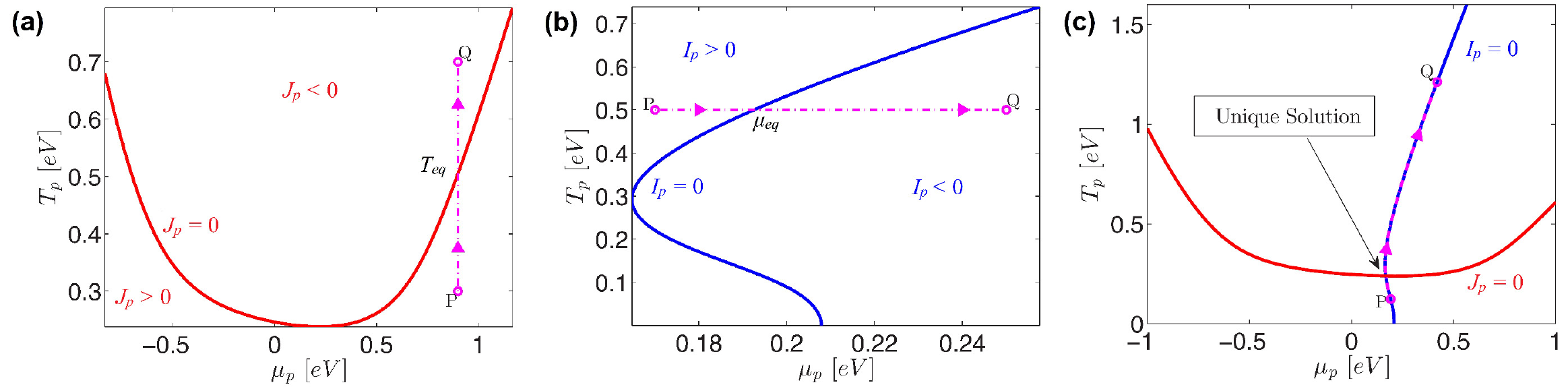}
  \caption
  {
   (a)
   Schematic diagram of the measurement of local temperature in a two-level system when $J_p = 0$.
   $T_p$ largely depends on $\mu_p$.
   The line from $P$ to $Q$ is at a constant electrochemical potential of the probe, and cuts the contour $J_p = 0$ at the point $T_{eq}$.
   In the point $P$, $J_p > 0$ means that the heat current flows into the probe,
   and in the point $Q$, $J_p < 0$ means that the heat current flows out of the probe.
   (b)
   Schematic diagram of the measurement of local electrochemical potential in a two-level system when $I_p = 0$.
   $\mu_p$ largely depends on $T_p$.
   The line from $P$ to $Q$ is at a constant temperature of the probe, and cuts the contour $I_p = 0$ at the point $\mu_{eq}$.
   In the point $P$, $I_p > 0$ means that the charge current flows into the probe,
   and in the point $Q$, $I_p < 0$ means that the charge current flows out of the probe.
   (c)
   Schematic diagram of the simultaneous measurement of $T^*$ and $\mu^*$ defined by \gls{zcc}.
   The contour $PQ$ along $I_p = 0$ cuts the contour $J_p$ once exactly, which implies the unique result of \Eq{def-zcc} when it exists.
   Reprinted with permission from \cite{Sta16155433}. Copyright 2016 American Physical Society.
   }
  \label{fig2-unique-T}
 \end{figure}

 Consider, for instance, a temperature probe weakly coupled to a two-level system sandwiched by two electron reservoirs under bias voltage \cite{Sta16155433}.
 The local temperature $T^*$ of the system is determined as $T^* = T_p$
 when the heat current between the system and the probe vanishes, namely $J_p(\mu_p, T_p) = 0$,
 so that a thermal equilibrium state is considered to be established between the system and the temperature probe.
 By keeping $J_p(\mu_p, T_p) = 0$ in the temperature measurement,
 the probe temperature $T_p$ can be expressed as a function of the electrochemical potential $\mu_p$ of the probe; see \Figure{fig2-unique-T}(a).
 When the probe is biased away from the thermal equilibrium state of the system,
 the probe temperature $T_p$ will become hotter or colder than $T_{eq}$, the value of the contour $J_p = 0$ at a specific $\mu_p$;
 see the line $PQ$ in \Figure{fig2-unique-T}(a).
 Consequently, a temperature gradient is established across the system-probe contact,
 and the heat current driven by this temperature gradient will flow through the contact.
 If $T_p$ is hotter than $T_{eq}$, such as the point $Q$ in \Figure{fig2-unique-T}(a),
 the heat current will flow from the probe into the system,
 namely $J_p < 0$, and vice versa.
 Analogously,
 when a voltage probe is weakly coupled to the two-level system \cite{Sta16155433},
 the local electrochemical potential of the system is obtained as $\mu^* = \mu_p$ in the case of $I_p(\mu_p, T_p) = 0$.
 \Figure{fig2-unique-T}(b) shows the dependence of the probe electrochemical potential $\mu_p$ on the probe temperature $T_p$.
 The charge current will spontaneously flow from the side with a high electrochemical potential to the side with a low one.
 From the above analysis, the direction in which heat/charge current flows through the system-probe contact has been pointed out:
 heat spontaneously flows from a hot body to a cold body, and charge spontaneously flows from a body with a high electrochemical potential to a body with a low one.
 Therefore, the local temperature/chemical potential defined by the \gls{zcc} is consistent with the Clausius's statement of the second law.

 Since $T_p$ largely depends on $\mu_p$ for the temperature measurement,
 the unique measurement of local temperature $T^*$ requires the knowledge of $\mu_p$.
 Similarly, a known $T_p$ is necessary for the unique measurement of local electrochemical potential $\mu^*$.
 In a word, to obtain a unique result of $T^*$ and $\mu^*$, it is necessary to measure both temperature and voltage simultaneously.
 Jacquet has analytically shown that the values of $T^*$ and $\mu^*$ are uniquely determined by simultaneous voltage and temperature measurements
 for a non-interacting electron system in the linear response regime \cite{Jac09709}.
 Stafford \emph{et al.} have analytically and numerically shown that
 the simultaneous measurement of temperature and electrochemical potential
 uniquely yields $T^*$ and $\mu^*$ in an interacting electron system arbitrarily away from equilibrium \cite{Sta16155433}.
 \Figure{fig2-unique-T}(c) shows the numerical simulation for simultaneous measurements of $T^*$ and $\mu^*$ defined by the \gls{zcc} in a two-level system.
 Only one point can be found along $I_p = 0$ that also satisfies $J_p = 0$,
 which implies a unique result of the \gls{zcc}.

 \noindent {\textsl{The Third Law} -}
 By carefully defining a local non-equilibrium entropy of a fermionic system in a non-equilibrium steady state,
 the local temperature has been found to be consistent with the Nernst's statement of the third law \cite{Sha15245417,Sta17092324,Sha19},
 that it is impossible for any process, no matter how idealized, to reduce the entropy of a system to its absolute-zero value in a finite number of operations \cite{Jos71}.
 The local non-equilibrium entropy $S(\bf x )$ is defined as follows \cite{Sha19}
\be
 S({\bf x}) = \sum_{\alpha} \int d\omega g_\alpha(\omega; {\bf x}) s(f_\alpha) + \sum_l | \psi({\bf x}) |^2 s(f_l).
 \label{local-entropy-system}
\ee
 Here, $g_\alpha(\omega; {\bf x}) = {\rm tr} \{\bm A_\alpha(\omega; {\bf x}) \}$ is the local \gls{dos} at the position $\bf x$ of the system
 due to scattering states incident on the system from reservoir $\alpha$. The spectral function $\bm A_\alpha(\omega) = \frac{1}{2 \pi} \bm G(\omega) \bm \Gamma(\omega) \bm G^\dagger(\omega)$ takes into account the system-reservoir tunneling matrix $\bm \Gamma$ and the \gls{negf} $\bm G$;
 $s(f_\alpha)$ is defined as $s(f_\alpha) = - k_B \{f_\alpha(\omega) \ln f_\alpha(\omega) + [1 - f_\alpha(\omega)] \ln [1 - f_\alpha(\omega)]\}$,
 where $f_\alpha$ is the Fermi-Dirac function of the reservoir $\alpha$;
 $f_l$ is the occupation number of the $l$-th localized state of the system,
 and $\psi_l({\bf x})$ is the wave-function of the $l$-th localized state.
 It has been analytically shown in \cite{Sha15245417,Sta17092324,Sha19}
 that the local entropy $S(\bf x)$ asymptotically converges to zero as $T^* \to 0$,
 consistent with the third law of thermodynamics.

\subsubsection{A scale of local thermal fluctuations and excitations} \label{subsubsec:scale}

 As already reviewed in \Sec{subsubsec:DF&FDR}, the \gls{fdr}-based definition of local temperature provides a measure of local fluctuations in a non-equilibrium nanosystem.
 Here, we will go beyond the \gls{fdr},
 focus on the attempts to establish a generic non-equilibrium \gls{fdt} in an open quantum system out of equilibrium,
 and analyze the applicability of the existing \gls{fdt} to define a local temperature.

 Prost \emph{et al.} have proposed a generic \gls{fdt} applicable for any system with Markovian dynamics in a non-equilibrium steady state \cite{Pro09090601}.
 The small variations of observables are applied around the steady state of a classical system
 which is defined by a set of observables denoted by $\bm c$, such as positions, velocities and other parameters.
 The state of the system is controlled by a set of parameters $\bm \lambda$.
 For fixed values of the control parameters $\bm \lambda$,
 a steady state is assumed which is characterized by its probability distribution function $\rho_{\rm SS}(\bm c; \bm \lambda)$.
 When small variations of the control parameters $ \delta \bm \lambda (t) \equiv \bm \lambda(t) - \bm \lambda^{\rm SS} $ are imposed on a steady state value $\bm \lambda^{\rm SS}$,
 the non-equilibrium \gls{fdt} reads \cite{Pro09090601}:
\be
  \bigg\la \frac{\partial \phi(\bm c(t); \bm \lambda^{\rm SS})} {\partial \lambda_\alpha} \bigg\ra = \int^{t}_{t_i} dt^\prime \chi_{\alpha \gamma}(t - t^\prime) \delta \lambda_\gamma (t^\prime),
 \label{fdt-Markov}
\ee
 where $\chi_{\alpha \gamma} (t - t^\prime)$ is the response function in the steady state of the system,
 and the potential $\phi(\bm c; \bm \lambda)$ is defined by $\phi(\bm c; \bm \lambda) = -\log[\rho_{\rm ss}(\bm c; \bm \lambda)]$.
 The \gls{fdt} of \Eq{fdt-Markov} has been generalized to any Markovian quantum system \cite{Meh1866}.
 Although the temperature is not explicitly involved in \Eq{fdt-Markov},
 the changes of temperature still have an effect on the parameters $\bm \lambda$ \cite{Pro09090601},
 and the response function may depend on temperature.

 Zhang \emph{et al.} have presented a general theory of non-Markovian dynamics for an open quantum system consisting of non-interacting particles \cite{Zha12170402,Zha191849}.
 A non-equilibrium \gls{fdt} has been proposed \cite{Zha12170402,Ali17033830}
\be
  v(\tau,t) = \int^{\tau}_{t_0} d\tau_1 \int^{t}_{t_0} d\tau_2  u(\tau,\tau_1)  {\tilde g(\tau_1,\tau_2)}  u^\dg(t,\tau_2),
 \label{fdt-non-markov}
\ee
 where the \gls{negf} $u$ describes the dissipation dynamics of the system,
 and the \gls{negf} $v$ characterizes the non-equilibrium fluctuations inside the open system.
 The non-Markovian interactions between the system and the environment are taken into account by the self-energy correction ${\tilde g}$.

 In the steady state limit $t \to \infty$,
 the function $v$ can be expressed as $v (t, t)|_{t \to \infty} \equiv \int d\omega  \chi(\omega)$ \cite{Lo159423,Xio1513353,Zha191849}.
 The non-equilibrium \gls{fdt} then reads \cite{Lo159423,Xio1513353}
\be
 \chi(\omega) = [{D_l}(\omega) + {D_d}(\omega)] f(\omega, T^*),
 \label{fdt-non-markov-steady-0}
\ee
 in a steady state.
 Here, the first term ${D_l}(\omega)$ on the r.h.s. of \Eq{fdt-non-markov-steady-0} arises from the particles in localized bound states of the system,
 and the dissipation spectrum ${D_d}(\omega)$ is the contribution from the particles moving into the environment.
 $f(\omega, T^*)$ is the Bose or Fermi distribution function of particles, depending on the system being made of bosons or fermions.
 If there is no particle in the localized states, $ {D_l}(\omega)$ will become zero, then \cite{Xio1513353,Zha191849}
\be
 \chi(\omega) = {D_d}(\omega) f(\omega, T^*).
 \label{fdt-non-markov-steady}
\ee
 Equation (\ref{fdt-non-markov-steady}) resembles the formulation of the equilibrium \gls{fdt} at temperature $T^*$ in the particle number representation \cite{Xio1513353,Zha191849}.
 $T^*$ in \Eq{fdt-non-markov-steady} then provides a scale of local fluctuations of a nanosystem in a non-equilibrium steady state.

\subsubsection{Local properties of nanosystems} \label{subsubsec:prop}

\begin{figure}[htbp]
  \centering
  \includegraphics[width=0.8\columnwidth]{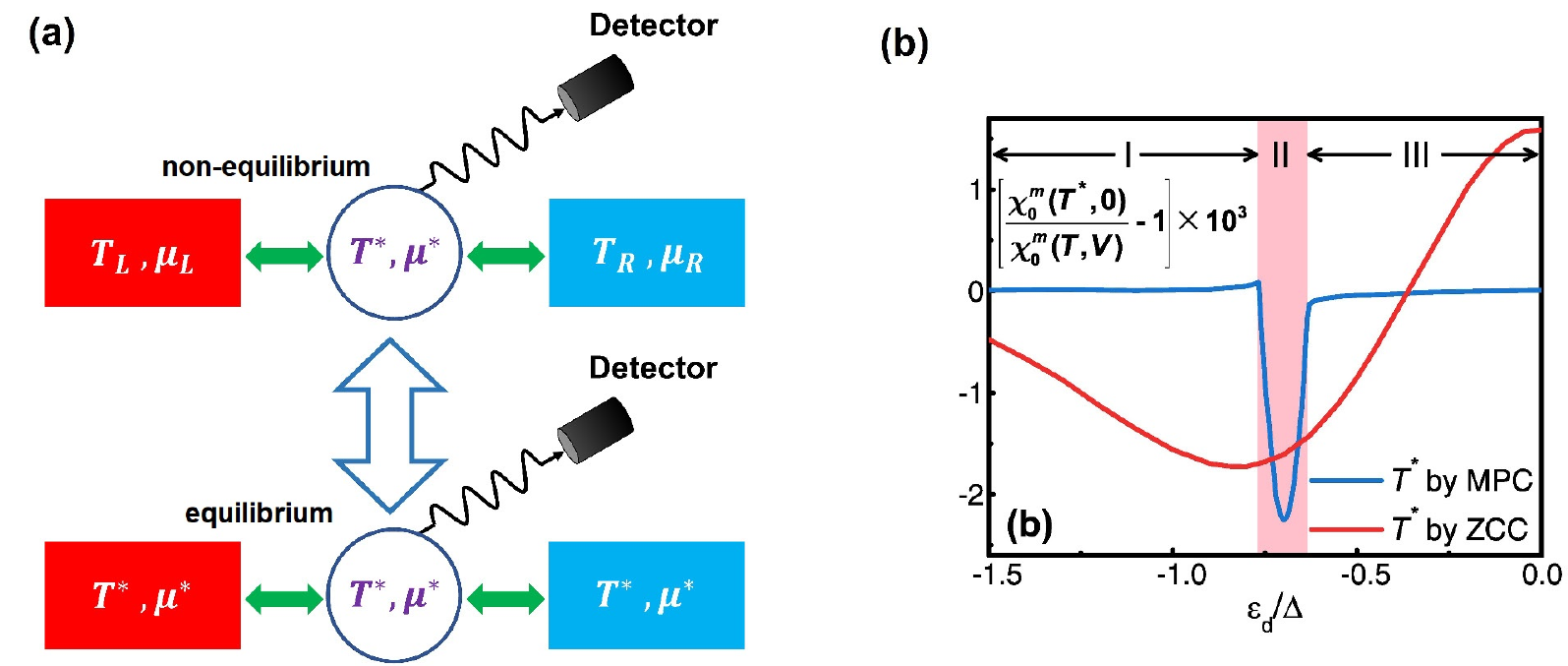}
  \caption
  {
   (a) Schematic illustration of the relation defined by \Eq{correspondence-relation-broad-band}.
   The local observable $\la O_o \ra$ of a non-equilibrium \gls{qd} (upper panel) can be made equivalent to that of a reference equilibrium \gls{qd} (lower panel),
   provided the two dots have the same local temperature $T^*$.
   (b) The relative deviation between $\chi^m (T,V) \equiv \frac{\partial \la \hat{m}_z (T,V) \ra} {\partial H_z}|_{H_z \to 0}$ and $\chi^m (T^*,0)$ versus $\epsilon_d$ for a non-interacting \gls{qd} under a bias voltage $V$.
   The temperature gradient across two leads is set to zero, $T_R = T_L = T$.
   Reprinted with permission from \cite{Ye16245105}. Copyright 2016 American Physical Society.
   }
  \label{fig3-correspondence-relation-noninteraction}
 \end{figure}

 Ye \emph{et al.} have studied the effect of bias-induced electronic excitations on a local observable $\la \hat{O} \ra$ \cite{Ye16245105}.
 For a non-interacting \gls{qd},
 it has been analytically shown in \cite{Ye16245105} that the \gls{mpc}-defined local temperature not only characterizes the magnitude of the electronic excitations,
 but also establishes a correspondence relation for $\la \hat{O} \ra$ between a non-equilibrium \gls{qd} and a reference equilibrium \gls{qd}
\be
 O_0 (T_L,\mu_L;T_R,\mu_R) = O_0 (T^*,\mu^*;T^*,\mu^*),
 \label{correspondence-relation-broad-band}
\ee
 in the broadband limit,
 provided that the zero perturbation condition
\be
 \frac{\delta O_p(T_p,\mu_p)}{\Delta_p} |_{T_p =T^*,\mu_p = \mu^*,\Delta_p \to 0} =0,
 \label{zero-perturbation-condition}
\ee
 can be satisfied.
 Here, $\Delta_p$ is the dot-probe coupling strength,
 and $O_o (T_L,\mu_L;T_R,\mu_R)$ is the expectation value of $\la \hat{O} \ra$ in the absence of the probe,
 with the lead $\alpha =L,R$ subjected to $T_\alpha$ and $\mu_\alpha$.
 As shown in Figure \ref{fig3-correspondence-relation-noninteraction}(a),
 the physical significance of $T^*$ is clarified as follows:
 the electronic excitations induced by a bias voltage or temperature gradient can be equivalently characterized as thermal excitations induced by a uniform equilibrium temperature.
 This relation provides a microscopic interpretation of the \gls{mpc}-based definition of local temperature.

 When the bandwidth of the leads is finite, the correspondence relation still holds under a small bias voltage $V = (\mu_R - \mu_L)$:
\be
 O_0 (T_L,\mu_L;T_R,\mu_R) = O_0 (T^*,\mu^*;T^*,\mu^*) + \mathcal{O}(V^2),
 \label{correspondence-relation-finite-band}
\ee
 provided the zero perturbation condition of \Eq{def-MPC-observable} is satisfied.
 Here, a voltage-dependent higher order term $\mathcal{O}(V^2)$ emerges due to the finite bandwidth.

 Figure \ref{fig3-correspondence-relation-noninteraction}(b) shows the numerical verification for the correspondence relation
 between a non-equilibrium \gls{qd} and a reference \gls{qd},
 by comparing the relative deviation for local magnetic susceptibility $\chi^m$ of the two systems.
 For $T^*$ determined by the \gls{mpc},
 the deviation appears to be vanishingly small in regions I and III;
 while in region II the deviation remains appreciable.
 The explanation of such large deviation is
 that the dot is in a near-resonance situation,
 local and non-local excitations could both take place,
 and the zero perturbation condition of \Eq{zero-perturbation-condition} is not achievable.

 \begin{figure}[htbp]
  \centering
  \includegraphics[width=0.7\columnwidth]{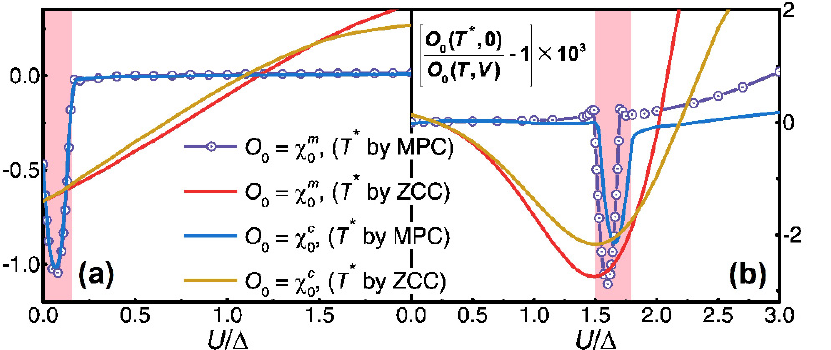}
  \caption
  { The relative deviations between $O_0 (T,V)$ and $O_0 (T^*,0)$ for two observables $O_0$ versus the Coulomb interaction energy $U$ for an interacting \gls{qd}
   under a bias voltage $V$ at the background temperature (a) $T = \Delta$ and (b) $T = 0.1\Delta$.
   The \gls{mpc} is imposed on the local magnetic susceptibility $\chi^c \equiv - \frac{\partial \la \hat{n}_d \ra} {\partial \epsilon_d}$ and local charge susceptibility $\chi^m$.
   The temperature gradient across two leads is set to zero, $T_R = T_L = T$.
   Reprinted with permission from \cite{Ye16245105}. Copyright 2016 American Physical Society.
   }
  \label{fig4-correspondence-relation-interaction}
 \end{figure}

 The correspondence relation for an interacting \gls{qd} has been verified by employing a numerical approach \cite{Ye16608}.
 The relative deviations between $O_0 (T,V)$ and $O_0 (T^*,0)$
 are shown in \Figure{fig4-correspondence-relation-interaction}
 for two choices of the local observable $\la \hat O \ra$--the local magnetic susceptibility $\chi^m$ and the local charge susceptibility $\chi^c$.
 The \gls{mpc}-defined $T^*$ leads to rather minor deviations so long as the zero perturbation condition can be achieved, except for the
 interaction strength in the shaded region of \Figure{fig4-correspondence-relation-interaction}.
 In that case, as for non-interacting electrons,
 the dot is in a near-resonance situation,
 and the local and non-local excitations could both take place,
 where the correspondence relation for local observables does not hold.

 In \Figure{fig4-correspondence-relation-interaction}(b), the deviation between $\chi^m (T,V)$ and $\chi^m (T^*,0)$ appears to be a small but finite value around the Coulomb energy $U = 3.0 \Delta$.
 This is because the Kondo resonant states inside the \gls{qd} start to emerge as $U$ increases.
 Under a bias voltage, the Kondo resonant states facilitate the electron co-tunneling processes,
 which can be understood as the concurrence of local spin-flip and non-local electron-transfer excitations.
 Therefore, as in the case of the shaded region, the correspondence relation for local observables does not hold
 since local and non-local excitations both take place.

\subsection{Practical applications} \label{subsec:app}

\subsubsection{Quantum oscillations of local temperature in nanostructures} \label{subsubsec:Temp-Osc}

 When a classical bulk system is in contact with two heat baths with different temperatures,
 heat will transfer from a hot bath, through the system, to a cold bath.
 The thermal transport finally establishes a temperature distribution across the system.
 When a bulk system is in the linear response regime,
 the Fourier's law provides a good description for the thermal transport and the temperature distribution.
 However, if the size of the system is reduced to the nanoscale, which is comparable to or even smaller than the wavelength or mean free path of its constituent particles,
 the energy transport process will be significantly different from that in a bulk system due to the emergence of quantum coherence effects.
 The quantum effects will then give rise to quantum oscillations of the local temperature in the entire nanostructure as predicted in \cite{Dub0997}.
 It has been found that the temperature oscillations in nanowires \cite{Dub0997,Cas10041301,Cas11165419} strongly violate the Fourier's law
 which predicts a linear temperature profile along a one-dimensional nanosystem \cite{Roy08062102,Dha08457,Mic03325,Gar015486}.

\begin{figure}[htbp]
  \centering
  \includegraphics[width=0.35\columnwidth]{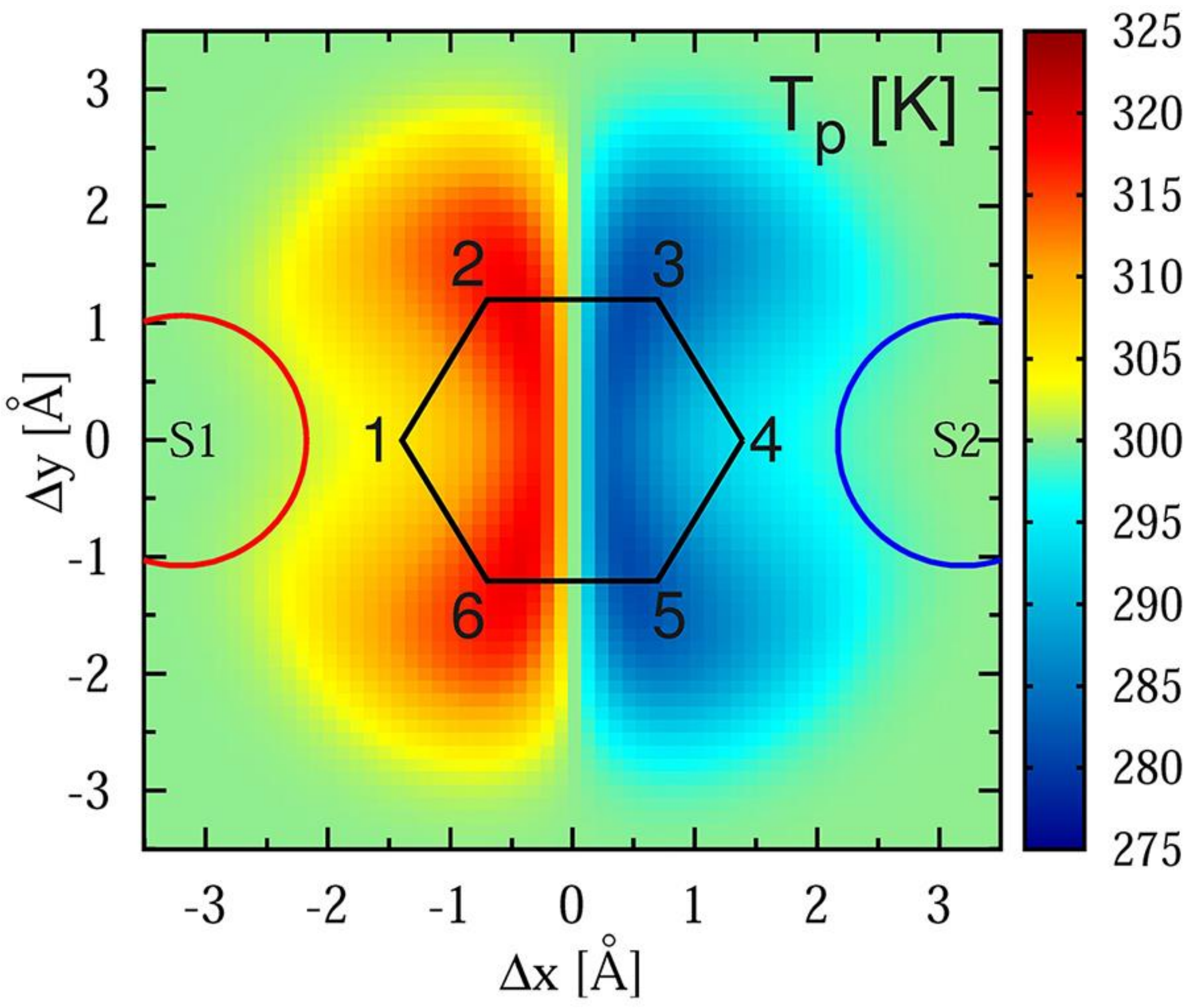}
  \caption
  {
  The calculated local temperature distribution of the BDT junction.
  Quantum oscillations of local temperature are visible in the vicinity of the molecule,
  which can be explained in terms of the charge transfer between the electrodes and the benzene ring.
  The ambient temperature is taken as $T_0 = 300$ K.
  Reprinted with permission from \cite{Ber134429}. Copyright 2013 American Chemical Society.
  }
  \label{fig5-temperature-oscillation-molecule}
\end{figure}

 Bergfield \emph{et al.} have investigated the local temperature distribution in a benzenedithiol junction subjected to a thermal bias \cite{Ber134429}.
 \Figure{fig5-temperature-oscillation-molecule} illustrates spatial oscillations of the local temperature around the molecule.
 The oscillations have been explained by the charge transfer from the electrodes with different temperatures into the benzene ring.
 The electrons from the electrodes with a specific temperature carry finite thermal energy,
 which can lead to local heating/cooling in nanostructures.
 Assuming that the left electrode is hotter than the ambient temperature and the right one is colder,
 the charge transfer mechanism between the electrodes and the molecule \cite{Ber134429} indicates that electrons from the hot side (S1) can move to atoms 2, 4, and 6,
 while electrons from the cold side (S2) can move to atoms 1, 3, and 5.
 Therefore, the atoms 3 and 5 appear colder than the ambient temperature and the atoms 2 and 4 appear hotter, as displayed in \Figure{fig5-temperature-oscillation-molecule}.
 Taking into account that the atoms 1 and 4 are in proximity to the hot and cold electrodes, respectively,
 the temperature of the electrodes washes out the temperature oscillations around the atoms 1 and 4.
 Therefore, local temperature oscillations could be understood as a result of the quantum interference of heat flow from the electrodes with different temperatures \cite{Ber134429} .

 Other numerical calculations have predicted the emergence of quantum oscillations of local temperature in various nanostructures \cite{Ber134429,Sta16155433,Sha15245417,Ber14235438}.
 However, the ranges of these temperature oscillations in nanosystems are usually less than 1 nm,
 which is beyond the spatial resolution of the present thermometry \cite{Kim124248}.
 To find a condition under which the range of the temperature oscillations can be brought within the spatial resolution of present techniques in thermal microscopy,
 theoretical efforts have been made to simulate the local temperature distribution in a realistic system, such as a graphene flake under a thermal bias \cite{Ber15125407}.
 It has been found that the wavelength of the temperature oscillations in a graphene flake is related to that of the Friedel oscillations,
 an equilibrium property that results from long-range interferences in the local \gls{dos}.
 The wavelength of the Friedel oscillations $\lambda$ in a graphene flake can be controlled by the Fermi energy $\mu$ of the flake,
 namely $\lambda \propto (\mu - \mu_{\rm Dirac})^{-1}$.
 As the Fermi energy $\mu$ approaches the Dirac point $\mu_{\rm Dirac}$ of the graphene,
 the wavelength of the temperature oscillations grows dramatically \cite{Ber15125407}.
\begin{figure}[htbp]
  \centering
  \includegraphics[width= \columnwidth]{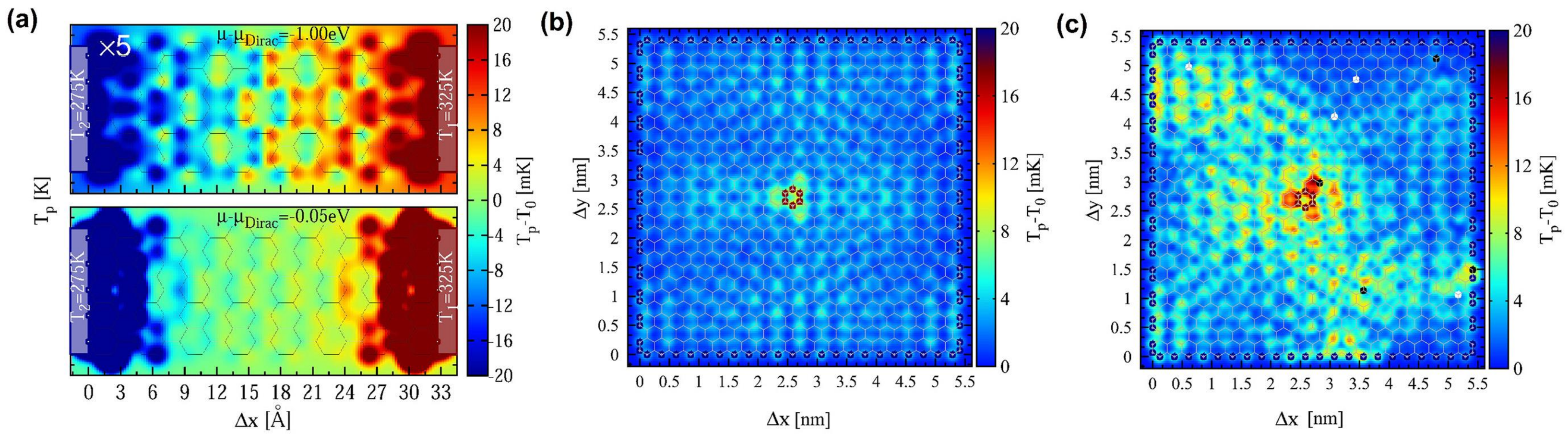}
  \caption
  {
  (a)
  The calculated spatial temperature profile for a graphene flake.
  The values in the upper panel are multiplied by a factor of five.
  By adjusting $\mu - \mu_{\rm Dirac}$ the temperature oscillation wavelength can be tuned.
  The ambient temperature is taken as $T_0 = 300$ K.
  (a) from \cite{Jus13}.
  The simulated temperature distribution of a graphene flake without (b) and with (c) impurities (white and black circles).
  A hot needle-like terminal (red circles) is located in the center of the flake,
  and the edge of the flake held at ambient temperature $T_0 = 300$ K by cold electrodes (blue circles).
  The existence of impurities significantly affects the temperature field.
  Reprinted with permission from \cite{Ber15125407}. Copyright 2015 American Physical Society.
  }
  \label{fig6-temperature-oscillation-graphene}
\end{figure}
 \Figure{fig6-temperature-oscillation-graphene}(a) shows that
 when $\mu - \mu_{\rm Dirac} = - 1.00$\,eV in the upper panel \Figure{fig6-temperature-oscillation-graphene}(a)
 the quantity $\lambda$ is evaluated as $3.8\,$\AA\ \cite{Jus13}, and there are relatively large oscillations in the middle of the graphene.
 When $\mu - \mu_{\rm Dirac} = - 0.05$ eV in the lower panel \Figure{fig6-temperature-oscillation-graphene}(a)
 the wavelength $\lambda$ increases to $\sim 10\,$\AA\ \cite{Ber15125407}.
 The temperature oscillations almost vanish in the middle of the graphene flake due to its finite size.
 In practice, the Fermi energy of a graphene flake can be continuously tuned by applying a gate voltage on the flakes \cite{Yu093430,Cra1142}.
 Thus, the local temperature oscillations may be observed in a realistic system by applying a proper gate voltage.
 In addition, the numerical simulation of local temperature has been carried out on a graphene flake doped with impurities and vacancies \cite{Ber15125407}.
 \Figure{fig6-temperature-oscillation-graphene}(b) and (c) suggest that these defects have a strong influence on the temperature distribution.
 Long-range temperature oscillations appear across the doped graphene.
 This result suggests that doping a graphene flake with defects may be an experimentally feasible method to observe local temperature oscillations in a realistic system.

\subsubsection{Recovering Fourier's law at the nanoscale} \label{subsubsec:Fourier-law}
\begin{figure}[htbp]
  \centering
  \includegraphics[width=0.85\columnwidth]{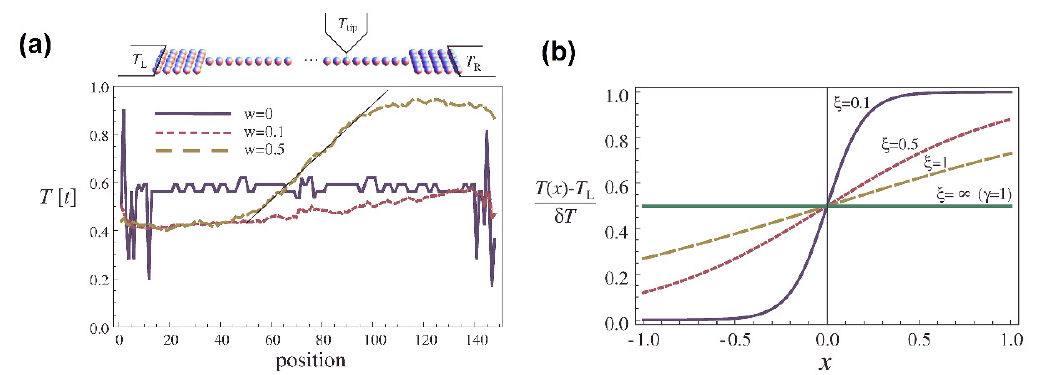}
  \caption
  {
  (a)
  Upper panel:
  Schematic representation of the calculation to determine the local temperature via the addition of a probe.
  A nanowire is coupled to the left and right leads (the solid lines at the edges).
  Lower panel:
  Local temperature as a function of the position $x$ along the wire.
  The local temperature is calculated for three different values of disorder strength $W$.
  For the wire with $W = 0$ the temperature hardly changes along the wire,
  while a uniform temperature gradient builds up as the disorder strength $W$ increases (showed by a thin solid line).
  Reprinted with permission from \cite{Dub09115415}. Copyright 2009 American Physical Society.
  (b)
  The normalized local temperature $[T(x) - T_L] / \delta T$ as a function of the position variable $x$ along the wire.
  The local temperature is calculated for three different values of dephasing length $L_\phi$.
  Here, $\xi$ in the figure is the $L_\phi$ in the main text.
  In this figure, the position variable is defined as $x \equiv \frac{2l}{L} - 1$, with the position $l$, and the length of the nanowire $L$.
  The value $x = -1$ is equivalent to $l = 0$ and $x = 1$ corresponds to $l = L$.
  $\delta T \equiv T_R - T_L$ is the temperature gradient across the nanowire.
  The temperature $T$ in the two figures is the local electron temperature defined by \Eq{MPC-density}.
  Reprinted with permission from \cite{Dub09042101}. Copyright 2009 American Physical Society.
  }
  \label{fig7-the-Fourier's-law-nanowire}
\end{figure}
 The above studies suggest that quantum effects can produce spatial oscillations of the local temperature in nanostructures, thus
 violating the Fourier's law.
 This motivates the search for the conditions under which the quantum temperature oscillations vanish, and the Fourier's law is
 restored.

 Efforts to understand how the Fourier's law in a one-dimensional model system may be recovered have been made previously by Dubi and Di Ventra \cite{Dub09115415,Dub09042101}.
 They considered a nanowire subjected to a thermal bias,
 and randomly tuned every on-site energy of the wires (for example, $\epsilon_i$ at the $i$ th point of the wires) in a limited range $[-W,W]$,
 with $W$ being the disorder strength \cite{Dub09115415}.
 The \gls{mpc}-defined local electron temperature is then calculated as a function of the positions along the wire under different $W$.
 The results in \Figure{fig7-the-Fourier's-law-nanowire}(a) illustrate that strong disorder strength gives rise to the Fourier's law at the nanoscale.

 This crossover from the ballistic (the Fourier's law is invalid) to diffusive (the Fourier's law is valid) regime could be interpreted in terms of electronic dephasing.
 To discuss the condition for the dephasing,
 the characteristic length $L_\phi$ is introduced,
 which describes the length over which the electronic wave-function retains its phase \cite{Dub09042101}.
 If the length $L_\phi$ is greater than the length of the nanowire $L$, namely $L_\phi \geq L$,
 the dephasing has no effect and thermal transport is ballistic,
 while for $L_\phi \leq L$ the thermal transport is in the diffusive regime due to the dephasing effect.
 \Figure{fig7-the-Fourier's-law-nanowire}(b) shows the local temperature distribution under different characteristic lengths.
 With increasing $L_\phi$ the temperature gradient across the system vanishes,
 and a uniform temperature profile emerges as $L_\phi \to \infty$.
 This change of the temperature profile corresponds to the crossover of thermal transport from the ballistic to the diffusive regime.
 The disorder in a nanowire breaks the quantum phases of the electronic wave-functions \cite{Her74117,Per06036801,Bil08891} and reduces the length $L_\phi$.
 When the disorder becomes sufficiently strong, $L_\phi \leq L$, and electronic dephasing occurs in the nanowire.
 The Fourier's law is thus retrieved in such a strongly-dephased system \cite{Dub09042101}.

 Since the dephasing mechanism has successfully explained how the Fourier's law can be recovered in some model systems,
 several theoretical efforts were made to explain the emergence of the Fourier's law in realistic systems by disorder or dephasing \cite{Ber13310}.
 However, retrieving the Fourier's law in realistic systems,
 such as molecular junctions, would require a very strong disorder or dephasing to make $L_\phi$ shorter than the scale of the junction.
 This, in turn, would likely destroy the covalent bonds of the molecule \cite{Ber13310,Inu184304}, effectively breaking the molecule.

 \begin{figure}[htbp]
  \centering
  \includegraphics[width=0.8\columnwidth]{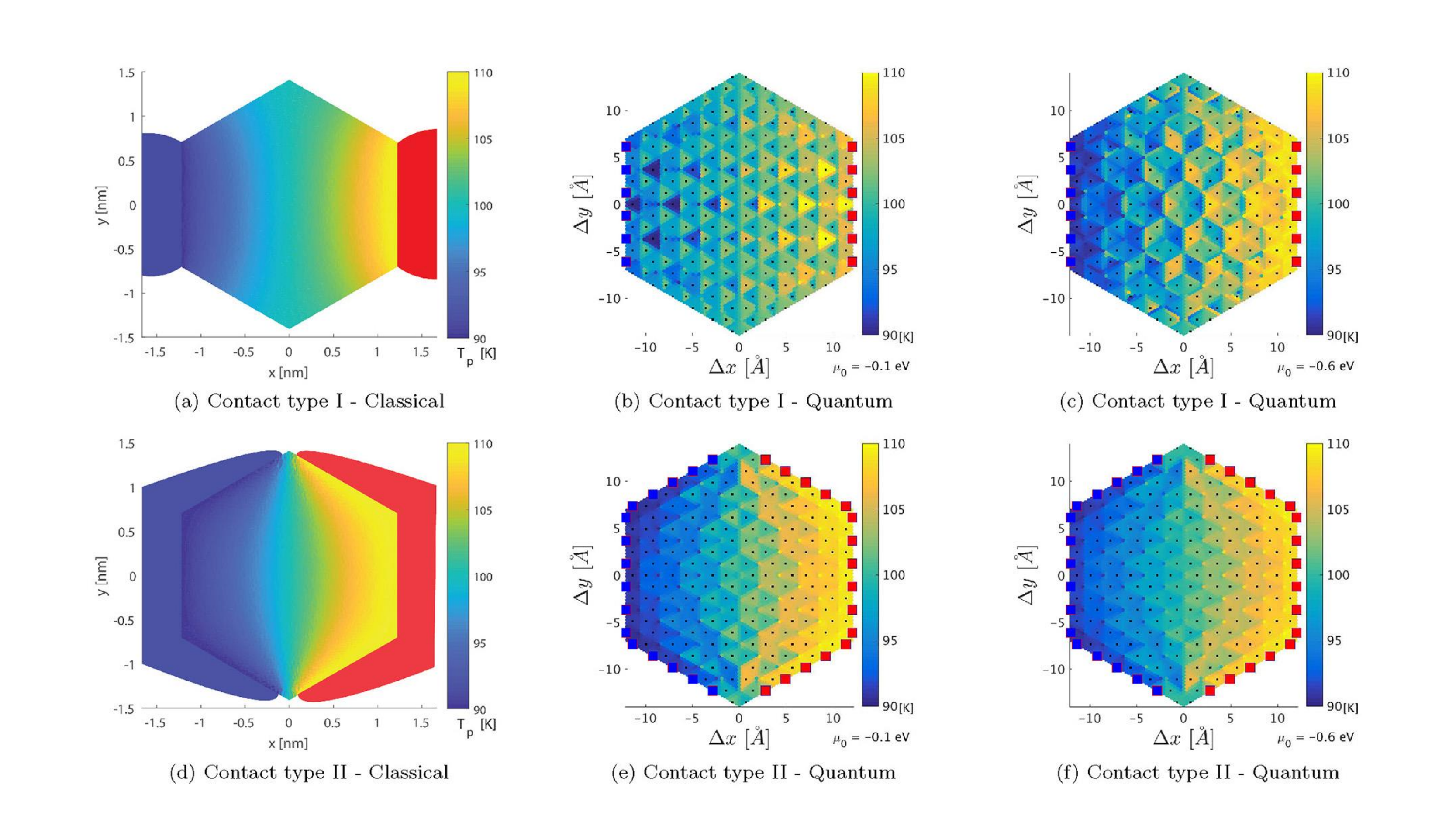}
  \caption
  {
  The classical (a, d) and local (b, c, e, f) temperature distribution of a graphene flake under thermal bias for two contact geometries.
  In type I contact (top panels), only the left and right edges of the flake couple to the electrodes.
  In type II contact, two electrodes cover three edges of the flake, leading to a stronger coupling.
  The classical temperature distribution is evaluated by the Fourier's law,
  and the local temperature is given by \Eq{def-zcc}.
  The local temperature calculations are at Fermi energies $\mu_0 - \mu_{Dirac} = -0.1$ eV (b, e) and $\mu_0 - \mu_{Dirac} = -0.6$ eV (c, f).
  The hot electrode (red) is held at 110 K, and the cold electrode (blue) is held at 90 K.
  The squares indicate the carbon atoms covalently bonded to the hot (red) and cold (blue) electrodes.
  The local temperature distributions for type I contact exhibit strong oscillations that depend on $\mu_0$,
  while for type II contact, the distributions resemble the classical distribution.
  Reprinted with permission from \cite{Inu184304}. Copyright 2018 American Chemical Society.
  }
  \label{fig8-the-Fourier's-law-graphene}
\end{figure}

 \begin{figure}[htbp]
  \centering
  \includegraphics[width=0.80\columnwidth]{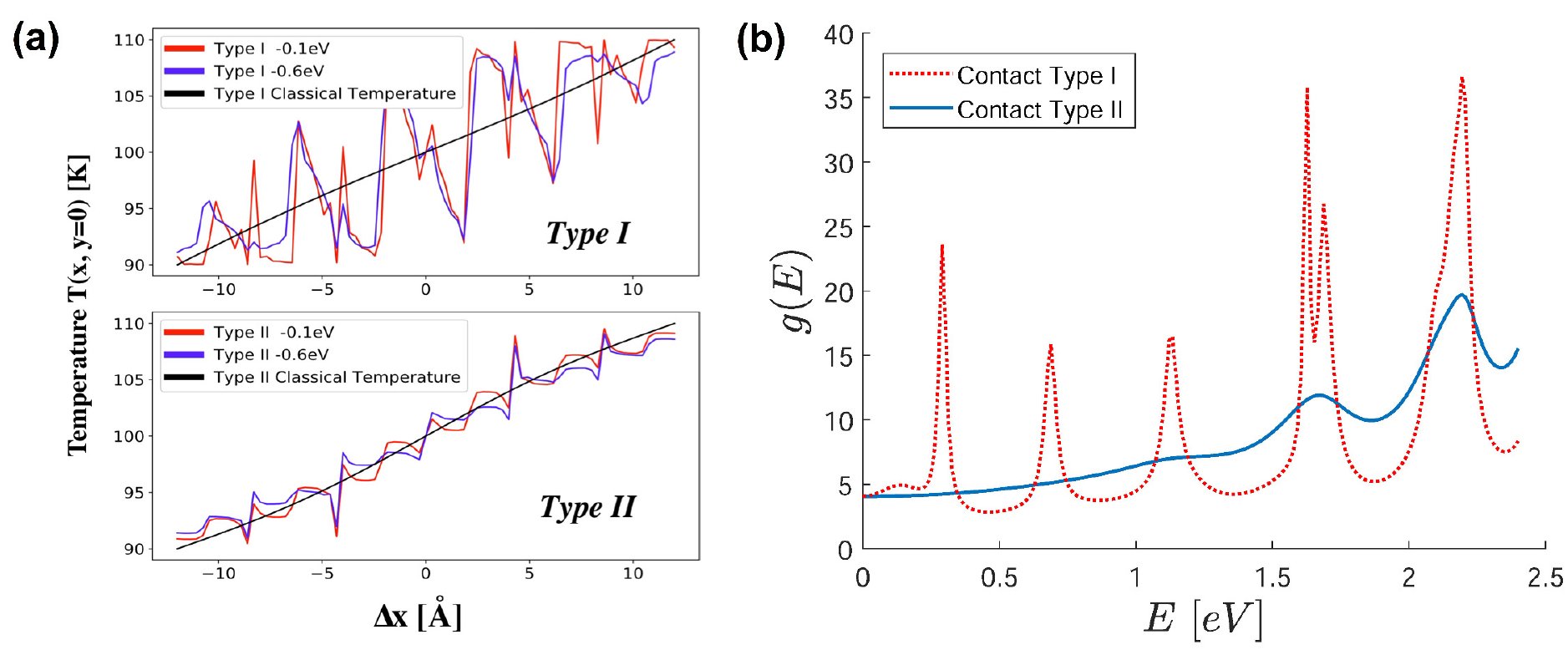}
  \caption
  {(a) Temperature profiles along the centerline of the system ($y = 0$) for each contact type and the Fermi energy
  shown in \Figure{fig8-the-Fourier's-law-graphene}.
  In the type II  contact, the temperature profile is much closer to that predicted by the Fourier's law.
  (b)
  Calculated \gls{dos}, $g(E)$, of a graphene flake junction for two different contact geometries.
  The \gls{dos} for the type I contact exhibits a sequence of sharp peaks,
  while the type II contact exhibits a smooth, nearly featureless \gls{dos}.
  Reprinted with permission from \cite{Inu184304}. Copyright 2018 American Chemical Society.
  }
  \label{fig9-the-Fourier's-law-graphene}
\end{figure}

 To address this issue, Inui \emph{et al.} have studied the Fourier's law in graphene flakes subjected to a thermal bias~\cite{Inu184304}.
 \Figure{fig8-the-Fourier's-law-graphene} shows the calculated local temperature distribution for two different graphene-electrode contact geometries \cite{Inu184304}. In the type I contact only the left and right edges of the flake couple to the electrodes, while for type II contact two electrodes cover three edges of the flake, leading to a stronger coupling.
 Compared with the temperature distribution for the type I contact,
 the temperature distribution for the type II contact is closer to the classical results predicted by the Fourier's law.
 This can be seen clearly from \Figure{fig9-the-Fourier's-law-graphene}(a) which compares the temperature profiles along the line $y = 0$ for each contact type.

 The different nature of thermal transport for the contact types I and II can be understood from the \gls{dos} of the system, $g(E)$, shown in \Figure{fig8-the-Fourier's-law-graphene}.
 For the type I contact, $g(E)$ exhibits a sequence of sharp peaks, corresponding to the energy eigenfunctions of the graphene flake.
 A sharply peaked \gls{dos} implies that the system is in the resonant-tunneling regime,
 where thermal transport is controlled by the wave-function of a single resonant state.
 In contrast, the type II contact has a smooth \gls{dos} which indicates that many quantum states contribute to thermal transport,
 so that temperature oscillations tend to average out.
 From the above analysis, these authors have put forward the following hypothesis for recovering Fourier's law in a realistic system \cite{Inu184304}:
 if sufficient quantum states participate in thermal transport across the system, the Fourier's law will emerge;
 when transport occurs in or near the resonant-tunneling regime, there is no classical behavior for the temperature distribution.

\subsubsection{Laser-induced ultrafast adsorption dynamics of molecules on metal surfaces} \label{subsubsec:adsorption}

 The dynamics of molecules from one adsorption site to another is the most elementary process of bond-breaking and -making on a surface,
 and is a key step in processes such as catalysis \cite{Rod148226}.
 The time when reactants are rearranged into products is merely of the order of a few hundred femto-seconds.
 Therefore, understanding ultrafast adsorption dynamics of these molecules on metal surfaces is of fundamental importance in the field of catalysis.
 Numerous experiments have studied the ultrafast elementary motions of adsorbates on metal surfaces by means of high time-resolution techniques \cite{Bon004653,Ste05236103,Bac051790,Lan06186105,Wat10241408,Ino16186101}.
 In these experiments, the molecular motion is generally triggered by femto-second laser pulses,
 which rapidly excite surface electrons to a high energy level.
 To characterize the magnitude of the thermal excitation for electrons or vibrational modes,
 the concept of the electron or phonon temperature has been used in adsorption dynamics.
 The temporal and spatial evolution of these two temperatures can be evaluated quantitatively by the two-temperature model \cite{Ani74375,All871460,Lin08075133},
 which describes the thermal response of a metal surface to a ultrashort laser pulse.
 The changes of two temperatures reflect the energy exchange between the laser-excited surface electrons and the vibrational modes of the adsorbates and surface lattice.

\begin{figure}[htbp]
  \centering
  \includegraphics[width=0.7\columnwidth]{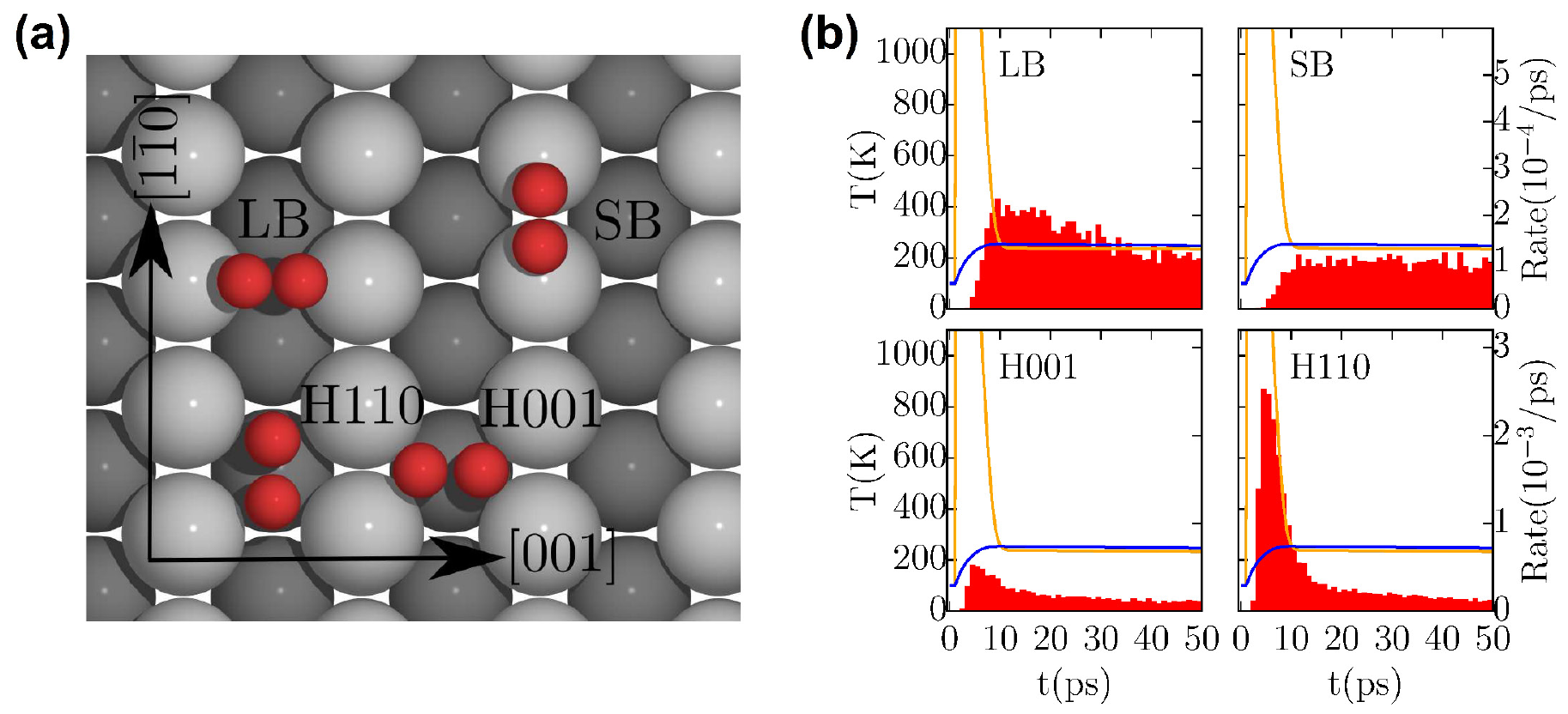}
  \caption
  {
  (a)
  Sketch of the position of the O$_2$ molecules (red ball) in the four adsorption sites:
  long bridge (denoted LB), short bridge (denoted SB),
  hollow along the $[1\bar{1}0]$ direction (denoted H110),
  and hollow along the [001] direction (denoted H001).
  (b)
  The electron (orange line) and phonon (blue line) temperatures calculated from the two-temperature model as a function of time from the four adsorption sites.
  The electronic temperature peaks at values larger than 6000 K.
  The desorption rates are also shown by the histograms.
  Reprinted with permission from \cite{Ivo16014301}. Copyright 2016 American Physical Society.
  }
  \label{fig10-absorption}
\end{figure}

 Based on the two-temperature model,
 Lon\v{c}ari\'{c} \emph{et al.} have theoretically investigated the laser-induced desorption of O$_2$ from Ag(110) \cite{Ivo16014301},
 using the Langevin equation to describe the dynamics of each atom of the molecule \cite{Spr94L57,Spr9673}.
 \Figure{fig10-absorption}(a) shows the model of the molecule position in the four adsorption sites,
 and \Figure{fig10-absorption}(b) shows the time dependence of the electron and phonon temperatures obtained by the two-temperature model,
 in comparison with the time evolution of the desorption rate for each adsorption site.
 Laser pluses imposed at the initial time rapidly excite electrons in the metal surface to several thousands Kelvin due to a relatively low electron heat capacity of metals.
 The electron-phonon interaction then transfers the energy from electrons to phonons
 which finally reach a thermal equilibrium state with each other.
 Compared with the evolution of the two temperatures,
 the behavior of the desorption rate of different positions illustrates that the desorption mechanism for O$_2$ depends on the adsorption configuration.
 The desorption rates for the hollow positions (H001 and H110) seem to follow the time evolution of the electron temperature, but with some time delay.
 These results suggest desorption from the hollow sites is mainly an electron mediated effect,
 where the energy transfer from the electrons excited by the laser pulse to the adsorbed molecule plays a dominant role \cite{Ivo16014301}.
 In contrast, the desorption rates from the bridge sites (SB and LB) do not seem to be very much affected by the high increase of electron temperature at the beginning.
 In these sites, the highest values of the desorption rates occur at longer times, once the electron and phonon temperatures are equilibrated.
 This suggests that the heating of electrons is not that important for desorption from the bridge sites,
 and that the laser mediated phonon excitation is the relevant mechanism in these sites \cite{Ivo16014301}.
 %

 Novko \emph{et al.} have also used to the two-temperature model to understand the laser-induced ultrafast dynamics of CO adsorbates on Cu(100) \cite{Nov19016806}.
 A \gls{dft} calculation has been employed to investigate the frequency and linewidth changes of the CO internal stretch mode on the metal surface.
 It has been found that two distinct processes give rise to these changes:
 electron-hole pair excitations result in the frequency shifts,
 while electron-mediated vibrational mode coupling gives rise to linewidth changes.
 Since the electron temperature affects the electron-hole pair excitations
 and the electron-mediated vibrational mode coupling depends on both electron and phonon temperatures \cite{Nov18156804},
 the electron and phonon temperatures play an important role in laser-induced ultrafast dynamics.
\subsubsection{Local temperature of Kondo systems} \label{subsubsec:Temp-Kondo-systems}

 Lobaskin and Kehrein have used a time-dependent local temperature to characterize the formation of the Kondo singlet \cite{Lob06301}.
 These authors have focused on a Kondo model system under a small periodic perturbation with frequency $\omega$.
 The local temperature $T^*(t_w)$ is defined by the \gls{fdr} expressed as follows \cite{Mit05121102,Lob06301}:
\be
 \lim_{\omega \to 0} \frac{{\rm Im} R(\omega, t_w)}{\omega} = \frac{1}{2 T^*(t_w)} C^{\rm (cum)}_{ \{A,A\} }(\omega = 0),
 \label{strong-correlation-Teff}
\ee
 in the zero frequency limit.
 Here, the imaginary part of the response function $R(\omega,t_w)$
\be
 R(\omega,t_w) = 2 i \int d\tau \theta(\tau) C_{ [A,A] }(t_w + \tau, t_w) e^{i \omega \tau}
 = i \int d\tau \theta(\tau) \la [\hat{A}(t_w + \tau) , \hat{A}(t_w)] \ra e^{i \omega \tau},
 \label{strong-correlation-Teff}
\ee
 describes the energy dissipated from the system after time $t_w$.
 $C^{\rm (cum)}_{ \{A,A\}}(\omega)$ is the Fourier transform of the second-order cumulant of the two-time correlation function $C_{ \{A,A\} }(t,t^\prime) = 1/2 \la \{ \hat{A}(t), \hat{A}(t^\prime) \} \ra$, and reads
\be
 C^{\rm (cum)}_{ \{A,A\}}(t,t^\prime) = C_{ \{A,A\} }(t,t^\prime) - \la \hat{A}(t) \ra \la \hat{A}(t^\prime) \ra,
 \label{strong-correlation-Teff}
\ee
 with an operator $\hat{A}$.

\begin{figure}[htbp]
  \centering
  \includegraphics[width=0.45\columnwidth]{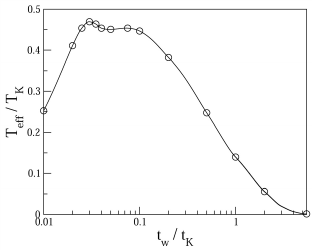}
  \caption
  {
  Local temperature $T^*$ as a function of time $t_w$ in a Kondo model.
  After a rapid increase to the maximum, $T^*$ eventually approaches zero.
  The size of the circles indicates the magnitude of the numerical error of the calculated data.
  $T_K$ is the Kondo temperature and $t_K$ is the time scale, defined as $t_K = 1/T_K$.
  $T_{\rm eff}$ is $T^*$ in this review.
  Reprinted with permission from \cite{Lob06301}. Copyright 2006 Springer Nature.
  }
  \label{fig11-Kondo-system}
\end{figure}

 Equation (\ref{strong-correlation-Teff}) has been applied to a Kondo model system at zero temperature.
 The local temperature is defined from the above \gls{fdr} imposed on the spin-spin correlation function
 $C^{\rm (cum)}_{\rm {S_z,S_z}}$ with $S_z$ being the impurity spin.
 The system is prepared in an initial state with a frozen impurity spin, then allowed to relax after $t_w = 0$.
 \Figure{fig11-Kondo-system} shows a relaxation of the non-equilibrium initial state towards an equilibrium state.
 The local temperature first rises up quickly as a function of time until it reaches a maximum of $T^* \simeq 0.45 T_K$ at $t_w = 0.03 t_K$.
 After that, $T^*$ decreases and nearly reaches zero.
 Here, $T_K$ is the Kondo temperature and $t_K$ is the time scale, defined as $t_K = 1/T_K$.
 This aging effect is interpreted in terms of the formation of the Kondo singlet.
 The Kondo singlet starts building up at $t_w = 0$.
 In the vicinity of the impurity,
 the conduction band electrons get locally heated up due to the release of the binding energy,
 when the Kondo singlet is being formed.
 After a sufficiently long time the Kondo singlet has been formed,
 the binding energy diffuses away, and the local temperature nearly becomes zero.

 Kehrein \emph{et al.} have calculated the local temperature defined by the \gls{fdr} imposed on various operators \cite{Kir10631,Rib1350001},
 and have presented a dependence of $T^*$ on the choice of the operator $\hat{A}$.
 It has been found that
 a large deviation among these local temperatures, calculated from different operators, starts to emerge when the system-bath coupling becomes strong \cite{Rib1350001}.

\subsubsection{Local electrochemical potential of two-dimensional networks} \label{subsubsec:local-mu}

\begin{figure}[htbp]
  \centering
  \includegraphics[width=0.7\columnwidth]{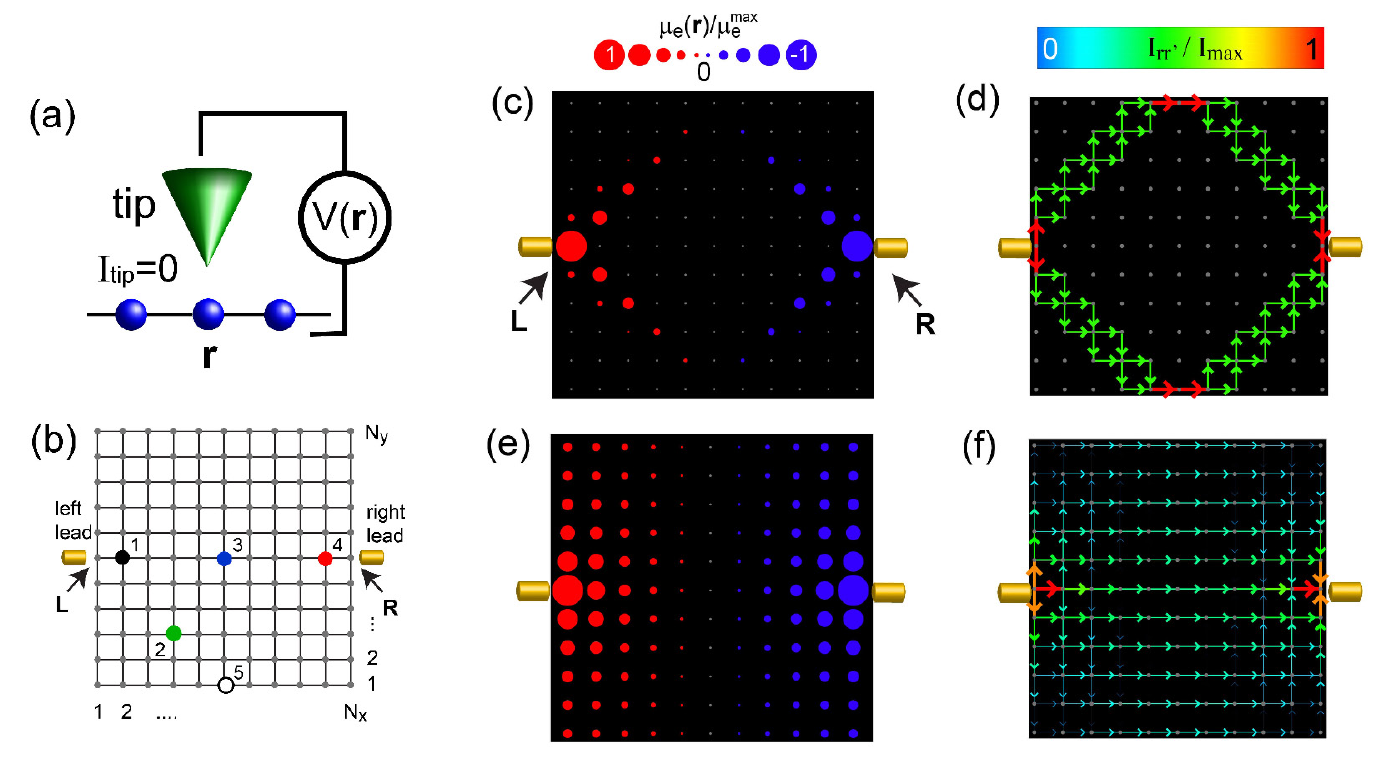}
  \caption
  {
  (a)
  Schematic representation of the principle of a scanning tunneling potentiometry: when the potentiometry probe is above a site $\bm r$ of the network,
  its potential $\mu_p(\bm r) = eV(\bm r)$ is adjusted so that there is a zero net current flowing between the probe and the network.
  Here $V(\bm r)$ is the voltage of the probe.
  (b)
  Schematic of the two-dimensional network of electronic sites that are coupled to two narrow leads with different voltages.
  (c-d) plot the normalized $\mu^*/\mu_{max}$ for electron-phonon interaction strength $g = 0.01t^2$ and $g = 500t^2$ with the energy scale $t$,
  and (e-f) plot the corresponding normalized current $I(\bm r, \bm r^\prime)/I_{max}$.
  Reprinted with permission from \cite{Mor17195162}. Copyright 2017 American Physical Society.
  }
  \label{fig12-STP-mu}
\end{figure}

 Analogous to the role of local temperature in thermal transport at the nanoscale,
 the knowledge of local electrochemical potential is not only important as a source of information about electrical properties of a nanosystem,
 but also is of fundamental importance to understand and explore charge transport at the nanoscale \cite{Mur86514,Bri96R5283,Voi18101101}.
 In practice, a local electrochemical potential of electrons can be measured by a scanning tunneling potentiometry with the spatial resolution down to several nanometers \cite{Cla137956,Wil156399,Wan13236802,Ji12114}.
 \Figure{fig12-STP-mu}(a) exhibits the principle of a scanning tunneling potentiometry:
 when the potentiometry probe is above the site $\bm r$ of the sample,
 the electrochemical potential of the probe $\mu_p(\bm r)$
 is adjusted such that there is a zero net electric current flowing between the probe and the network \cite{Mur86514}.

 Referring to the principle of a scanning tunneling potentiometry,
 a theoretical approach based on the \gls{zcc} has been proposed to
 image the local electrochemical potential distribution of a two-dimensional network \cite{Mor17195162,Mor1619,Wan11,Bev14415701}:
 one weakly couples a voltage probe to a two-dimensional network shown in \Figure{fig12-STP-mu}(b),
 and shifts the $\mu_p(\bm r)$ at every site
 such that the current $I_p(\bm r)$ between the probe and a site $\bm r$ in the network vanishes.
 By keeping $I_p(\bm r) = 0$, the local electrochemical potential $\mu^*(\bm r)$ is determined to be identical $\mu_p(\textsl{\bf r})$.
 Both the network and the probe are assumed in thermal equilibrium with the environment,
 and heat current between the probe and the network is not considered in \cite{Mor17195162}.

 Morr has studied the relation between the \gls{zcc}-defined local electrochemical potential and the electric current inside a two-dimensional network
 from the quantum to classical charge transport regime \cite{Mor17195162}.
 In the quantum regime, the electron-phonon interaction strength $g = 0$ describes a fully coherent system with an infinitely large elastic mean free path of electrons.
 The increasing strength of the electron-phonon interaction strength $g$ reduces the mean free path to a value much smaller than the scale of the network.
 A strong electron-phonon interaction leads to the electronic dephasing,
 and the charge transport in the network is in the classical regime when $g \to \infty$ \cite{Mor1619}.
 Therefore, one can tune the network's charge transport properties from the quantum to classical regime by increasing $g$.
 The current $I_p(\bm r)$ \cite{Bev14415701},
 and the current $I_{\bm r, \bm r^\prime}$ between the sites $\bm r$ and $\bm r^\prime$ of the network \cite{Ram86323,Car71916} are both computed using the \gls{negf} formalism.
 The spatial distributions of $\mu^*(\bm r)$ and $I_{\bm r, \bm r^\prime}$ are shown in \Figure{fig12-STP-mu}(c-f) under different $g$.
 It has been found that the Ohm's law relates
 $\mu^*(\bm r)$ to $I_{\bm r, \bm r^\prime}$,
 $I_{\bm r, \bm r^\prime} = \sigma(\bm r, \bm r^\prime) [\mu^*(\bm r) - \mu^*(\bm r^\prime)]$,
 with $\sigma(\bm r, \bm r^\prime)$ being the electric conductivity between two neighboring sites \cite{Gri99}.
 For a small $g$, the charge transport is in the quantum regime and $\sigma(\bm r, \bm r^\prime)$ is site-dependent,
 while for a large $g$, the charge transport is in the classical regime and $\sigma(\bm r, \bm r^\prime)$ becomes a constant.

\subsubsection{Local chemical potential of non-equilibrium magnons} \label{subsubsec:local-mu-magnon}


 Magnons, also known as the quanta of spin waves,
 are the quasi-particles of collective excitations representing electronic spins in a crystal lattice \cite{Ser10264002,Bau12391,Chu15453}.
 The creation and annihilation operators of a magnon fulfill the boson commutation relations.
 Being massless bosons, magnons at thermal equilibrium
 obey the Bose-Einstein distribution
 and the chemical potential is zero,
 since the magnon energy and particle number are not conserved \cite{Dem06430,Dem08045029,Cor16014412}.
 In contrast, magnons in a non-equilibrium state can be characterized by a nonzero chemical potential.
 For instance, in a ferromagnet
 a magnon may interact with other magnons and with the crystal lattice.
 The interaction between the magnons and the lattice
 can conserve \cite{Spa64} or change \cite{Str19184442} the number of magnons,
 depending on the specific form of the interaction.
 The relaxation times corresponding to different types of interactions are quite different.
 Take the ferromagnetic \gls{yig} thin films as an example \cite{Dem06430}.
 The thermal relaxation time
 due to two- or four-magnon scattering mechanisms can be as short as 100-200 ns,
 while the magnon-lattice relaxation time can be longer than 1 $\mu s$.
 Consequently, when an external pumping is applied to a sample,
 the non-equilibrium magnons generated by pumping will thermalize rapidly
 by magnon-conserving interactions \cite{Cor16014412},
 and these thermalized magnons
 are practically decoupled from the crystal lattice \cite{Dzy0764}.
 When the rate of magnon generation balances the magnon decay rate,
 the number of magnons will eventually reach a constant,
 and thus a quasi-equilibrium state is established for magnons.
 Such a quasi-equilibrium state can be described by
 a Bose-Einstein distribution with a characteristic temperature and a finite chemical potential \cite{Dem06430,Dzy0709C103,Dem08045029,Dem171579}.
 Here, the nonzero chemical magnon potential is associated with the frequency of pumped magnons \cite{Dzy0764}.

 From all this it is clear that the concept of chemical potential
 is crucially important for the investigation of the \gls{bec} of magnons \cite{Dzy0764}.
 As mentioned above,
 the population of magnons at a quasi-equilibrium state obeys a Bose-Einstein distribution
 %
 \be
 n(\epsilon_m,\mu_m,\beta_m) = \frac{1}{e^{\, \beta_m \left( \epsilon_m-\mu_m \right)}-1},
 \label{N0-magnon-1}
 \ee
 where $\epsilon_m$ is the energy of magnons, $\beta_m$ is the inverse magnon temperature,
 and $\mu_m$ is the magnon chemical potential.
 The chemical potential $\mu_m$ satisfies the following relation with the density of the magnons
 \be
 N(\mu_m,\beta_m) = \int^\infty_{\epsilon_{\rm min}} D(\epsilon_m) n(\epsilon_m,\mu_m,\beta_m) d \epsilon_m.
 \label{N0-magnon-2}
 \ee
 Here, $D(\epsilon_m)$ is the magnon \gls{dos},
 and $\epsilon_{\rm min}$ is the energy of the lowest state.
 As $n$ and $N$ increase at
 a given temperature $\beta_m$,
 the chemical potential $\mu_m$ increases as well.
 It can been seen from \Eq{N0-magnon-1} that the magnon chemical potential $\mu_m$ cannot be larger than $\epsilon_{\rm min}$
 because the magnon number cannot be negative.
 Thus, \Eq{N0-magnon-2} with the condition $\mu_m = \epsilon_{\rm min}$ defines a critical density $N_c$.
 When $\mu_m$ reaches the minimal energy state $\epsilon_{\rm min}$
 and $N$ is larger than $N_c$,
 a large number of magnons occupy the minimal energy state leading to the formation of \gls{bec} \cite{Dzy0709C103,Dzy0764}.
 Instead, if $\mu_m < \epsilon_{\rm min}$, $N$ will be smaller than $N_c$ and the condensation does not occur.

 In practice,
 microwave pumping has been employed to generate magnons \cite{Dem06430}.
 The number of magnons generated $\delta N$ is proportional
 to the increase of the internal energy $\delta E$, i.e.,  $\delta E = h \nu_p \delta N$,
 where $\nu_p$ is the frequency of the pumped magnons.
 The increase of $\mu_m$ is associated with $\nu_p$ \cite{Dzy0764}.
 If the elevated $\mu_m$ closely approaches to $\epsilon_{\rm min}$,
 and $N$ exceeds the critical value $N_c$,
 the \gls{bec} will reappear \cite{Dzy0709C103}.

 Demidov \emph{et al.} have reported that
 non-equilibrium magnons can be generated by pure spin current,
 since the flow of the angular momentum provided by the spin current can be converted into
 magnons by the spin system of the ferromagnet \cite{Dem11107204,Dem171579}.
 Their experimental work \cite{Dem171579} has shown that a
 large spin current can drive magnons into a quasi-equilibrium state described
 by the Bose-Einstein statistics with a nonzero chemical potential.

\begin{figure}[t]
 \centering
 \includegraphics[width=0.9\columnwidth]{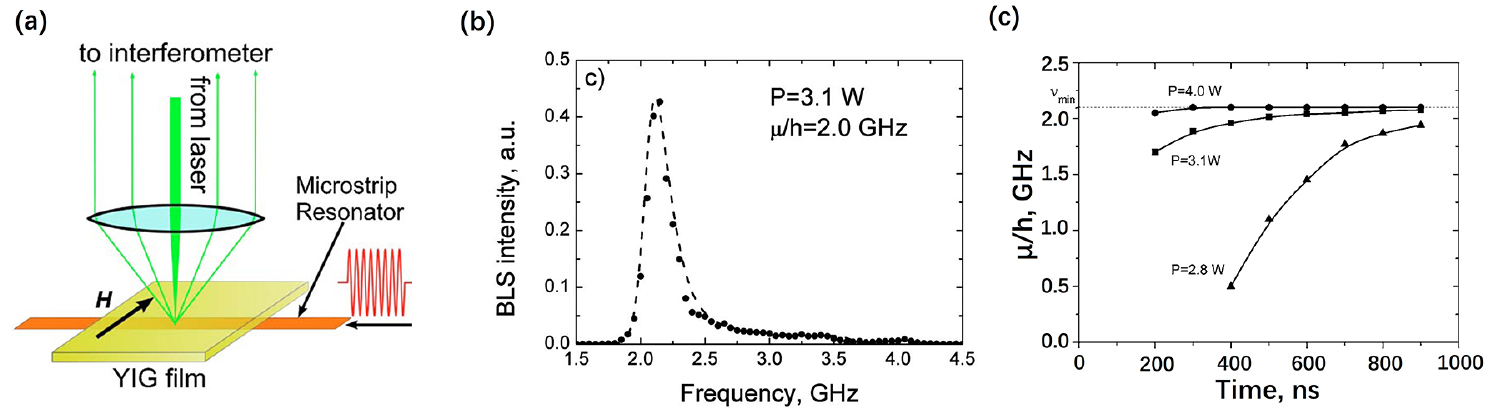}
 \caption
  {
  (a)
  Schematic of the experimental set-up for the measurement of the magnon chemical potential $\mu_m$.
  The incident laser beam is focused onto the resonator.
  The beam passing through the \gls{yig} film is reflected by the resonator, and passes through the film again.
  Then the light is collected by a wide-aperture objective lens and sent to
  the interferometer for analyzing the frequency of photons inelastically scattered by the magnons.
  A microstrip resonator attached to the \gls{yig} film creates a microwave pumping field.
  (b)
  Measured \gls{bls} spectra of magnons.
  The circles show data points recorded at the pumping power of 3.1 W.
  The dash line is the best fit of the spectra based on \Eq{local-mu-magnon-1} with $\mu_m$ being a fitting parameter.
  (c)
  Measured $\mu_m$ versus time at different values of pumping power.
  Reprinted with permission from \cite{Dzy0709C103}. Copyright 2007 American Institute of Physics.
  }
 \label{fig13-local-mu-magnon}
\end{figure}

 Using the microwave pumping, Demokritov \emph{et al.} have realized the \gls{bec} of quasi-equilibrium magnons at room temperature
 in \gls{yig} thin films \cite{Dem06430,Dzy0709C103,Dzy0764,Dem08045029}.
 They have applied the \gls{bls} technique to study the chemical potential of magnons \cite{Dzy0709C103}.
 \Figure{fig13-local-mu-magnon}(a) shows the experimental set-up for the measurement of $\mu_m$.
 An incident laser beam is focused on a local region of the sample,
 and the reflected light is collected to analysis the frequencies of photons inelastically scattered by the magnons.
 The measured \gls{bls} intensity signal is proportional to the spectral density of magnons \cite{Dem06430}
 \be
 \rho (\nu) = D(\nu)n(\nu) = \frac{D(\nu)}{e^{ {(h\nu - \mu_m)}/{k_B T_0} } - 1},
 \label{local-mu-magnon-1}
 \ee
 where $\nu$ is the magnon frequency,
 $n(\nu)$ is the Bose-Einstein distribution function with room temperature $T_0$,
 and $D(\nu)$ is the magnon \gls{dos} and can be determined prior to the measurement of $\mu_m$.
 %
%
 The dashed line in \Figure{fig13-local-mu-magnon}(b)
 is the best fit of measured signal to \Eq{local-mu-magnon-1} with $\mu_m$ being the fitting parameter.
 \Figure{fig13-local-mu-magnon}(c) illustrates the time evolution of $\mu_m$ with increasing pumping power.
 When $\mu_m$ reaches the minimal energy level $\mu_m = h \nu_{\rm min}$ the \gls{bec} is formed.
 Note that the measured \gls{bls} intensity carries the information only of a local region,
 and thus the determined $\mu_m$ is, in principle, a local chemical potential of magnons.

 The chemical potential of magnons not only can characterize the formation of Bose-Einstein condensation,
 it also plays an important role in the spin transport of magnetic insulators.
 For example,
 Cornelissen \emph{et al.} have reported that the long-range spin and heat transport in \gls{yig} films
 is mainly driven by the magnon chemical potential at room temperature \cite{Cor16014412}.

 Du \emph{et al.} have developed a single-spin magnetometery for characterizing the magnon chemical potential \cite{Du17195}.
 Since the magnon chemical potential is inherently related to spin fluctuations,
 the $\mu_m$ can be quantitatively determined by measuring the magnetic fields generated by these fluctuations.
 This is realized by using a nitrogen-vacancy center in diamond
 as a single-spin sensor to measure the vacancy spin relaxation rates $\Gamma_{\pm}$ to magnetic field fluctuations
 at the electron-spin resonance frequencies $\nu_{\pm}$.
 The magnon chemical potential can then be determined as
 \be
 \mu_m = h \nu_{\pm} \left[ 1 - \frac{\Gamma_{\pm}(0)}{\Gamma_{\pm}(\mu_m)} \right].
 \label{local-mu-magnon-2}
 \ee
 %
 Since the local magnetic fields generated by spin fluctuations are so weak,
 one has to place the magnetic field sensor in nanoscale proximity to the sample.
 Thus, this approach realizes a local measurement of magnon chemical potential with a nanoscale resolution
 determined by the distance between the sensor spin and the system under study \cite{Du17195}.


\section{Summary and perspectives} \label{sec:Sum}

 In this review, we have surveyed recent theoretical and experimental efforts
 on a topic that is of both fundamental and practical importance for the
 advancement of nanoscience: the local temperatures of nanoscopic (or mesoscopic) systems out of equilibrium.
 The contents of this review have focused on physical implications of
 the notion of local temperature in the following three aspects.
 (1) The local temperature lays the foundations for the generalization of the
 thermodynamic laws and relations to non-equilibrium scenarios.
 (2) It provides an energy scale for local thermal fluctuations and excitations, and (3) it facilitates the characterization of local properties of a
 given system
 under external driving conditions.

 In Section \ref{sec:Def}, we have reviewed a variety of theoretical definitions
 of local temperature based on some of the above physical considerations.
 In Section \ref{sec:Experiment}, we have reviewed the existing experimental strategies
 for measuring local temperatures,
 as well as some state-of-the-art techniques and instruments.
 The reviewed work covers a broad spectrum of non-equilibrium nanosystems,
 which range from single molecular junctions to two-dimensional nanosheets
 with sizes of tens or hundreds of nanometers;
 from near-equilibrium situations to those very far from equilibrium;
 from stationary states to those undergoing time evolution;
 and from systems subjected to mechanical forces or thermal gradients
 to those driven by bias voltage or laser fields.
 In Section \ref{sec:Imp&App}, we have finally discussed numerous examples
 which exemplify the physical implications and
 practical applications of the concept of local temperature.

 Despite all the exciting and encouraging progress,
 there is still a noticeable gap between theory and experiments.
 From the theoretical perspective, more attention should be paid to
 the experimental feasibility of the theoretically proposed
 definitions or protocols to determine a local temperature.
 At the same time, it is crucial to go beyond simple models
 and apply the theoretical definitions or protocols to
 predict local temperatures in realistic nano-devices under typical experimental conditions.
 From the experimental perspective,
 it is appealing, yet rather challenging to break the conventional/classical
 limits such as the local Boltzmann distribution and the Fourier's law,
 and to resolve the characteristic quantum effects in nano-sized systems. More work towards increasing the spatial (and temporal) resolution of the
 instruments and approaches discussed in this review would definitely unravel quantum features of local temperatures out of equilibrium.

 Finally, we shall mention several emerging frontiers
 where the concept of local temperature may play an essential role.
 First, quantitative predictions of the local temperature distribution
 based on first-principles simulations are expected to
 provide useful information in understanding
 novel phenomena and mechanisms in emerging quantum devices \cite{Eic17063001}.
 Second, the accurate determination of such a quantity may
 offer a new way to understand and characterize
 the formation and evolution of strongly correlated electronic states,
 which are at the heart of many exotic or complex materials \cite{Lob06301,Kir10631,Rib1350001}.
 Third, the notion of local temperature could also serve as a key quantity
 in theories of quantum thermodynamics \cite{Ali18,Ali1635568,Ali88918,Joh09041119,Bar93196,Fel03016101,Wei0830008}.
 The challenge here is to accurately account for
 the contribution of system-environment entanglement
 to the thermodynamic energy and entropy, so that
 a local temperature can be properly defined
 for an open quantum system. We thus hope this review will stimulate
 both the theoretical and experimental communities in pursuing these directions of study that are poised to lead to further breakthroughs in
 these exciting new frontiers and even beyond.

\addcontentsline{toc}{section}{Acknowledgments}

\section*{Acknowledgments}

The support from the Ministry of Science and Technology of China
(Grant Nos. 2016YFA0400900 and 2016YFA0200600)
and the National Natural Science Foundation of China (Grant Nos. 21573202 and 21973086)
is gratefully acknowledged.


\addcontentsline{toc}{section}{References}

\section*{References}

\bibliography{bibrefs}
\bibliographystyle{elsarticle-num}
\biboptions{sort&compress}



\end{document}